\documentclass[aps,prx,superscriptaddress,reprint,notitlepage]{revtex4-2}

\usepackage{bm,color}
\usepackage{graphicx}
\usepackage{amsmath, amssymb}
\usepackage{braket}
\usepackage{comment}
\usepackage[compat=1.1.0]{tikz-feynman}
\usepackage{txfonts}

\usepackage[utf8]{inputenc}

\usepackage[hidelinks]{hyperref}

\usepackage{hyperref}
\usepackage{xcolor}
\hypersetup{colorlinks=true}
\usepackage{graphicx}
\usepackage{bm}
\usepackage{amsmath}
\usepackage{amssymb}
\usepackage{xspace}
\usepackage{algorithmic}
\usepackage{algorithm}
\usepackage[capitalise]{cleveref}
\usepackage{txfonts}
\usepackage{physics}
\usepackage{siunitx}
\usepackage{booktabs}
\usepackage{chemformula}
\usepackage{ulem}
\newcommand{\subfigref}[2]{Fig.~\hyperref[#1]{\ref*{#1}#2}}
\newcommand{\pluseq}{\mathrel{+}=}

\begin{document}
\title{Fate of Topological Dirac Magnons in van der Waals Ferromagnets at Finite Temperature
}
\author{Rintaro Eto}
\email{rintaro.eto@tum.de}
\affiliation{Technical University of Munich, TUM School of Natural Sciences, Physics Department, 85748 Garching, Germany}
\affiliation{Munich Center for Quantum Science and Technology (MCQST), Schellingstraße 4, 80799 München, Germany}
\affiliation{Institute of Physics, Johannes Gutenberg University Mainz, Staudingerweg 7, 55128 Mainz, Germany}
\affiliation{Department of Applied Physics, Waseda University, Okubo, Shinjuku-ku, Tokyo 169-8555, Japan}
\author{Ignacio Salgado-Linares}
\affiliation{Technical University of Munich, TUM School of Natural Sciences, Physics Department, 85748 Garching, Germany}
\affiliation{Munich Center for Quantum Science and Technology (MCQST), Schellingstraße 4, 80799 München, Germany}
\author{Masahito Mochizuki}
\affiliation{Department of Applied Physics, Waseda University, Okubo, Shinjuku-ku, Tokyo 169-8555, Japan}
\author{Johannes Knolle}
\affiliation{Technical University of Munich, TUM School of Natural Sciences, Physics Department, 85748 Garching, Germany}
\affiliation{Munich Center for Quantum Science and Technology (MCQST), Schellingstraße 4, 80799 München, Germany}
\affiliation{Blackett Laboratory, Imperial College London, London SW7 2AZ, United Kingdom}
\author{Alexander Mook}
\affiliation{University of Münster, Institute of Solid State Theory, 48149 Münster, Germany}
\affiliation{Institute of Physics, Johannes Gutenberg University Mainz, Staudingerweg 7, 55128 Mainz, Germany}
\date{\today} 
\begin{abstract}
    Dirac magnons, the bosonic counterparts of Dirac fermions in graphene, provide a versatile platform to explore symmetry-protected band crossings and quantum geometry in magnetic insulators, while promising high-velocity, low-dissipation spin transport for next-generation magnonic technologies. However, their stability under realistic, finite-temperature conditions remains an open question. Here, we develop a microscopic theory of thermal magnon-magnon interactions in van der Waals honeycomb ferromagnets, focusing on both gapless and gapped Dirac magnons. Using nonlinear spin-wave theory with magnon self-energy corrections and a T-matrix resummation that captures two-magnon bound states, we quantitatively reproduce temperature- and momentum-dependent energy shifts and linewidths observed experimentally in the gapless Dirac magnon material CrBr$_3$, even near the Curie temperature. Our approach provides a consistent interpretation of theoretical predictions and experiment shedding light on the role of bound states in enhancing magnon damping at low temperatures. For gapped Dirac magnon materials such as CrI$_3$, CrSiTe$_3$, and CrGeTe$_3$, we find a thermally induced reduction of the topological magnon gap, while no indication of thermally driven topological phase transitions is observed within the considered parameter range. Classical atomistic spin dynamics simulations corroborate the gap’s robustness up to the Curie temperature. Furthermore, we establish a practical criterion for observing topological gaps by determining the minimum ratio of Dzyaloshinskii-Moriya interaction to Heisenberg exchange required to overcome thermal broadening throughout the ordered phase, typically around 5\%. Taken together, our results elucidate the interplay between thermal many-body effects and topology in low-dimensional magnetic systems and provide a controlled framework for the interpretation of spectroscopic measurements.
\end{abstract}
\maketitle

\section{Introduction}

Quasiparticles with linear dispersion relations—such as Dirac fermions in graphene or Weyl fermions in topological semimetals—have left a lasting mark on modern condensed matter physics. More recently, their bosonic analogs, known as \textit{Dirac magnons}, have come to the forefront as intriguing manifestations of relativistic physics in magnetic materials \cite{fransson2016, Pershoguba2018, Chen2018CrI3, yuan2020dirac, Scheie2022, Nikitin2022, Do2022}. These collective spin excitations feature linear band crossings at high-symmetry points in the Brillouin zone, which can give rise to nontrivial Berry curvature effects and topological magnon bands \cite{Katsura2010, zhang2013, mook2014edge, owerre2016first, Kim2016honey, Malki2020}. Beyond enriching our understanding of quantum magnetism, Dirac and topological magnons are also viewed as potential building blocks for low-dissipation information processing in magnonic devices \cite{Shindou13}. For a recent overview, see Ref.~\cite{mcclarty2022ar}.

Layered van der Waals (vdW) ferromagnets such as CrBr$_3$, CrI$_3$, CrSiTe$_3$, and CrGeTe$_3$ \cite{Du2015, Williams2015, Kuo2016, Lin2016, Huang2017, Gong2017, Wang2022} provide an ideal platform to explore these ideas. Their honeycomb-based magnetic structures naturally host Dirac magnon points \cite{fransson2016, Pershoguba2018}, while spin-orbit coupling can induce small band gaps, opening the door to topological magnon edge states \cite{owerre2016first, Kim2016honey}. Their 2D nature allows for monolayer isolation and inelastic electron tunneling spectroscopy experiments \cite{Klein2018, Ghazaryan2018, Kim2019}, while the bulk materials remain amenable to spectroscopic probes such as inelastic neutron scattering (INS). The latter has identified CrBr$_3$ \cite{Nikitin2022} as a virtually gapless Dirac magnon material and CrI$_3$ \cite{Chen2018CrI3}, and CrSiTe$_3$ and CrGeTe$_3$ \cite{Zhu2021topmagnon} as gapped Dirac magnon materials. We note that the finite Dirac magnon gap of CrBr$_3$ identified in Ref.~\cite{Cai2021CrBr3} was later attributed to extrinsic effects in Refs.~\cite{Nikitin2022} and \cite{Do2022}, with the latter study also showing gapless Dirac magnons in the easy-plane ferromagnet CrCl$_3$.

Although linear spin-wave theory offers a good starting point to describe magnon spectra in these systems, it does not capture the full picture. Many-body effects, that is, magnon-magnon interactions, become increasingly important at finite temperatures, leading to renormalizations of the magnon dispersion, finite lifetimes, and a redistribution of spectral weight \cite{Dyson1956, Harris1971, Kaganov1987InteractingMagnons, zhitomirskychernyshev2013}. Thus, although many-body \textit{quantum} fluctuations at zero temperature are suppressed in the ground state of ferromagnets with spin rotation symmetry and one does not have to worry about spontaneous decay \cite{zhitomirskychernyshev2013, mook2021, Gohlke2023PRL, Bao2025cuprates}, \textit{thermal} fluctuations play a central role in interpreting experimental data, especially from INS, which probes the dynamical structure factor throughout the entire Brillouin zone \cite{Bayrakci2006, Bayrakci2013, Nikitin2022}.

Thermal interaction effects in CrBr$_3$ and CrI$_3$ have previously been explored both theoretically and experimentally~\cite{Pershoguba2018,Lu2021,Nikitin2022,Sun2023,Banerjee2025}. For CrBr$_3$, which exhibits essentially gapless Dirac magnons within experimental resolution, Ref.~\cite{Pershoguba2018} pointed out a key distinction between Dirac magnons and Dirac electrons in terms of thermal scaling. That work also predicted momentum-dependent renormalization of the magnon dispersion and compared it to  experimental INS data collected in 1970s \cite{Cobb1971, Yelon1971}. More recent INS experiments \cite{Nikitin2022} confirmed the expected quadratic temperature dependence of the overall dispersion, but did not observe the predicted momentum dependence---resulting in quantitative discrepancies up to several hundred percent.
Refs.~\cite{Sun2023,Banerjee2025} also addressed the origin of thermal broadening, focusing primarily on qualitative trends rather than a detailed quantitative comparison to experiment.
Thus, a fully quantitative theoretical description of the thermal evolution of magnon dispersion and lifetime in CrBr$_3$ remains an open challenge.

In CrI$_3$, which hosts gapped, topologically nontrivial Dirac magnons \cite{Chen2018CrI3}, the theoretical picture is less settled. One study \cite{Lu2021} proposed that thermal interactions could drive a closing and reopening of the magnon gap---effectively a temperature-driven topological transition. However, this result depends sensitively on the approximations used \cite{CommentLu2021, ReplyLu2021}, and it remains an open question whether such a transition reflects an intrinsic physical effect or is sensitive to the underlying theoretical approximations. Similar predictions have since been extended to other systems \cite{Li2023, Zhu2024inter, Shi2024}, calling for a robust justification of the adopted approach.

Overall, a consistent and quantitative understanding of how two-dimensional magnons evolve with temperature is still lacking. Theoretical treatments often employ approximations whose range of validity can be challenging to assess systematically---for example, by simplifying scattering matrix elements, using low-temperature expansions outside their valid regime, or neglecting higher-order effects while invoking self-consistency for lower-order ones. These \textit{ad hoc} elements can lead to widely varying predictions and hinder reliable comparison with experiment. To date, Ref.~\cite{Nikitin2022} provides the most comprehensive experimental characterization of the thermal evolution of the magnon spectrum across the full Brillouin zone in a vdW ferromagnet---and key aspects of these data remain unexplained.

In this work, we present a quantitative microscopic theory of thermal magnon-magnon interactions in vdW ferromagnets such as CrBr$_3$, CrI$_3$, CrSiTe$_3$, and CrGeTe$_3$. Employing nonlinear spin-wave theory at several levels of approximation, we compute magnon self-energies and extract temperature-dependent shifts and broadenings of the magnon spectrum. Our results reproduce the main features of the INS data for bulk CrBr$_3$ \cite{Nikitin2022} throughout the entire Brillouin zone.
After benchmarking our approach with CrBr$_3$, we apply it to the gapped Dirac systems CrI$_3$, CrSiTe$_3$, and CrGeTe$_3$, where the gap arises from spin-orbit coupling, which enters in the form of Dzyaloshinskii-Moriya interaction (DMI) \cite{Chen2021DMI, Zhu2021topmagnon}. We find no evidence of a thermally driven topological transition: in all three materials, the magnon gap remains open up to the magnetic transition temperature. We further identify a critical ratio of DMI to nearest-neighbor Heisenberg exchange above which the magnon gap exceeds the lifetime broadening throughout the ordered phase up to the Curie temperature. A DMI strength of approximately $5\,\%$ relative to the exchange coupling is sufficient to meet this condition. This criterion is roughly satisfied in CrI$_3$, CrSiTe$_3$, and CrGeTe$_3$, suggesting that the topologically enforced chiral edge magnons within the gap should produce clear, albeit broadened, spectral signatures in experiments, distinct from the bulk states.
Note the difference to the case of quantum damping induced by magnon number-nonconserving two-magnon scattering processes, in which the decay rate of chiral edge magnons is proportional to square of the magnon number-nonconserving DMI strength $D_\perp$~\cite{Habel2024}.

\begin{figure}
    \centering
    \includegraphics[scale=1.0]{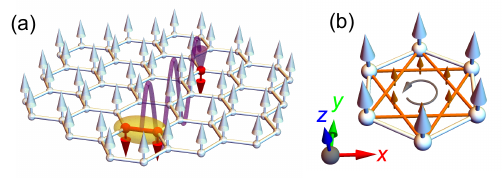}
    \caption{Sketch of van der Waals ferromagnets on a honeycomb lattice. (a) Honeycomb-lattice ferromagnet with a single spin-flip and a bound pair of spin-flip excitations. When delocalized over the entire lattice, these correspond to single-magnon excitations and two-magnon bound states, respectively. The wiggly line indicates that these excitations interact with each other at finite temperatures, as captured by a T-matrix resummation. (b) Unit hexagon of the honeycomb lattice. Orange bonds and arrows represent the next-nearest-neighbor Dzyaloshinskii-Moriya interaction, where, by definition, all Dzyaloshinskii-Moriya vectors (small orange arrows) point upward for counterclockwise circulation.} 
    \label{Fig01}
\end{figure}

The Cr-based materials studied here host spins of $S = 3/2$, which reduces quantum fluctuations and places the system near the classical limit—yet still calls for a careful assessment of quantum effects. To complement our spin-wave analysis, we first perform classical Landau-Lifshitz spin dynamics simulations. These support our overall findings but, by construction, neglect quantum corrections. To incorporate the residual quantumness of the excitations---even in the presence of a fully polarized ground state---our approach goes beyond simple perturbative treatments of the magnon self-energy, which are known to be unreliable at short wavelengths \cite{Silberglitt1967}. Specifically, we apply a \textit{non-perturbative $T$-matrix resummation} inspired by seminal work from the 1960s and 1970s \cite{Silberglitt1967, Silberglitt1968, Cobb1971, Yelon1971} and their recent applications~\cite{Nomura2001,Nagao2007,Fauseweh2015}, which effectively captures two-magnon bound states and resonances within the continuum. These many-body features open additional thermal scattering channels for single magnons [see Fig.~\ref{Fig01}(a)], leading to distinct signatures in both lifetimes and energy renormalizations at low temperatures. While these effects are most pronounced in systems with $S=1/2$, since smaller $S$ causes a more efficient binding of magnons, our results show that even in CrBr$_3$ with $S=3/2$, interactions with two-magnon bound states may lead to characteristic linewidth broadening signatures over a broad temperature range. 

In summary, by addressing previously noted discrepancies between theory and experiment  for CrBr$_3$, our work provides a reliable microscopic foundation for interpreting magnon spectra in current and future van der Waals magnets. This foundation clarifies the stability of Dirac and topological magnons at finite temperature and provides a rule of thumb for how large the spin-orbit interaction must be to render chiral edge magnons sufficiently robust throughout the entire ordered phase. These findings deepen our understanding of interacting bosonic quasiparticles, and help guide future efforts in magnon-based information technologies and spin caloritronics.

The rest of this paper is organized as follows.
In Sec.~\ref{sec:Model_and_Methods}, after introducing a spin model that describes magnetic excitations in vdW honeycomb ferromagnets (Sec.~\ref{sec:spin_model}), we present the interacting spin-wave theory (Secs.~\ref{sec:HolsteinPrimakoff}--\ref{sec:spectral_function}) with several different levels of approximation (Sec.~\ref{sec:list_of_approximations}), including the T-matrix resummation (Sec.~\ref{sec:resummation}). 

Two-magnon bound states, effectively captured by the nonperturbative T-matrix resummation, are shown to contribute to the renormalization of single-magnon excitations in CrBr${}_3$ (Sec.~\ref{sec:CrBr3}). We begin with the kinematics of unbound multimagnon excitations (Sec.~\ref{sec:CrBr3_multiparticle_continua}), and then discuss how two-magnon bound states can renormalize single-magnon spectra (Sec.~\ref{sec:CrBr3_twomagnonBS}).
By comparing INS data for bulk CrBr${}_3$ reported in Ref.~\cite{Nikitin2022} across a wide temperature range with the results of our interacting spin-wave analysis including the T-matrix resummation, we uncover possible contributions from bound states (Secs.~\ref{sec:CrBr3_T_ll_1} and \ref{sec:CrBr3_T_lesssim_1}).
Moreover, the robustness of the Dirac points even in the vicinity of $T_\textrm{C}$ is examined (Sec.~\ref{sec:CrBr3_Dirac_points}).
Some of our findings based on the interacting spin-wave theory are further corroborated within the level of classical atomistic spin dynamics (Sec.~\ref{sec:CrBr3_classical_spin_dynamics}).

CrI${}_3$ serves as an example demonstrating that the topological gap induced by the DMI remains robust up to the Curie temperature (Sec.~\ref{sec:CrI3}). After recapping previous disputes over the presence or absence of thermally induced topological transitions (Sec.~\ref{sec:CrI3_recap}), we present a comprehensive analysis based on the T-matrix resummation, which supports their absence (Sec.~\ref{sec:CrI3_absence_of_gap_closing}). Besides, a realistic criterion to observe the robust topological gap up to $T_\textrm{C}$ is estimated (Sec.~\ref{sec:CrI3_observability_of_gap}).

Additional discussions toward more precise $T_\textrm{C}$ predictions and other promising candidate materials are presented in Sec.~\ref{sec:discussion}. We conclude by summarizing this work in Sec.~\ref{sec:summary}. Technical details of the theoretical frameworks used are provided in the Appendices~\ref{Appendix:SpinBosontransform}--\ref{Appendix:Hartree_fullSunset_tempev}.
\section{Model and Methods}
\label{sec:Model_and_Methods}
\subsection{Spin model}
\label{sec:spin_model}
\begin{figure}
    \centering
    \includegraphics[scale=1.0]{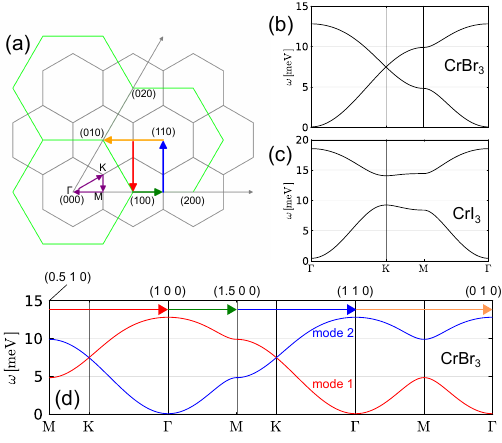}
    \caption{(a) High-symmetry momentum paths used in this work, inspired by the experimental path of Ref.~\cite{Nikitin2022}. Small hexagons represent the repeated Brillouin zones of the honeycomb lattice, while the larger green hexagons indicate the extended Brillouin zones reflecting the spatial periodicity of the structure factors due to the non-Bravais nature of the honeycomb lattice. The purple arrows depict the conventional $\Gamma$-K-M-$\Gamma$ path. (b,c) Linear spin-wave dispersion relations of (b) CrBr$_3$ and (c) CrI$_3$, with parameters given in the main text. (d) Schematic illustration defining modes 1 and 2, used in the discussion of neutron scattering results for CrBr$_3$ in Ref.~\cite{Nikitin2022}. Red, green, blue, and orange arrows correspond to the momentum paths of the same colors shown in (a).} 
    \label{Fig02}
\end{figure}
To describe the vdW ferromagnets CrBr$_3$, CrI$_3$, CrSiTe$_3$, and CrGeTe$_3$ we adopt an established spin model on the 2D honeycomb lattice \cite{Chen2018, Zhu2021topmagnon, Nikitin2022}. As indicated in Fig.~\ref{Fig01}(b), the spin Hamiltonian is given by
\begin{align}
\begin{aligned}
    \mathcal{H}_{\rm Spin} =& \sum_{n=1}^3 J_n \sum_{\langle\mathbf{r},\mathbf{r}'\rangle\in n{\rm NN}} \mathbf{S}_{\mathbf{r}}\cdot\mathbf{S}_{\mathbf{r}'} - A\sum_\mathbf{r} \left(S_\mathbf{r}^z\right)^2 \\
    &\quad + \sum_{\langle\mathbf{r},\mathbf{r}'\rangle\in {\rm 2NN}} 
    D_z \nu_{ij} \mathbf{e}_z \cdot \mathbf{S}_\mathbf{r}\times\mathbf{S}_{\mathbf{r}'},
    \label{eq:SpinHm}
\end{aligned}
\end{align}
where $\mathbf{S}_\mathbf{r}=\left(S_\mathbf{r}^x,S_\mathbf{r}^y,S_\mathbf{r}^z\right)^\top$ denotes the vectorial spin-$S$ operators on an atomic site $\mathbf{r}$. The first term denotes the isotropic exchange interactions on the $n$-th nearest-neighbor ($n$NN) bonds. Here we consider up to the 3NN bonds so that $n=1,2,3$. The second term with $A>0$ denotes the easy-axis single-ion anisotropy along the perpendicular direction to the layer plane. The last term describes the 2NN Dzyaloshinskii-Moriya interaction (DMI), where $\nu_{ij} = \pm 1$. Positive (negative) sign applies to going in counter-clockwise (clockwise) around a hexagon. The ferromagnetically aligned, fully polarized magnetic order along the perpendicular direction to the layer plane is the exact ground state, i.e., $\ev{S_\mathbf{r}}=S(0,0,1)^\top$ independent of $\mathbf{r}$ (the superscript ``$\top$'' denotes transposition).
\subsection{Spin-to-boson transformation}
\label{sec:HolsteinPrimakoff}
The Hamiltonian $\mathcal{H}_{\rm Spin}$ in Eq.~\eqref{eq:SpinHm} is bosonized by applying the Holstein-Primakoff (HP) transformation \cite{holsteinprimakoff1940} given by
\begin{equation}
\begin{aligned}
    \mathbf{S}_\mathbf{r} &= \mathbf{e}^0_\mathbf{r} \left( S-\hat{a}^\dagger_\mathbf{r}\hat{a}_\mathbf{r} \right) \\ &\quad
    + \mathbf{e}^-_\mathbf{r} \sqrt{S-\frac{\hat{a}^\dagger_\mathbf{r}\hat{a}_\mathbf{r}}{2}} \hat{a}_\mathbf{r} 
    + \mathbf{e}^+_\mathbf{r} \hat{a}^\dagger_\mathbf{r} \sqrt{S-\frac{\hat{a}^\dagger_\mathbf{r}\hat{a}_\mathbf{r}}{2}},
    \label{eq:HPtransform}
\end{aligned}
\end{equation}
which is an expansion around the ground state. For the fully polarized out-of-plane ferromagnetic order, i.e., $\ev{S_\mathbf{r}}=S(0,0,1)^\top$, the unit vectors are given by $\mathbf{e}_\mathbf{r}^\pm=(1,\pm i,0)/\sqrt{2}$ and $\mathbf{e}_\mathbf{r}^0=(0,0,1)$ independent of $\mathbf{r}$. The bosonic operators obey the usual commutation relation $[\hat{a}_\mathbf{r},\hat{a}^\dagger_{\mathbf{r}'}]=\delta(\mathbf{r}-\mathbf{r}')$. A Taylor expansion of the square roots in Eq.~(\ref{eq:HPtransform}) followed by normal ordering of operators gives a $1/S$ series expression of the Hamiltonian
\begin{equation}
    \mathcal{H}_\textrm{Magnon} = \sum_{n=0}^\infty S^{2-n/2} \mathcal{H}^{(2n)},
    \label{eq:HmSeries}
\end{equation}
where $\mathcal{H}^{(2n)}$ consists of terms with $n$ creation and $n$ annihilation operators. 
All odd-order contributions $\mathcal{H}^{(2n-1)}$ $(n\in\mathbb{Z})$ are absent because of expanding around a stable state ($\mathcal{H}^{(1)}=0$) and the $SO(2)$ symmetry around the $z$-axis of both $\mathcal{H}_\mathrm{Spin}$ and its ground state. As a result, there is a $U(1)$ symmetry for the magnons and their number is conserved.
\subsection{Linear spin-wave theory}
\label{sec:LSW}
To study the magnon excitation spectrum of $\mathcal{H}_\mathrm{Spin}$, we first focus on the bilinear term $\mathcal{H}^{(2)}$ in Eq.~(\ref{eq:HmSeries}), where we recover results known from Refs.~\cite{Kim2016honey,owerre2016first, Pershoguba2018}. A momentum-space representation of $\mathcal{H}^{(2)}$ is given by
\begin{equation}
\begin{aligned}
    \mathcal{H}^{(2)} &= \sum_\mathbf{k} \hat{\varphi}^\dagger_\mathbf{k} \mathcal{H}^{(2)}_\mathbf{k} \hat{\varphi}_\mathbf{k} \\
    &=\sum_\mathbf{k} \hat{\varphi}^\dagger_\mathbf{k}
    \left( \begin{array}{cc}
        J_\mathrm{intra}(\mathbf{k})+D(\mathbf{k}) & J_\mathrm{inter}(\mathbf{k}) \\ J_\mathrm{inter}^*(\mathbf{k}) & J_\mathrm{intra}(\mathbf{k})-D(\mathbf{k})
    \end{array} \right)
    \hat{\varphi}_\mathbf{k},
    \label{eq:H2_matrix}
\end{aligned}
\end{equation}
where $\hat{\varphi}_\mathbf{k}=\left( \hat{a}_{\mathbf{k}\mathrm{A}} \ \hat{a}_{\mathbf{k}\mathrm{B}} \right)^\top$, and $\mathbf{k}=(k_x,k_y)^\top$ denotes a crystal moment. The matrix entries of $\mathcal{H}^{(2)}_\mathbf{k}$ are given by
\begin{equation}
\begin{aligned}
    J_\mathrm{intra}(\mathbf{k}) &= -3J_1-3J_3-2J_2\left(3-\sum_{i=1}^3\cos(\mathbf{k}\cdot{\bm \zeta}_i)\right) + A\left(2-\frac{1}{S}\right), \\
    J_\mathrm{inter}(\mathbf{k}) &= J_1 \sum_{i=1}^3 e^{i\mathbf{k}\cdot{\bm \delta}_i} + J_3 \sum_{i=1}^3 e^{i\mathbf{k}\cdot{\bm \xi}_i}, \\
    D(\mathbf{k}) &= 2D_z\sum_{i=1}^3 \sin(\mathbf{k}\cdot{\bm \zeta}_i).
\end{aligned}
\end{equation}
Note that ${\bm \delta}_i$, ${\bm \zeta}_i$, and ${\bm \xi}_i$ $(i=1,2,3)$ denote directions of 1NN, 2NN, and 3NN bonds on the 2D honeycomb lattice, respectively given by
\begin{equation}
\begin{aligned}
    {\bm \delta}_1 &= \left(0,1/\sqrt{3}\right)^\top, \\
    {\bm \delta}_2 &= \left(-1/2,-1/2\sqrt{3}\right)^\top, \\
    {\bm \delta}_3 &= \left(1/2,-1/2\sqrt{3}\right)^\top,
\end{aligned}
\end{equation}
and
\begin{equation}
\begin{aligned}
    {\bm \zeta}_1 &= \left(1,0\right)^\top, \\
    {\bm \zeta}_2 &= \left(-1/2,\sqrt{3}/2\right)^\top, \\
    {\bm \zeta}_3 &= \left(-1/2,-\sqrt{3}/2\right)^\top,
\end{aligned}
\end{equation}
and
\begin{equation}
\begin{aligned}
    {\bm \xi}_1 &= \left(1,1/\sqrt{3}\right)^\top, \\
    {\bm \xi}_2 &= \left(-1,1/\sqrt{3}\right)^\top, \\
    {\bm \xi}_3 &= \left(0,-2/\sqrt{3}\right)^\top.
\end{aligned}
\end{equation}
We diagonalize the quadratic Hamiltonian given by Eq.~(\ref{eq:H2_matrix}) by applying a unitary transformation from the atomic basis $\hat{\varphi}_\mathbf{k}$ to the normal basis $\hat{\psi}_\mathbf{k}=\mathcal{U}_\mathbf{k}^\dagger\hat{\varphi}_\mathbf{k}$ where $\hat{\psi}_\mathbf{k}=(\hat{b}_{\mathbf{k}-} \ \hat{b}_{\mathbf{k}+})^\top$. Eigenvalues $\mathcal{E}_\mathbf{k}=\mathcal{U}^\dagger_\mathbf{k}\mathcal{H}^{(2)}_\mathbf{k}\mathcal{U}_\mathbf{k}=\textrm{diag}(\varepsilon_{\mathbf{k}-},\varepsilon_{\mathbf{k}+})$ are given in the form
\begin{equation}
    \omega_{\mathbf{k}\mp} = S  \varepsilon_{\mathbf{k}\mp} = S \left(J_\mathrm{intra}(\mathbf{k}) \mp \sqrt{\left| J_\mathrm{inter}(\mathbf{k}) \right|^2 + D(\mathbf{k})^2} \right).
\end{equation}
\par Along the high-symmetry path indicated in Fig.~\ref{Fig02}(a), which was chosen to stay as close as possible to the experiment in Ref.~\cite{Nikitin2022}, the linear spin-wave spectrum of CrBr$_3$ is shown in Fig.~\ref{Fig02}(b), where the follwing parameter set was employed:
\begin{equation}
\begin{aligned}
    S &= 3/2, \\
    J_1 &= -1.485 \ [\textrm{meV}], \\
    J_2 &= -0.077 \ [\textrm{meV}], \\
    J_3 &= +0.068 \ [\textrm{meV}], \\
    A &= +0.028 \ [\textrm{meV}]. \\
\end{aligned}
\end{equation}
The exchange parameters $J_1$, $J_2$, $J_3$ were obtained from fitting of the INS data in Ref.~\cite{Nikitin2022}. This reference has also estimated the single-ion anisotropy $A$ from previous ferromagnetic resonance measurements in Ref.~\cite{Alyoshin1997}. A linear crossing of magnon bands, i.e., a Dirac magnon, is observed at the K point in Fig.~\ref{Fig02}(b).
\par The linear magnon bands of CrI$_3$,  plotted in Fig.~\ref{Fig02}(c) with parameter set~\cite{Chen2018,Lu2021}
\begin{equation}
\begin{aligned}
    S &= 3/2, \\
    J_1 &= -2.01 \ [\textrm{meV}], \\
    J_2 &= -0.16 \ [\textrm{meV}], \\
    D_z &= -0.31 \ [\textrm{meV}], \\
    A &= +0.22 \ [\textrm{meV}], \\
\end{aligned}
\end{equation}
exhibit a sizable topological gap of approximately 5~meV associated with nonzero $D_z$. The lower and upper bands have a finite Chern number of $\pm 1$, and chiral edge magnons are expected for open boundary conditions (e.g., see Refs.~\cite{owerre2016first, Kim2016honey}). Note that for both materials, any parameters not explicitly mentioned are set to zero.
\subsection{Many-body perturbation theory}
\label{sec:perturbation_theory}
We apply the unitary transformation discussed in the previous subsection to $\mathcal{H}^{(4)}$, and get its normal form given by
\begin{equation}
\begin{aligned}
\mathcal{H}^{(4)} &= \sum_{\mathbf{q}_1\mathbf{q}_2\mathbf{q}_3} \sum_{\nu_1\nu_2\nu_3\nu_4} \mathcal{Q}_{\mathbf{q}_1,\mathbf{q}_2\leftrightarrow\mathbf{q}_3,\mathbf{q}_4}^{\nu_1,\nu_2\leftrightarrow\nu_3,\nu_4} \hat{b}^\dagger_{\mathbf{q}_1\nu_1}\hat{b}^\dagger_{\mathbf{q}_2\nu_2}\hat{b}_{\mathbf{q}_3\nu_3}\hat{b}_{\mathbf{q}_4\nu_4} \\
&\quad\quad \times \delta(\mathbf{q}_1+\mathbf{q}_2-\mathbf{q}_3-\mathbf{q}_4).
\end{aligned}
\end{equation}
The analytical expression for the four-magnon vertex $\mathcal{Q}_{\mathbf{q}_1,\mathbf{q}_2\leftrightarrow\mathbf{q}_3,\mathbf{q}_4}^{\nu_1,\nu_2\leftrightarrow\nu_3,\nu_4}$ is provided in Appendix~\ref{Appendix:T-matrix}. Note that all terms in Eq.~(\ref{eq:SpinHm}), i.e., $J_n$ $(n=1,2,3)$, $D_z$, and $A$, contribute to the vertex and that the momentum conservation is to be understood modulo a reciprocal lattice vector. \par
\begin{figure}
    \centering
    \includegraphics[scale=1.0]{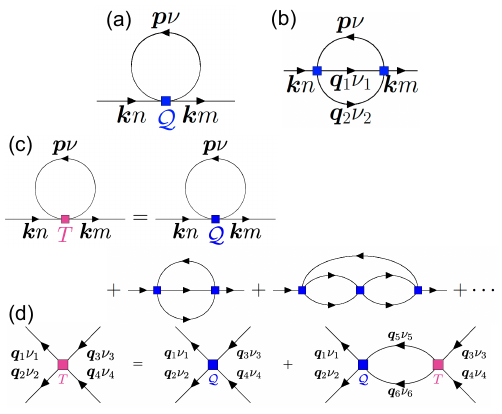}
    \caption{Overview of nonlinear spin-wave approximations. (a) The Hartree diagram of order $1/S$. (b) The sunset diagram of order $1/S^2$. (c,d) Diagramatic expressions of (c) the resummation of the self-energy and (d) the Bethe-Salpeter equation given by Eq.~\eqref{eq:BSE}.}
    \label{Fig03}
\end{figure}
We consider a many-body perturbation theory with the nonperturbative term $\mathcal{H}^{(2)}$ and the perturbation $\mathcal{H}^{(4)}$. Note that, unless stated otherwise, the higher-order terms like $\mathcal{H}^{(6)}$ are neglected. We introduce a textbook temperature Green's function given by \cite{Rastelli2011}
\begin{align}
\begin{aligned}
    \mathcal{G}_{\mathbf{k},mn}(\tau) &= -\sum_{j=0}^\infty \left( -\frac{1}{\hbar} \right)^j\frac{1}{j!} \int_0^{\beta\hbar} d\tau_1 \cdots \int_0^{\beta\hbar} d\tau_j \\
    &\quad \times\Braket{\mathcal{T}\left[ \mathcal{H}_4(\tau_1) \cdots \mathcal{H}_4(\tau_j) \hat{b}_{\mathbf{k}m}(\tau)\hat{b}^\dagger_{\mathbf{k}n}(0) \right]}_0^{\rm c},
    \label{eq:TemperatureGreenF}
\end{aligned}
\end{align}
where $\mathcal{T}$ orders imaginary times $\tau_j$, and $\beta=1/k_\mathrm{B}T$ denotes inverse temperature, for which we adopt the convention $k_\mathrm{B}=1$ hereafter. In the following, we evaluate the time-ordered ensemble of the connected diagrams $\ev{\cdots}_0^c$ in Eq.~(\ref{eq:TemperatureGreenF}) using several approximations at different levels of the $1/S$ expansion.
\par \textit{Hartree diagram.-----}
First, we consider the leading-order $1/S$ contribution to the self-energy, which is described by the Hartree diagram presented in Fig.~\ref{Fig03}(a). The corresponding frequency-independent Hartree self-energy is given by
\begin{equation}
    \Sigma_{\mathbf{k},mn}^{(\textrm{Hartree})} = \frac{4}{\hbar} \sum_\mathbf{p} \sum_\nu \mathcal{Q}_{\mathbf{k},\mathbf{p}\leftrightarrow\mathbf{k},\mathbf{p}}^{m,\nu\leftrightarrow n,\nu} n_{\mathbf{p}\nu}^{(0)}, \label{eq:Hartree}
\end{equation}
where $n^{(0)}_{\mathbf{k}\nu}=\left[\exp(\beta\omega_{\mathbf{k}\nu})-1\right]^{-1}$ denotes the Bose-Einstein distribution function. Notably, the Hartree self-energy is purely real, indicating that it contributes to the frequency shift of magnons, without affecting their lifetime.
\par \textit{Sunset diagram.-----} Next we consider subleading $1/S^2$ contributions to the self-energy. While there are several scattering processes with contributions of order $1/S^2$, arising either from a second-order process in $\mathcal{H}^{(4)}$ or a first-order process in $\mathcal{H}^{(6)}$, the only one that includes a first-order contribution of the Bose-Einstein distribution function is given by the sunset diagram shown in Fig.~\ref{Fig03}(b). The explicit form of the frequency-dependent sunset self-energy is given by
\begin{equation}
\begin{aligned}
    \Sigma_{\mathbf{k},mn}^\textrm{(Sunset)}(i\omega_s) &= \frac{8}{\hbar^2} \sum_\mathbf{p}\sum_\nu\sum_{\mathbf{q}_1,\mathbf{q}_2}\sum_{\nu_1\nu_2} \mathcal{Q}_{\mathbf{k},\mathbf{p}\leftrightarrow\mathbf{q}_1,\mathbf{q}_2}^{m,\nu\leftrightarrow \nu_1,\nu_2}
    \mathcal{Q}_{\mathbf{q}_1,\mathbf{q}_2\leftrightarrow\mathbf{k},\mathbf{p}}^{\nu_1,\nu_2\leftrightarrow n,\nu} \\
    &\quad \times \frac{n^{(0)}_{\mathbf{p}\nu}\left(1+n^{(0)}_{\mathbf{q}_1\nu_1}+n^{(0)}_{\mathbf{q}_2\nu_2}\right)-n^{(0)}_{\mathbf{q}_1\nu_1}n^{(0)}_{\mathbf{q}_2\nu_2}}{i\omega_s-\omega_{\mathbf{q}_1\nu_1}-\omega_{\mathbf{q}_2\nu_2}+\omega_{\mathbf{p}\nu}} \\
    &\quad \times \delta(\mathbf{k}+\mathbf{p}-\mathbf{q}_1-\mathbf{q}_2),
    \label{eq:SE_sunset}
\end{aligned}
\end{equation}
where $\omega_s=2\pi s/\beta$ $(s\in\mathbb{Z})$ denotes the bosonic Matsubara frequency. This sunset self-energy is usually complex and thus accounts not only for frequency shift but also for finite lifetimes of magnons. Importantly, the sunset self-energy is the leading contribution to the magnon lifetimes and the only contribution at order $1/S^2$. Consequently, the imaginary part of the self-energy can be evaluated exactly up to this order by considering only the sunset diagram. 
\par \textit{Reduced sunset diagram.}----- In Ref.~\cite{Pershoguba2018} a low-temperature approximation to efficiently evaluate $\Sigma_{\mathbf{k},mn}^{(\textrm{Sunset})}$given in Eq.~(\ref{eq:SE_sunset}) was suggested: omitting the $(\mathbf{p},\nu)$-sum and instead taking only the $(\mathbf{p}\approx\mathbf{0},\nu=-)$-contribution into account. Within this approximation and the isotropic limit ($\omega_{\mathbf{0}-}\rightarrow0$, i.e., $A\rightarrow0$), one gets the \textit{reduced} sunset self-energy given by \cite{Pershoguba2018}
\begin{equation}
    \Sigma_{\mathbf{k},mn}^\textrm{(R-Sunset)}(i\omega_s) = \frac{8\tilde{C}T^2}{\hbar^2} \sum_{\mathbf{q}}\sum_{\nu_1\nu_2} \frac{\left[\mathsf{Q}_{\mathbf{k};\mathbf{q},\mathbf{k}-\mathbf{q}}^{m;\nu_1,\nu_2}\right]^* \mathsf{Q}_{\mathbf{k};\mathbf{q},\mathbf{k}-\mathbf{q}}^{n;\nu_1,\nu_2}}{i\omega_s-\omega_{\mathbf{q}\nu_1}-\omega_{\mathbf{k}-\mathbf{q}\nu_2}}
    \label{eq:SE_reduced_sunset}
\end{equation}
where $\mathsf{Q}_{\mathbf{k};\mathbf{q}_1,\mathbf{q}_2}^{\nu;\nu_1,\nu_2}$ is a \textit{reduced} four-magnon vertex (see Ref.~\cite{Pershoguba2018} for details and Ref.~\cite{Liu2023MnBi2Te4} for an application to ferromagnets on the triangular lattice). $\tilde{C}$ is a constant prefactor.
The imaginary part of the \textit{reduced} sunset self-energy is further simplified to
\begin{equation}
\begin{aligned}
    &\quad \textrm{Im} \left[\Sigma_{\mathbf{k},nn}^\textrm{(R-Sunset)}(\omega+i0^+) \right] \\
    &= -\frac{8\pi\tilde{C}T^2}{\hbar^2} \sum_{\mathbf{q}}\sum_{\nu_1\nu_2} \left| \mathsf{Q}_{\mathbf{k};\mathbf{q},\mathbf{k}-\mathbf{q}}^{n;\nu_1,\nu_2} \right|^2
    \delta(\omega-\omega_{\mathbf{q}\nu_1}-\omega_{\mathbf{k}-\mathbf{q}\nu_2})
    \label{eq:SE_reduced_sunset_2MDOS}
\end{aligned}
\end{equation}
after analytic continuation $i\omega_s\rightarrow\omega+i0^+$.
The delta function part in Eq.~(\ref{eq:SE_reduced_sunset_2MDOS}) indicates that the reduced sunset diagram accounts for interactions between magnons and the two-magnon density of states given by
\begin{equation}
    \mathcal{D}^{(2)}_\mathbf{k}(\omega) = \frac{1}{N_\mathrm{muc}} \sum_\mathbf{q}\sum_{\nu_1\nu_2} \delta\left( \omega-\omega_{\mathbf{q}\nu_1}-\omega_{\mathbf{k}-\mathbf{q}\nu_2} \right),
    \label{eq:2magDOS}
\end{equation}
where $N_\mathrm{muc}$ denotes the number of magnetic unit cells in the system considered. Consequently, this approximation is by construction expected to imprint van Hove singularities in the two-magnon density of states onto the single-magnon spectra.

\subsection{Resummation}
\label{sec:resummation}
Typically, Feynman diagrammatic contributions to the magnon self-energy of order $(1/S)^n$ involve at most $n$ Bose distribution factors. This implies that the conventional $1/S$ expansion tends to converge at low temperatures, where $n^{(0)}_{\mathbf{k}\nu} \ll 1$ holds for all momenta $\mathbf{k}$ and band indices $\nu$. In such regimes, a \textit{truncated} expansion up to a finite order in $1/S$ generally provides reliable results.
However, at higher temperatures, there exist momenta $\mathbf{k}$ and band indices $\nu$ where $n^{(0)}_{\mathbf{k}\nu} \ll 1$ no longer holds. This tendency is expected to be more significant in smaller spin systems such as those with $S=1/2$ because the total magnon bandwidth and critical temperature are smaller. As a result, the truncated expansion---such as the Hartree or Hartree+sunset approximations---may become inaccurate, particularly by overestimating the self-energies. This overestimation can be especially problematic for the imaginary part of the self-energy, since all diagrammatic contributions to it are strictly negative due to causality constraints.
This issue of convergence can be partially mitigated by employing the resummation scheme introduced in this Section and schematically shown in Fig.~\ref{Fig03}(c). Moreover, comparing the results obtained via truncated expansions and the resummation offers a useful diagnostic for assessing the convergence behavior of the $1/S$ expansion. In particular, examining discrepancies in the imaginary part of the self-energy is especially informative, as it directly reflects the cumulative effect of contributions with a fixed negative sign.
To improve the accuracy of the spin-wave approximation, we consider a systematic summation of an infinite number of diagrams that obey certain rules, independent of the order of $1/S$. Following the seminal works of the 1960s and 1970s \cite{Silberglitt1967, Silberglitt1968, Cobb1971, Yelon1971}, this is achieved by employing a resummation technique based on the Bethe-Salpeter equation (BSE) for the ladder diagrams shown in Fig.~\ref{Fig03}(d):  
\begin{align}
\begin{aligned}
    &\quad T_{\mathbf{q}_1,\mathbf{q}_2\leftrightarrow\mathbf{q}_3,\mathbf{q}_4}^{\nu_1,\nu_2\leftrightarrow\nu_3,\nu_4}\left(z\right) 
    = \mathcal{Q}_{\mathbf{q}_1,\mathbf{q}_2\leftrightarrow\mathbf{q}_3,\mathbf{q}_4}^{\nu_1,\nu_2\leftrightarrow\nu_3,\nu_4} \\
    & + \frac{2}{\hbar} \sum_{\mathbf{q}_5}\sum_{\nu_5,\nu_6} \mathcal{Q}_{\mathbf{q}_1,\mathbf{q}_2\leftrightarrow\mathbf{q}_5,\mathbf{q}_6}^{\nu_1,\nu_2\leftrightarrow\nu_5,\nu_6} \frac{1+n^{(0)}_{\mathbf{q}_5\nu_5}+n^{(0)}_{\mathbf{q}_6\nu_6}}{z-\omega_{\mathbf{q}_5\nu_5}-\omega_{\mathbf{q}_6\nu_6}} T_{\mathbf{q}_5,\mathbf{q}_6\leftrightarrow\mathbf{q}_3,\mathbf{q}_4}^{\nu_5,\nu_6\leftrightarrow\nu_3,\nu_4}\left(z\right),
    \label{eq:BSE}
\end{aligned}
\end{align}
where $\mathbf{q}_4 = \mathbf{q}_1+\mathbf{q}_2-\mathbf{q}_3$, $\mathbf{q}_6 = \mathbf{q}_1+\mathbf{q}_2-\mathbf{q}_5$, and $z\in\mathbb{C}$. The explicit solution of the BSE, i.e., the T-matrix $T_{\mathbf{q}_1,\mathbf{q}_2\leftrightarrow\mathbf{q}_3,\mathbf{q}_4}^{\nu_1,\nu_2\leftrightarrow\nu_3,\nu_4}\left(z\right)$ in Eq.~(\ref{eq:BSE}), can be derived analytically, which is presented in Appendix~\ref{Appendix:T-matrix} (see also Ref.~\cite{Rastelli2011}).
Finally, the T-matrix self-energy within the short-wavelength limit $k_\mathrm{B}T\ll\hbar\omega_{\mathbf{k}\nu}$ is given by
\begin{equation}
    \Sigma^{{\rm (T)}}_{\mathbf{k},mn}(i\omega_s) \approx
    \frac{4}{\hbar}\sum_\mathbf{p}\sum_\nu T_{{\mathbf{k},\mathbf{p}\leftrightarrow\mathbf{k},\mathbf{p}}}^{m,\nu\leftrightarrow n,\nu}(i\omega_s+\omega_{\mathbf{p},\nu}) n^{(0)}_{\mathbf{p}\nu}. 
    \label{eq:SE_T-matrix_SWL}
\end{equation}
It should be noted that an exact evaluation without invoking the short-wavelength limit is possible only for the imaginary part of the diagonal self-energy~\cite{Rastelli2011}:
\begin{equation}
\begin{aligned}
    & \textrm{Im} \left[ \Sigma^{{\rm (T)}}_{\mathbf{k},nn}(i\omega_s) \right] \\
    &= \textrm{Im} \left[\frac{4}{\hbar}\sum_\mathbf{p}\sum_\nu T_{{\mathbf{k},\mathbf{p}\leftrightarrow\mathbf{k},\mathbf{p}}}^{n,\nu\leftrightarrow n,\nu}(i\omega_s+\omega_{\mathbf{p},\nu}) \left( n^{(0)}_{\mathbf{p}\nu} - n^{(\mathrm{c})}(i\omega_s+\omega_{\mathbf{p}\nu}) \right) \right],
    \label{eq:SE_T-matrix_ImLWL}
\end{aligned}
\end{equation}
where $n^\textrm{(c)}(z)=[e^{\beta z}-1]^{-1}$ $(z\in\mathbb{C})$ denotes the complex Bose-Einstein distribution function.
Since the T-matrix is the solution of the BSE, it by definition contains information about two-magnon \textit{bound} states and resonances~\cite{Cobb1971,Yelon1971,Silberglitt1967,Silberglitt1968,Rastelli2011}.

The T-matrix resummation systematically incorporates all diagrams containing exactly one backward-propagating line~\cite{Silberglitt1967,Rastelli2011}. For example, a single loop, such as that appearing in the Hartree diagram shown in Fig.~\ref{Fig03}(a), corresponds to one backward-propagating line. This selection criterion is diagrammatically well defined and can be motivated within a low-temperature expansion. In particular, the diagrams retained under this criterion correspond to those contributing at the leading nontrivial order---first order---in the bare Bose distribution $n^{(0)}_{\mathbf{q}\nu}$ entering the self-energy. By resumming these contributions to all orders, while omitting higher-order insertions that contribute at subleading order in $n^{(0)}_{\mathbf{q}\nu}$, the T-matrix approach provides a controlled description of the dominant low-temperature interaction effects.

\subsection{List of different approximations}
\label{sec:list_of_approximations}
We summarize several different levels of interacting spin-wave approximations for evaluating magnon-magnon interaction effects in van der Waals ferromagnets:
\begin{enumerate}
\item \textbf{Hartree approximation} \\ The self-energy is given by
\begin{equation}
    \Sigma_{\mathbf{k},mn}(i\omega_s) \approx \Sigma_{\mathbf{k},mn}^{(\textrm{Hartree})},
\end{equation}
which is frequency-independent and purely real. $\Sigma_{\mathbf{k},mn}^{(\textrm{Hartree})}$ is given in Eq.~\eqref{eq:Hartree}. Although this self-energy is most commonly evaluated in a non-self-consistent manner, an alternative approach—known as the self-consistent Hartree approximation—is occasionally employed. This method leverages the Hermitian nature of the self-energy and incorporates it into the unitary eigenvalue problem of the bilinear Hamiltonian. In this formulation, the Hamiltonian is modified as 
\begin{equation}
    \mathcal{H}^\textrm{eff} = \sum_\mathbf{k} \left[ \mathcal{H}^{(2)}_\mathbf{k} + \mathcal{U}_\mathbf{k} \Sigma_{\mathbf{k}}^{(\mathrm{Hartree})} \mathcal{U}_\mathbf{k}^\dagger \right]. \label{eq:selfconHartree}
\end{equation}
The resulting eigenvalue problem with respect to $\mathcal{H}^\textrm{eff}$ is solved self-consistently.
\item \textbf{``Hartree+reduced sunset" approximation} \\ The self-energy is given by
\begin{equation}
    \Sigma_{\mathbf{k},mn}(i\omega_s) \approx \Sigma_{\mathbf{k},mn}^{(\textrm{Hartree})} + \Sigma_{\mathbf{k},mn}^{(\textrm{R-Sunset})}(i\omega_s),
\end{equation}
which is frequency-dependent and complex. The Hartree self-energy $\Sigma_{\mathbf{k},mn}^{(\textrm{Hartree})}$ is provided in Eq.~\eqref{eq:Hartree} and the reduced sunset self-energy $\Sigma_{\mathbf{k},mn}^{(\textrm{R-Sunset})}(i\omega_s)$ in Eq.~\eqref{eq:SE_reduced_sunset}. This approximation highlights repulsive interaction between single magnons and \textit{bare} two-magnon states built from two non-interacting single magnons.
\item \textbf{``Hartree+full sunset" approximation} \\ The self-energy is given by
\begin{equation}
    \Sigma_{\mathbf{k},mn}(i\omega_s) \approx \Sigma_{\mathbf{k},mn}^{(\textrm{Hartree})} + \Sigma_{\mathbf{k},mn}^{(\textrm{Sunset})}(i\omega_s), \label{eq:full-perturb}
\end{equation}
with $\Sigma_{\mathbf{k},mn}^{(\textrm{Hartree})}$ in Eq.~\eqref{eq:Hartree} and the frequency-dependent and complex sunset self-energy $\Sigma_{\mathbf{k},mn}^{(\textrm{Sunset})}(i\omega_s)$ in Eq.~\eqref{eq:SE_sunset}. The imaginary part of the self-energy in Eq.~\eqref{eq:full-perturb} is exact up to order $1/S^2$.
\item \textbf{Resummation (T-matrix approximation)} \\ The self-energy is given by
\begin{equation}
    \Sigma_{\mathbf{k},mn}(i\omega_s) \approx \Sigma^{(\mathrm{T})}_{\mathbf{k},mn}(i\omega_s),
\end{equation}
which is the frequency-dependent and complex T-matrix self-energy $\Sigma^{(\mathrm{T})}_{\mathbf{k},mn}(i\omega_s)$ in Eq.~\eqref{eq:SE_T-matrix_SWL}. As discussed in Ref.~\cite{Silberglitt1967}, since the T-matrix contains information about two-magnon bound states, this approximation effectively captures the repulsive interaction between a single magnon and the two-magnon bound states or at least two-magnon resonances. 
\end{enumerate}

We close this subsection by highlighting advantages of the T-matrix resummation:
\begin{itemize}
    \item \textbf{Capacity to account for two-particle binding as a  nonperturbative quantum effect} \\
    As repeatedly mentioned, since the T-matrix is a solution of the BSE given by Eq.~(\ref{eq:BSE}), it inherently captures two-magnon bound states. To describe binding effects which are nonperturbative quantum phenomena, it is essential to incorporate an infinite series of connected bubble diagrams. The T-matrix resummation can be regarded as one of the simplest realizations of this approach.
    \item \textbf{Accuracy in the short-wavelength region} \\
    Ref.~\cite{Silberglitt1967} discusses the breakdown of the Hartree approximation, i.e., the first Born approximation, in the short-wavelength region. The T-matrix resummation captures not only the bound states themselves but also their resonance within the unbound continua whose effects are strongly highlighted in the short-wavelength regime, offering reasonable accuracy to capture a variety of short-wavelength excitations.
    \item \textbf{Compatibility with the single-particle picture} \\
    While the BSE is an equation for two-particle Green's functions, its solution — the T-matrix — enters single-particle self-energy diagrams.
    This enables us to describe hybridization between different particle-number sectors of the Hilbert space in an efficient manner. While this work solely focuses on \textit{thermal} hybridization between single-partice and two-particle sectors, it potentially describes \textit{quantum} hybridization between them in particle-number-nonconserving systems~\cite{mook2021,Mook2023boundmagnon}, as briefly mentioned in Sec.~\ref{sec:discussion}.
    \item \textbf{Ability to describe non-perturbative finite temperature effects} \\
    The T-matrix resummation preserves the transparent quasiparticle picture and complements other \textit{non-perturbative} quantum solvers, such as density matrix renormalization group, projected entangled pair states, and variational Monte Carlo, whose extensions to finite-temperature physics are still under active development.
    While exact diagonalization can capture thermal effect by combining itself with thermal pure quantum state technique, its accessible spatial (momentum space) resolution is largely limited. Quantum Monte Carlo can sample various physical quantities of unfrustrated systems at finite temperatures, but it is generally difficult to obtain statistically reliable results at ultralow temperatures, where the effects of bound states become prominent. Under these circumstances, although its ability to capture quantum entanglement is inferior to the above mentioned quantum solvers as it neglects three- or more-magnon scattering processes, the T-matrix resummation would be a powerful tool to describe \textit{non-perturbative} thermal many-body interactions.
\end{itemize}

\subsection{Spectral function}
\label{sec:spectral_function}
We calculate the single-particle spectral function of magnons given by
\begin{equation}
\begin{aligned}
    A(\mathbf{k},\omega) &= \textrm{Tr} \left[ \mathcal{A}(\mathbf{k},\omega) \right], \\
    \mathcal{A}(\mathbf{k},\omega) &= -\frac{1}{\pi} \textrm{Im} \left[\mathcal{G}_\mathbf{k}(\omega+i0^+)\right],
    \label{eq:Aqw}
\end{aligned}
\end{equation}
where
\begin{equation}
    [\mathcal{G}_\mathbf{k}(z)]^{-1}_{mn} = \delta_{mn}(z - \omega_{\mathbf{k}n}) - \Sigma_{\mathbf{k},mn}(z)
\end{equation}
denotes the interacting single-particle Green's function. Note that an analytic continuation $i\omega_s\rightarrow\omega+i0^+$ is applied in Eq.~(\ref{eq:Aqw}). We evaluate frequency-dependent self-energies in two different methods: {\it off-shell} and {\it on-shell} approximations. In the off-shell method, we compute Eq.~\eqref{eq:Aqw} without any approximations, indicating that it can capture weakly nonperturbative corrections to the lineshape which deviates from an ideal Lorentzian. In the on-shell method, we incorporate only the diagonal components of the self-energy and neglect off-diagonal ones. Moreover, self-energies are evaluated at the non-interacting magnon energy: $\Sigma_{\mathbf{k},nn}(\omega+i0^+)\rightarrow\Sigma_{\mathbf{k},nn}(\omega_{\mathbf{k}n}+i0^+)$. The on-shell spectral function is then given by
\begin{equation}
\begin{aligned}
    & A_\textrm{on-shell}(\mathbf{k},\omega) \\
    &= - \frac{1}{\pi} \sum_n \frac{\Sigma''_{\mathbf{k},nn}(\omega_{\mathbf{k}n})}{\left(\omega-\omega_{\mathbf{k}n}-
    \Sigma'_{\mathbf{k},nn}(\omega_{\mathbf{k}n})\right)^2+\left(\Sigma''_{\mathbf{k},nn}(\omega_{\mathbf{k}n})\right)^2},
    \label{eq:Aqw_onshell}
\end{aligned}
\end{equation}
where $\Sigma'_{\mathbf{k},nn}(\omega_{\mathbf{k}n}) = \mathrm{Re}\Sigma_{\mathbf{k},nn}(\omega_{\mathbf{k}n}+i0^+)$ and $\Sigma''_{\mathbf{k},nn}(\omega_{\mathbf{k}n}) = \mathrm{Im}\Sigma_{\mathbf{k},nn}(\omega_{\mathbf{k}n}+i0^+)$.
As seen in Eq.~(\ref{eq:Aqw_onshell}), the on-shell spectral function corresponds to a sum of Lorentzian peaks centered at the renormalized energies.
\begin{figure}[tbh]
    \centering
    \includegraphics[width=0.48\textwidth]{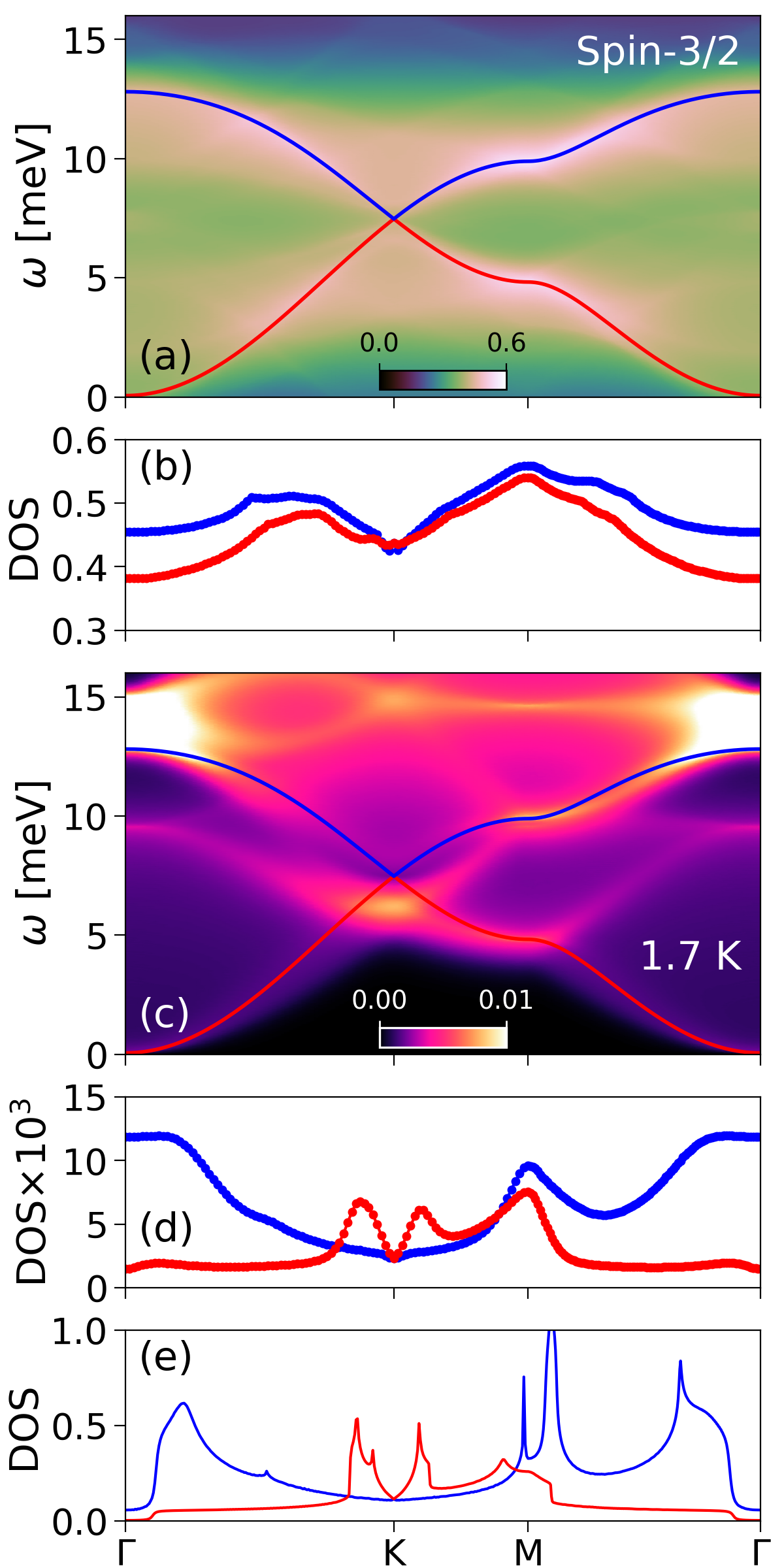}
    \caption{Multimagnon continua exhibited by the spin model for CrBr$_3$ with $S=3/2$. (a) Sunset density of states $\mathcal{D}^\textrm{(Sunset)}_\mathbf{k}(\omega)$ and (b) its on-shell trace $\mathcal{D}^\textrm{(Sunset)}_\mathbf{k}(\omega_{\mathbf{k}n})$ $(n=+,-)$. 
    (c) Temperature-weighted sunset DOS $\mathcal{B}^\textrm{(Sunset)}_\mathbf{k}(\omega)$ and (d) its on-shell trace $\mathcal{B}^\textrm{(Sunset)}_\mathbf{k}(\omega_{\mathbf{k}n})$ at 1.7~K. 
    (e) On-shell two-magnon density of states $\mathcal{D}^{(2)}_\mathbf{k}(\omega_{\mathbf{k}n})$. Red (blue) lines and markers indicate data for lower (upper) bands $n=-$ ($+$). Magnitudes of all the density of states are given with arbitrary units.}
    \label{Fig04}
\end{figure}
\begin{figure*}[tbh]
    \centering
    \includegraphics[width=\textwidth]{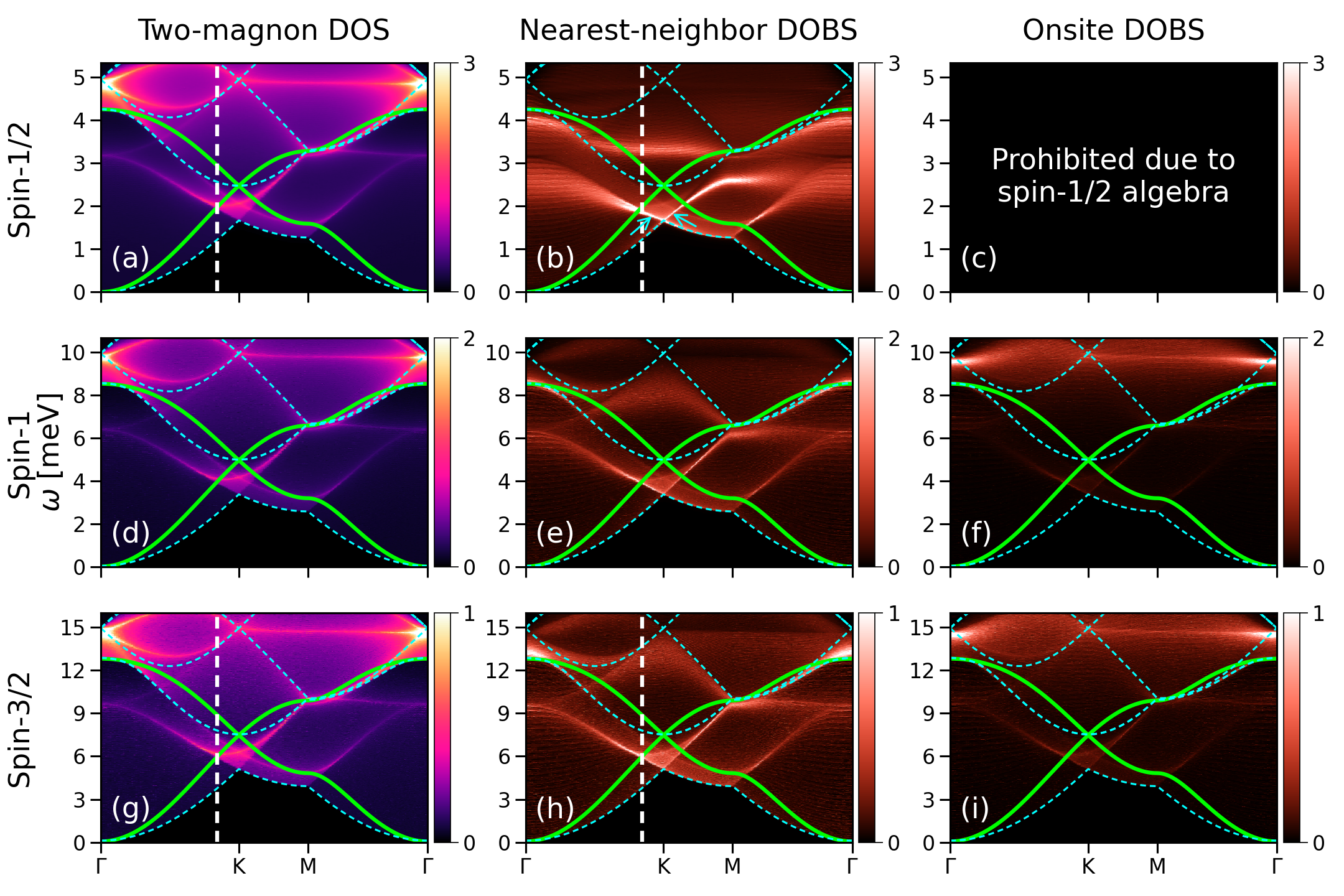}
    \caption{Two-particle excitation density of states in CrBr$_3$. (Left column) Two-magnon density of states $\mathcal{D}_\mathbf{k}(\omega)$, (middle column) nearest-neighbor density of two-magnon bound states, and (right column) onsite density of two-magnon bound states. The latter two quantities are calculated using exact diagonalization for a system with $2\times72 \times 72$ sites. The model parameters $\{J_1,J_2,J_3,A\}$ are taken from CrBr$_3$, while the spin length was set to (top row) $S=1/2$, (middle row) $S=1$, and (bottom row) $S=3/2$. Note that the onsite anisotropy $A$ does not play any role in spin-$1/2$ systems. Panels (g)-(i) correspond to the actual material CrBr$_3$. Green lines denote linear spin-wave spectra. Cyan dashed lines show lower and upper bounds of the band-resolved bare two-magnon continua, i.e., $\min_\mathbf{q}[\omega_{\mathbf{q}\pm}+\omega_{\mathbf{k}-\mathbf{q}\pm}]$ and $\max_\mathbf{q}[\omega_{\mathbf{q}\pm}+\omega_{\mathbf{k}-\mathbf{q}\pm}]$. Linecuts on the white dashed lines in panels (a), (b), (g), and (h) are plotted in Fig.~\ref{Fig06}.
    Cyan arrows in (b) indicate some of two-magnon bound state \textit{resonances} within the two-magnon continuum.}
    \label{Fig05}
\end{figure*}
\begin{figure}[tbh]
    \centering
    \includegraphics[width=0.50\textwidth]{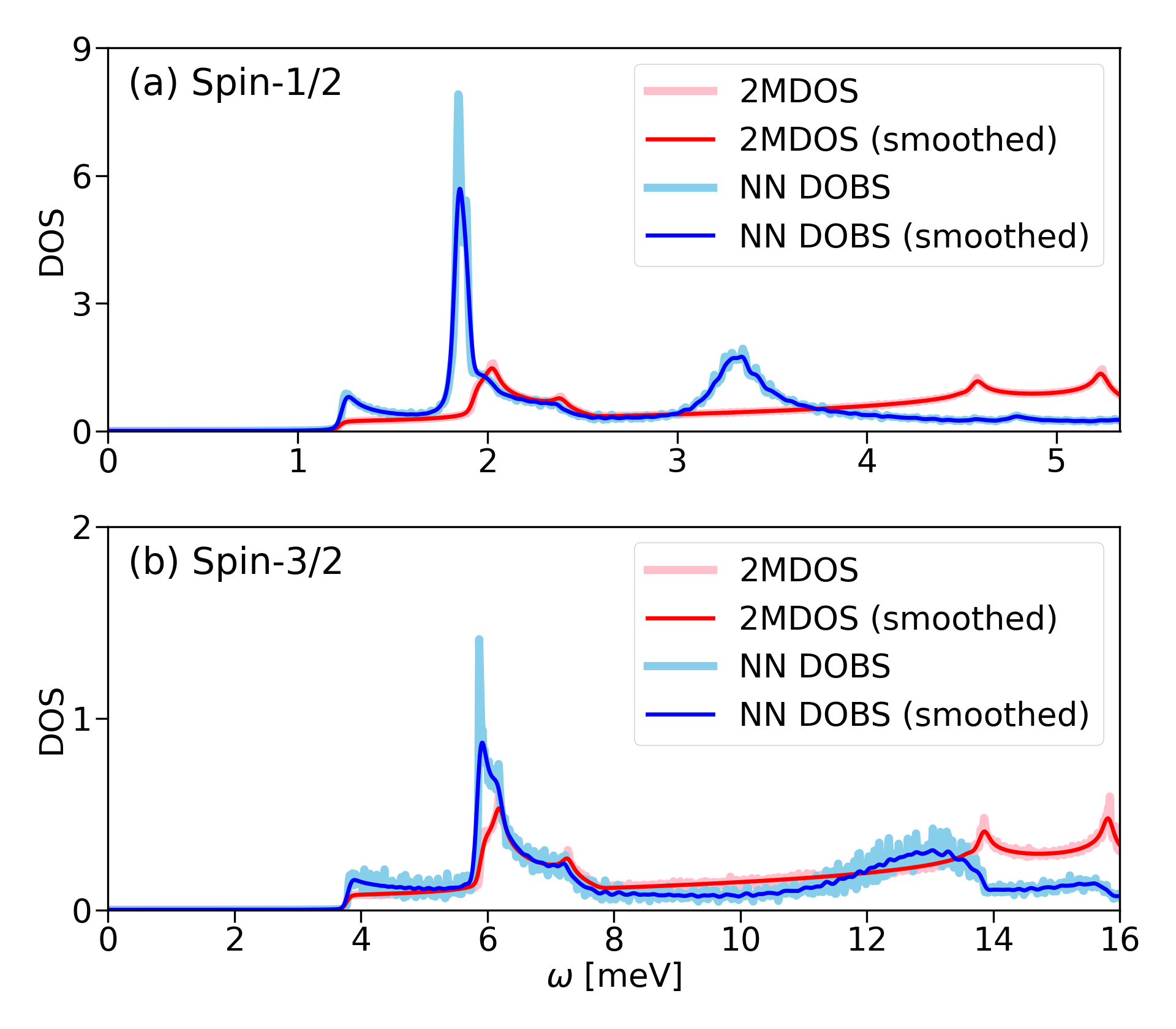}
    \caption{(a) Density of states linecuts of the spin-$1/2$ model, along the white dashed lines in Figs.~\ref{Fig05}(a) and~\ref{Fig05}(b). (b) Those for the spin-$3/2$ model describing CrBr$_3$, along the white dashed lines in Figs.~\ref{Fig05}(g) and~\ref{Fig05}(h).
    A Gaussian kernel with a half-width-at-half-maximum of 0.16~meV was used for smoothing.
    }
    \label{Fig06}
\end{figure}
\section{Results}
\label{sec:results}
\subsection{Gapless Dirac magnons in CrBr$_3$}
\label{sec:CrBr3}
In this Section, we study the temperature-induced renormalization of the magnon spectrum in CrBr$_3$ hosting gapless Dirac magnons.
\subsubsection{Multiparticle continua}
\label{sec:CrBr3_multiparticle_continua}
Before discussing entire self-energy contributions and the effects of binding of two magnons, we first focus on the underlying multiparticle excitation structures, i.e., multimagnon density of states (DOS) that influence the renormalization of single magnons at finite temperatures. According to the imaginary part of the sunset self-energy in Eq.~\eqref{eq:SE_sunset}, we consider the sunset density of states given by
\begin{equation}
    \mathcal{D}^\textrm{(Sunset)}_\mathbf{k}(\omega) = \frac{1}{N_\textrm{muc}^2} \sum_{\mathbf{p}\mathbf{q}} \sum_{\nu\nu_1\nu_2} \delta\left( \omega+\omega_{\mathbf{p}\nu}-\omega_{\mathbf{q}\nu_1}-\omega_{\mathbf{k}+\mathbf{p}-\mathbf{q}\nu_2} \right),
\label{eq:bare_SDOS}
\end{equation}
and its temperature-weighted version defined as
\begin{equation}
\begin{aligned}
    \mathcal{B}^\textrm{(Sunset)}_\mathbf{k}(\omega) &= \frac{1}{N_\textrm{muc}^2} \sum_{\mathbf{p}\mathbf{q}} \sum_{\nu\nu_1\nu_2} \delta\left( \omega+\omega_{\mathbf{p}\nu}-\omega_{\mathbf{q}\nu_1}-\omega_{\mathbf{k}+\mathbf{p}-\mathbf{q}\nu_2} \right) \\
    &\quad \times \left[ n^{(0)}_{\mathbf{p}\nu}\left( 1+n^{(0)}_{\mathbf{q}\nu_1}+n^{(0)}_{\mathbf{k}+\mathbf{p}-\mathbf{q}\nu_2} \right) -n^{(0)}_{\mathbf{q}\nu_1}n^{(0)}_{\mathbf{k}+\mathbf{p}-\mathbf{q}\nu_2} \right].
\label{eq:temp_weighted_SDOS}
\end{aligned}
\end{equation}
While $\mathcal{D}^\textrm{(Sunset)}_\mathbf{k}(\omega)$ only accounts for kinematically possible decay channels, $\mathcal{B}^\textrm{(Sunset)}_\mathbf{k}(\omega)$ additionally incorporates the Bose factor weights.
\par Figure~\ref{Fig04}(a) shows $\mathcal{D}^\textrm{(Sunset)}_\mathbf{k}(\omega)$ for CrBr$_3$ along the high-symmetry path given by the purple arrows in Fig.~\ref{Fig02}(a). Evaluating it on-shell, that is, evaluating $\mathcal{D}^\textrm{(Sunset)}_\mathbf{k}(\omega = \omega_{_\mathbf{k} \pm})$, reveals a very weak momentum dependence, as depicted in Fig.~\ref{Fig04}(b). This weak dependence suggests that kinematics alone cannot be responsible for momentum-dependent features of magnon lifetimes. This is an important difference to spontaneous decays, which often come with threshold effects \cite{zhitomirskychernyshev2013}.

A stronger momentum dependence is found when accounting for Bose factor weights: In Fig.~\ref{Fig04}(c), we plot $\mathcal{B}^\textrm{(Sunset)}_\mathbf{k}(\omega)$ at a temperature of 1.7~K, which exhibits a more pronounced structure than $\mathcal{D}^\textrm{(Sunset)}_\mathbf{k}(\omega)$ in Fig.~\ref{Fig04}(a). This is because scattering processes featuring thermally activated magnons at low energies come with a large Bose factor $n^{(0)}_{\mathbf{p}\nu}$ in Eq.~\eqref{eq:temp_weighted_SDOS}. As expected, when taken on-shell, that is $\mathcal{B}^\textrm{(Sunset)}_\mathbf{k}(\omega = \omega_{_\mathbf{k} \pm})$, a rich momentum dependence is identified, as shown in~\ref{Fig04}(d). While it can be expected that this dependence will get imprinted onto the magnon lifetimes, additional momentum dependent factors appear in the scattering vertex $\mathcal{Q}_{\mathbf{k},\mathbf{p}\leftrightarrow\mathbf{q}_1,\mathbf{q}_2}^{m,\nu\leftrightarrow \nu_1,\nu_2}
\mathcal{Q}_{\mathbf{q}_1,\mathbf{q}_2\leftrightarrow\mathbf{k},\mathbf{p}}^{\nu_1,\nu_2\leftrightarrow n,\nu}$ in Eq.~\eqref{eq:SE_sunset}, which set the full momentum dependence of the self-energy.
\par We emphasize that there are similarities between the temperature-weighted sunset DOS and the \textit{bare} two-magnon DOS $\mathcal{D}^{(2)}_\mathbf{k}(\omega)$ defined in Eq.~\eqref{eq:2magDOS}, where the latter relates to the reduced sunset self-energy given by Eq.~(\ref{eq:SE_reduced_sunset}). For instance, two peaks of $\mathcal{B}_\mathbf{k}^\textrm{(Sunset)}(\omega_{\mathbf{k}-})$ near the K point in Fig.~\ref{Fig04}(d) are mimicked by cusp-like enhancements of the on-shell two-magnon DOS $\mathcal{D}^{(2)}_\mathbf{k}(\omega_{\mathbf{k}+})$ shown in Fig.~\ref{Fig04}(e). Such cusp-like enhancements seen in $\mathcal{D}^{(2)}_\mathbf{k}(\omega_{\mathbf{k}n})$ $(n=\pm)$ are associated with van Hove singularities of the two-magnon DOS. While there are some similarities, 
we also observe stark differences, as illustrated by the presence (absence) of an enhancement of $\mathcal{B}_\mathbf{k}^\textrm{(Sunset)}(\omega_{\mathbf{k}+})$ $\left(\mathcal{D}_\mathbf{k}^\textrm{(2)}(\omega_{\mathbf{k}+})\right)$ toward the $\Gamma$ point. Taken together, these observations indicate that the reduced sunset self-energy originally introduced in Ref.~\cite{Pershoguba2018} is not necessarily sufficiently accurate to reproduce significant features of the \textit{full} sunset self-energy at low temperatures.
\subsubsection{Two-magnon bound states}
\label{sec:CrBr3_twomagnonBS}
While in the previous section we have argued that the \textit{bare} two-magnon DOS $\mathcal{D}_\mathbf{k}^\textrm{(2)}$ [defined in Eq.~\eqref{eq:2magDOS}] may not be the relevant quantity to study for analysing experimental data, two-magnon physics itself, in particular two-magnon binding, does play an important role in the T-matrix resummation, as suggested by the denominator of in Eq.~\eqref{eq:BSE}.

The left column of Fig.~\ref{Fig05} presents the \textit{bare} two-magnon DOS $\mathcal{D}_\mathbf{k}^\textrm{(2)}(\omega)$ in CrBr$_3$. Note that it is built from the sum of two free magnons, ignoring corrections from magnon-magnon interactions. We show results for $S = 1/2$, $S = 1$, and $S=3/2$ in Fig.~\ref{Fig05}(a), Fig.~\ref{Fig05}(d), and Fig.~\ref{Fig05}(g), respectively, of which the latter corresponds to the actual spin quantum number of CrBr$_3$ and the other cases are shown to highlight differences to the quantum limit. Importantly, $S$ only appears as a scaling factor for $\mathcal{D}_\mathbf{k}^\textrm{(2)}(\omega)$ because the single-magnon energies are of order $S$.

In contrast, in the central column of Fig.~\ref{Fig05}, we show the nearest-neighbor (NN) density of two-magnon bound states (DOBS). It is calculated from an exact diagonalization in the two-magnon sector---as done in Ref.~\cite{Reklis1974,Kecke2007,Mook2023boundmagnon} for the square lattice and explained in Appendix~\ref{Appendix:2magnonHilbertspace}---which explicitly takes magnon-magnon interactions into account. In particular, it accounts for magnon binding. A two-magnon bound state would separate from the continuum due to the binding energy \cite{Wortis1963} and its wavefunction has largest weight on states with two spin flips at neighboring sites because the binding derives from the nearest-neighbor exchange interaction. However, on the honeycomb lattice it is known that the binding energy is too weak to produce clearly separated bound states~\cite{Wada1975}. Instead, a projection of the wavefunction onto the three states with nearest-neighbor spin flips (along the three legs of the honeycomb) reveals two-magnon bound state \textit{resonances} within the two-magnon continuum, as indicated by cyan arrows in Fig.~\ref{Fig05}(b). These resonances can be thought of as lifetime broadened quasiparticles whose spectral weight gets diluted by hybridization with the continuum. Consequently, they show up in the spectral function as peaks of varying sharpness. 
Note that such a diluted two-magnon bound state in the one-dimensional Heisenberg ferromagnet was studied using time-dependent thermal density matrix renormalization group in Ref.~\cite{Nayak2022}.
Since the magnon binding is a quantum phenomenon (as the binding energy is a factor of $1/S$ smaller than the bare magnon energies \cite{Wortis1963}), the $S=1/2$ case in Fig.~\ref{Fig05}(b) exhibits the most pronounced effects, with two-magnon bound state resonances visible at the lower edge of the continuum close to the K point~\cite{Wada1975}. Note that these resonances do \textit{not} coincide with van Hove singularities of the continuum [cf.~the magenta v-shaped feature close to the K point in Fig.~\ref{Fig05}(a)], although they appear very close in energy. The two-magnon binding decreases for increasing $S$, such that the resonance features are considerably less pronounced for $S = 1$ and $S=3/2$ in Fig.~\ref{Fig05}(e) and Fig.~\ref{Fig05}(h). Nonetheless, when compared to the van Hove singularity features in Fig.~\ref{Fig05}(d) and Fig.~\ref{Fig05}(g), the bound state resonances still lead to a ``sharper'' lower edge of the v-shaped features at the K point.

Exploring the differences between the bare two-magnon DOS and the NN DOBS in greater detail, we take a closer look at the $\Gamma$--K path in Fig.~\ref{Fig05}(b). The peak in the NN DOBS penetrates into the two-magnon continuum at the K point, originating from the lower edge of the continuum around $\omega \approx 1.7$~meV. In this region, the bound state appears to be diluted by the continuum, suggesting a partial loss of its coherence. Compared to the van Hove singularity of the two-magnon DOS along the same path---also beginning at the K point near $\omega \approx 1.7$~meV and shown in Fig.~\ref{Fig05}(a)---the diluted bound state is slightly shifted to lower energies. This shift is clearly seen in the linecuts along the white dashed lines in Figs.~\ref{Fig05}(a) and~\ref{Fig05}(b) presented in Fig.~\ref{Fig06}(a). Moreover, surprisingly, despite of the partial loss of coherence by dilution, the NN DOBS near $\omega\approx1.9$ meV exhibits a sharper lineshape than the spike-like van Hove singularity of the two magnon DOS near $\omega\approx2$ meV. Similar trends including energy shifts and lineshapes in the two-magnon DOS and/or diluted bound states is also observed along the K--M path in Figs.~\ref{Fig05}(a) and~\ref{Fig05}(b). Beyond these, substantial structural differences between the two-magnon DOS and the NN DOBS are also observed across the entire energy and momentum ranges shown in Figs.~\ref{Fig05}(a) and~\ref{Fig05}(b).
\par As stated before, for larger spins $S=1$ and $S=3/2$ with weaker quantum effects, differences between NN DOBS and two-magnon DOS are less pronounced compared to those when $S=1/2$, as seen in Figs.~\ref{Fig05}(d),~\ref{Fig05}(e),~\ref{Fig05}(g), and~\ref{Fig05}(h). However, linecuts in Fig.~\ref{Fig06}(b) for $S=3/2$ still indicate appreciable difference between them, such as the spectral shapes and peak heights near $\omega\approx6$\,meV. Further differences are found at higher energies.
\par Finally, we also briefly mention the onsite DOBS presented in Figs.~\ref{Fig05}(f) and~\ref{Fig05}(i) for completeness. Single-ion or ``on-site'' two-magnon bound states arise from the single-ion anisotropy and physically correspond to putting two spin deviations on the same lattice site \cite{Silberglitt1970SIA, Tonegawa1970}, which is only possible for $S>1/2$ due to the spin-$1/2$ algebra [see Fig.~\ref{Fig05}(c)]. For $S=1$ and $S=3/2$, the intensity distribution of the onsite DOBS shown in Figs.~\ref{Fig05}(f) and \ref{Fig05}(i) is markedly different from the NN DOBS in Figs.~\ref{Fig05}(e) and \ref{Fig05}(h). Importantly, no clear features of single-ion bound states (or resonances) can be made out because the single-ion anisotropy $A$ is very weak and much smaller than the exchange energy. This situation can be contrasted with systems such as FeI$_2$ where the anisotropy is the leading energy scale \cite{legros2021}. We conclude that single-ion two-magnon bound states play no role in the considered systems.
\subsubsection{$T/T_\mathrm{C} \ll 1$}
\label{sec:CrBr3_T_ll_1}
\begin{figure*}[tbh]
    \centering
    \includegraphics[scale=1.0]{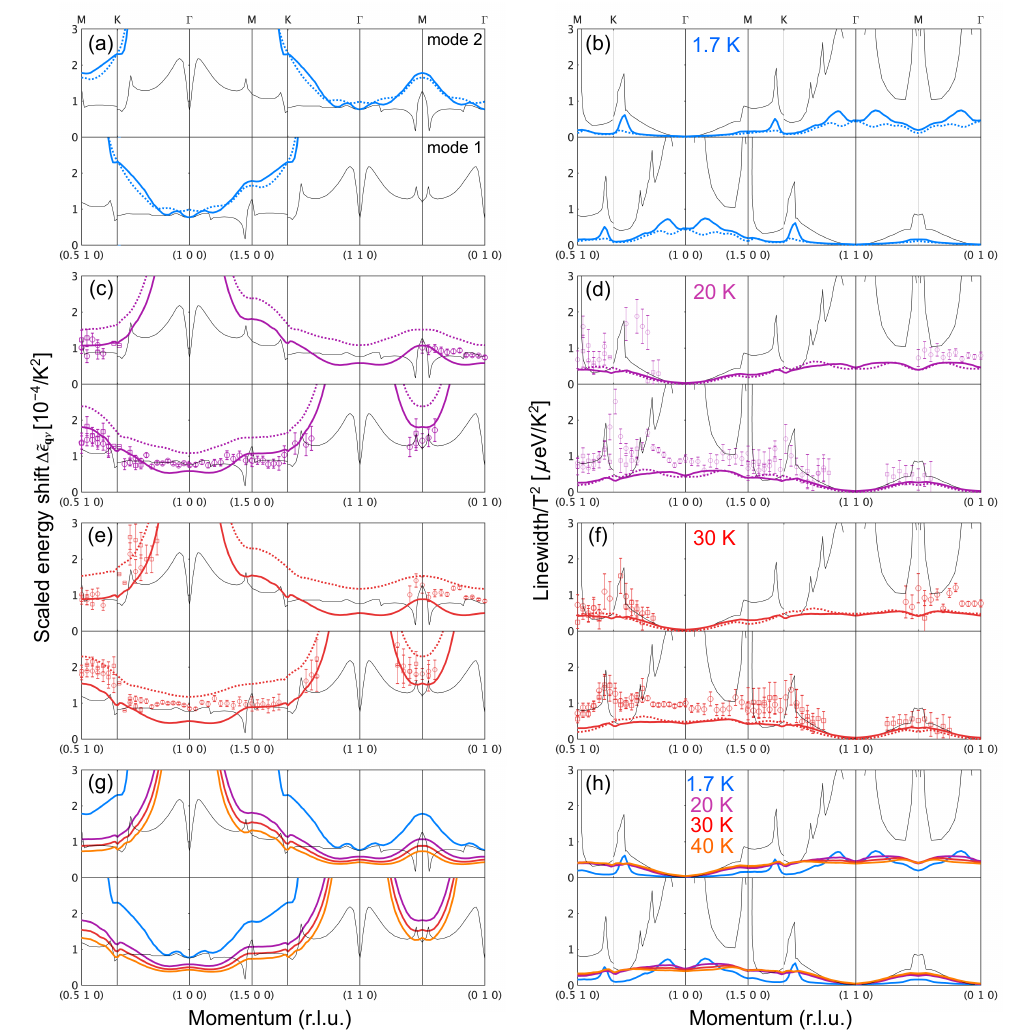}
    \caption{Comparison of the experimentally detected and theoretically predicted thermal evolution of the magnon spectrum of CrBr$_3$. (a,c,e) Relative magnon energy shift $\Delta\tilde{\epsilon}_{\mathbf{q}\nu}(T)$ as defined in Eq.~(\ref{eq:rescaled_energy_shift}) at (a) 1.7~K, (c) 20~K, and (e) 30~K, respectively. Square (circle) markers with error bars denote inelastic neutron scattering data obtained with incident energy of 15 (30)~meV, which are reproduced from Ref.~\cite{Nikitin2022}. The experimental data are quantitatively compared with predictions based on nonlinear spin-wave theory at different levels of approximation. The thin black line shows the result of the Hartree+reduced sunset approximation, also reproduced from Ref.~\cite{Nikitin2022}. The colored dashed line corresponds to the Hartree+full sunset approximation, while the colored solid line represents the resummation result. (b,d,f) Linewidth (half width at half maximum of the Lorentzian) of magnons divided by $T^2$. Again, the experimental data was reproduced from Ref.~\cite{Nikitin2022} and is compared with theoretical predictions. (g) Rescaled energy shift and (h) linewidth predicted by the resummation at selected temperatures.}
    \label{Fig07}
\end{figure*}

\begin{figure*}[tbh]
    \centering
    \includegraphics[width=0.85\textwidth]{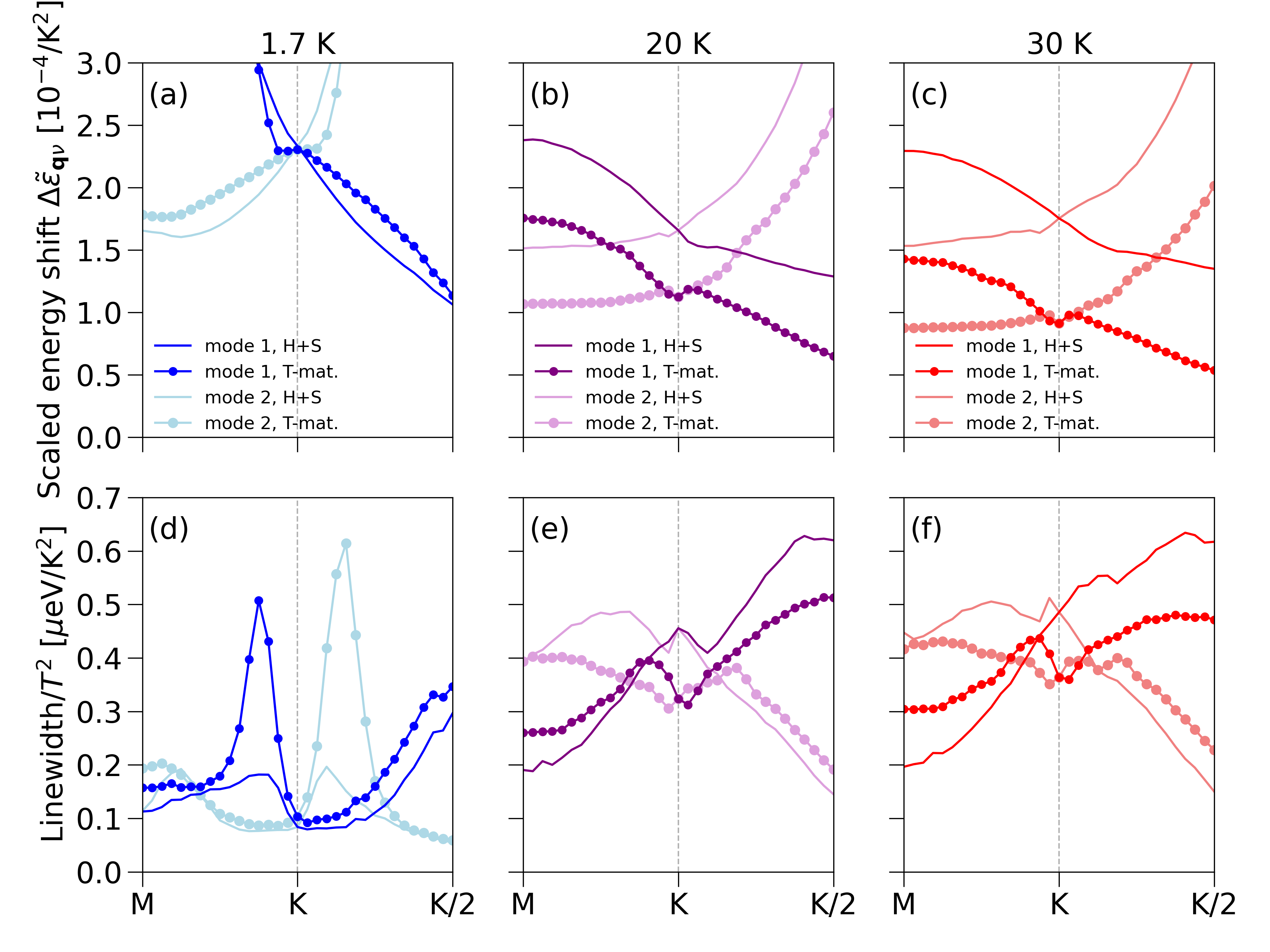}
    \caption{ 
    (a-c) Enlarged views of the scaled magnon energy shift of $\textrm{CrBr}_3$ estimated by the Hartree+full sunset approximation (H+S) and the T-matrix resummation (T-mat.) along the M--K--K/2 [(0.5 1 0)--(2/3 2/3 0)--(5/6 1/3 0)] path at (a)~1.7~K, (b)~20~K, and (c)~30~K, respectively.
    (d-f) Those of the scaled linewidth. Note that both the scaled energy shifts and the linewidths of the two bands coincide at the K point for both methods, indicating that the Dirac cones of the harmonic theory stay intact.
    }
    \label{Fig08}
\end{figure*}
We turn to the renormalization and linewidth of the magnon dispersion in the spin model of CrBr$_3$ at small temperatures far below the critical temperature $T_\mathrm{C}$, and discuss the qualitative and quantitative differences arising from the various levels of approximation introduced in Sec.~\ref{sec:Model_and_Methods}.

To comply with literature \cite{Nikitin2022}, we define the rescaled energy shift
\begin{equation}
    \Delta\tilde{\epsilon}_{\mathbf{q}\nu}(T) = \frac{\epsilon_{\mathbf{q}\nu}(0)-\epsilon_{\mathbf{q}\nu}(T)}{\epsilon_{\mathbf{q}\nu}(0)T^2},
    \label{eq:rescaled_energy_shift}
\end{equation}
where $\epsilon_{\mathbf{q}\nu}(T)$ is the magnon energy at temperature $T$. The motivation for introducing $\Delta\tilde{\epsilon}_{\mathbf{q}\nu}(T)$ derives from Ref.~\cite{Pershoguba2018} where a $T^2$ dependence of the magnon energy renomalization was predicted.
\par Figure~\ref{Fig07}(a) shows  $\Delta\epsilon_{\mathbf{q}\nu}(T)$ along high-symmetry paths at $T = 1.7$ K. 
The employed high-symmetry momentum path across multiple Brillouin zones and the definition of modes 1 and 2 in Fig.~\ref{Fig07} are depicted in Figs.~\ref{Fig02}(a) and \ref{Fig02}(d), respectively.
We identify pronounced differences among the three approximations: the Hartree+reduced sunset (shown as a black line), the Hartree+full sunset (shown as dashed colored line), and the resummation (shown as thick solid colored line).
The resummation predicts sensitively momentum-dependent features near the $\Gamma$ point for mode 1, which are absent in the Hartree+reduced sunset and the Hartree+full sunset predictions, highlighting the importance of including higher-order diagrams, which are systematically taken into account in the Bethe-Salpeter equation given by Eq.~\eqref{eq:BSE}.
Note further that $\Delta\varepsilon_{\mathbf{q}-}(T)$ is largest for the acoustic magnon close to the Brillouin zone origin. This is because there the largest relative change in energy happens as $\varepsilon_{\mathbf{0}-}(T) \to 0$ when approaching the Curie temperature. 
\par We next examine the predicted linewidths at low temperature and their sensitivity to the level of approximation.
As shown in Fig.~\ref{Fig07}(b), the linewidths differ significantly among the three approximations. In the Hartree+reduced sunset approximation (thin black line), several spike-like features appear where the linewidth is drastically enhanced. However, more than half of these spike-like structures disappear in both of the improved approximations---namely, the Hartree+full sunset and the resummation---indicating that they are artifacts of the low-temperature approximation used to derive the reduced sunset self-energy. Although such spikes are often attributed to repulsive interactions between a single magnon and the bare two-magnon van Hove singularity, their absence in the improved approximations suggests that this interpretation is not valid. This is consistent with the general experimental finding in Ref.~\cite{Nikitin2022}, where discrepancies between the Hartree+reduced sunset approximation and experimental data were reported. 
Unfortunately, no experimental linewidth data are available at this low a temperature because the experimental energy resolution is insufficient to resolve the expected $\mu$eV-scale linewidths.

Furthermore, there are substantial qualitative and quantitative differences between the Hartree+full sunset (dashed line) and the resummation results (solid line) across the entire Brillouin zone in Fig.~\ref{Fig07}(b).
Along the (1 0 0)--(1.5 0 0) [$\Gamma$--M] path, the resummation predicts significantly larger linewidth of mode 1 than the Hartree+full sunset, which the Hartree+full sunset approximation fails to capture, again underscoring the importance of higher-order diagrams.
Similarly, at a momentum point near the K point, namely (1.3 0.4 0), a drastic enhancement in the linewidth of mode 1 is observed in the resummation, while missing in the Hartree+full sunset approximation.
Within the Hartree+full sunset approximation---which describes the interaction between single magnons and the thermally weighted sunset DOS $\mathcal{B}^\textrm{(Sunset)}_\mathbf{k}$ depicted in Fig.~\ref{Fig04}(d)---only weak momentum-dependent features of the linewidth are predicted.
On the contrary, the resummation, which accounts for spectral redistribution in the sunset DOS because of magnon binding, 
describes strongly momentum-dependent magnon linewidth enhancement associated with two-magnon binding effect
[recall the diluted two-magnon bound state near $\omega \approx 6$ meV presented in Fig.~\ref{Fig06}(b)].
More precisely, within resummation, the linewidth enhancement appears because a magnon collides with a thermally excited magnon and scatters into a two-magnon bound resonance. Similar enhancements of linewidth associated with interactions of the lower magnon branch with bound magnons are found at (0.65 0.7 0) of mode 1, and at (0.7 0.6 0) of mode 2.
These differences between the Hartree+full sunset approximation and the T-matrix resummation are more clearly visible in Fig.~\ref{Fig08}(d), showing an enlarged view near the K point.
Moreover, not spike-like but broad enhancements of linewidth of mode 1 at (0.9 0.2 0) and (1.15 0 0) seem to be associated with interactions of the upper magnon branch and bound-magnon resonance near the $\Gamma$ point indicated in Fig.~\ref{Fig05}(h).
Taken together, these results demonstrate a fundamental difference between the two approaches.
The Hartree+full sunset approximation is insensitive to the fine structure in the momentum-frequency distribution of the background two-magnon states that mediate the scattering of thermally excited single magnons, and therefore predicts an overly smoothed, weakly momentum-dependent linewidth.
In contrast, the resummation essentially captures the interaction between single magnons and the two-magnon bound states and their resonances---an intrinsic quantum effect---and thus reliably predicts the large momentum dependence of the linewidth at low temperatures.

It is worth noting that the linewidth peak positions near the K point predicted by the Hartree+reduced sunset and by the resummation are comparable, even though their absolute magnitudes differ. This suggests that the Hartree+reduced sunset approximation---which represents the lowest-order approach for evaluating linewidths---is not entirely inadequate, as it can still provide a reasonable approximation in certain momentum regions at the lowest temperature. Nevertheless, the underlying physics captured by the Hartree+reduced sunset and the resummation are fundamentally different, and only the latter is technically justified. This underscores the importance of comparing results from different levels of approximation to achieve reliable predictions of the linewidth.

Altogether, the linewidths predicted by the highest-level approximation—the resummation—differ both qualitatively and quantitatively from those obtained using lower-level approximations, indicating that two-magnon bound states, their resonances, and the general spectral redistribution within the sunset DOS due to magnon-magnon interactions play an essential role in single-magnon excitations, even at the lowest temperatures. 
The predicted $\mu$eV-scale enhancement of the linewidth near the K point at low temperatures might be observable with Modulation of IntEnsity with Zero Effort (MIEZE)~\cite{Franz2019MIEZE}, a specialized variation of neutron spin-echo spectroscopy, given adequate sample volume.

\subsubsection{$T/T_\mathrm{C} \lesssim 1$}
\label{sec:CrBr3_T_lesssim_1}
\begin{figure}[tbh]
    \centering
    \includegraphics[scale=1.0]{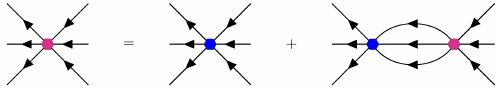}
    \caption{Three-magnon scatterings, which need to be included for a more quantitatively accurate description of the magnon linewidth at high temperatures near $T_\textrm{C}$.}
    \label{Fig09}
\end{figure}
Next we discuss thermal magnon renormalizations at higher temperatures, i.e., 20~K and 30~K, where a direct comparison between the theory and experimental data reported in Ref.~\cite{Nikitin2022} is possible. At 20~K, the rescaled energy shift of magnons predicted by the resummation shows remarkable agreement with the experimental data as seen in Fig.~\ref{Fig07}(c), indicating reasonable accuracy of this approximation in capturing temperature-induced renormalization. On the other hand, the lower level approximations are worse than the resummation. Especially the Hartree+full sunset approximation overestimates the energy shifts across the entire region of the BZ. At 30~K in Fig.~\ref{Fig07}(e), which is slightly below $T_\textrm{C}\approx32$~K, the resummation still fits experimental data qualitatively, but slightly underestimates the energy shifts, while the Hartree+full sunset still overestimates them.

Figures~\ref{Fig07}(d) and~\ref{Fig07}(f) show the linewidth of magnons at 20~K and 30~K, respectively.
Although the overall momentum dependence is well reproduced, both the Hartree+full sunset and the resummation tend to underestimate the linewidth of both modes 1 and 2 roughly by a factor of two. This quantitative discrepancy is possibly explained by higher-order diagrams that are not accounted for in either approximation. A likely scenario is that contributions from three-particle excitations---indicated by the diagrams given in Fig.~\ref{Fig09}---or from more-than-three multimagnon excitations could reduce the discrepancy, as temperatures close to $T_\textrm{C}$ activate such higher-order scattering processes.

In addition to this, the Kramers-Kronig relation, which connects the real and imaginary parts of the self-energy, provides another important insight. At 20 K, the T-matrix resummation reproduces the magnon energy shift in a reasonably accurate manner, while underestimating the linewidth. This indicates that the calculated real part of the self-energy coincides with the experimental data, whereas the imaginary part does not.
Although the imaginary part of the self-energy is constrained by its real part through the Kramers-Kronig relation, this constraint applies to the full frequency dependence of the self-energy—i.e., through an integral over a broad energy range—rather than to its on-shell imaginary part. Therefore, the remaining discrepancy in the on-shell linewidth does not contradict the Kramers-Kronig relation. Instead, it likely reflects additional intrinsic damping channels not captured in the present model---such as further magnon-magnon interaction processes at higher energies or incoherent scattering mechanisms---which could predominantly enhance the linewidth while leaving the renormalized dispersion largely unaffected.

Beyond these intrinsic magnon-magnon interaction effects which we targeted to study in this work, extrinsic processes like impurity- and/or lattice defect-induced scattering processes could be possible to contribute to enhance the linewidth, which are highly sensitive to the sample quality.
It is noteworthy that, due to the causality constraint of the self energy, all such extrinsic contributions \textit{enhance} the linewidth.
We also point out that limitations of the experimental setup tend to increase the observed linewidth.
The instrumental resolution in neutron scattering and the integration effect over small momentum-energy windows can have a momentum-dependent influence on the observed magnon linewidth, as the resolution function varies across the Brillouin zone depending on the scattering geometry and the steepness of the magnon dispersion.
Differences in incident neutron energy also affect energy resolution.

These considerations indicate that the remaining quantitative discrepancy in the linewidth does not signify a fundamental shortcoming of the present theoretical framework. Rather, it reflects the fact that both the theoretical treatment of intrinsic magnon-magnon interactions and the experimental determination of linewidth still involve uncertainties that can be further improved.

\begin{table}[htbp]
  \caption{Approximated power-law dependence of the temperature scaling of the magnon renormalization $f(T):=T^2\exp(-2SA/T)$ in the presence of the Goldstone gap $2SA=0.97$~K.}
  \begin{tabular}{rrc} \hline
      Temperature [K] & $l_f(T)=2+2SA/T$ & Effective scaling \\ \hline
      0.97 & 3.00 & $T^{3   }$ \\
      1.7  & 2.57 & $T^{2.57}$ \\
      5    & 2.19 & $T^{2.19}$ \\
      10   & 2.10 & $T^{2.10}$ \\
      20   & 2.05 & $T^{2.05}$ \\
      30   & 2.03 & $T^{2.03}$ \\ \hline
  \end{tabular}
  \label{table:T-scaling}
\end{table}

Comparing the theoretical predictions at low and high temperatures yields another important insight. 
As we discussed above, at 1.7~K, the T-matrix resummation predicts clear enhancements in the linewidth near the $\mathrm{K}$ point associated with magnon-bound magnon interactions. 
By contrast, these features are significantly renormalized at 20~K and 30~K, in reasonable agreement with the experimental data, while underestimating the absolute linewidths by roughly a factor of two, in a largely momentum-independent manner.
This non-monotonic temperature dependence of the linewidth challenges the simple $T^2$-dependence predicted by the Hartree+reduced sunset approximation for an ideal isotropic system with a quadratically-dispersive gapless Goldstone mode~\cite{Pershoguba2018}, although such a dependence was indeed experimentally observed at relatively high temperatures in Ref.~\cite{Nikitin2022}.
This deviation from the $T^2$ law in the lowest temperature region is presumably due to the presence of the Goldstone gap $2SA\approx0.97$~K.
An expected temperature dependence of the magnon band renormalization in the presence of the Goldstone gap $2SA$ is $f(T):=T^2\exp(-2SA/T)$, rather than the simple $T^2$. By taking the logarithmic derivative $l_f(T):=\frac{d\ln f}{d\ln T}$, the \textit{approximated} power-law dependence of $f(T)$ is given by $f(T)\approx T^{(2+2SA/T)}$ at each $T$.
When $T=2SA$, the approximated scaling is $T^3$, which completely differs from the ideal $T^2$.
As increase in temperature and it reaches to $T\gg2SA$, the effect of the gap becomes negligible and the $T^2$ behavior is gradually recovered, as indicated in Table~\ref{table:T-scaling}.
Indeed, as shown in Fig.~\ref{Fig07}(h), while the calculated linewidth using the resummation at 20~K, 30~K, and even 40~K (which is above $T_\mathrm{C}\approx32$~K) follow the $T^2$-law closely, the one at 1.7~K clearly deviates from this trend.
Moreover, as seen in Fig.~\ref{Fig07}(g), calculated magnon energy shift using the resummation also deviates from the $T^2$-law at lower temperatures.
We note that similar trends can be also seen in the Hartree+full sunset self-energy presented in Appendix~\ref{Appendix:Hartree_fullSunset_tempev}.

We close this section by highlighting a \textit{possible} trace of repulsive interactions between magnons and two-magnon bound resonances observed at the highest measured temperature. The T-matrix resummation predicts a tiny but finite enhancement of the linewidth of mode 1 at (0.65 0.7 0) at 20~K and 30~K, as shown in Figs.~\ref{Fig07}(d) and \ref{Fig07}(f), which are more pronounced in their enlargements Figs.~\ref{Fig08}(e) and~\ref{Fig08}(f).
Note the absence of such enhancements in the Hartree+full sunset prediction highlights their pure \textit{quantum} origins, which are correctly captured only by the T-matrix resummation.
Notably, the experimental data at 30~K in Fig.~\ref{Fig07}(f) exhibit a clear indication of such enhancement, in good qualitative agreement with the prediction by the T-matrix resummation. On the contrary, it is a bit of a pity that the 20~K data are considerably noisier near the K point, and discrepancies between datasets obtained with different incident neutron energies (15~meV and 30~meV) hinder any reliable conclusion at that temperature. We consider that the linewidth enhancement at 30~K may be a trace of magnon--\textit{bound} magnon repulsive interaction, which is more pronounced at 1.7~K. This indicates the possibility of observation of interaction effects between different particle-number states even at high temperatures near $T_\textrm{C}$, while their magnitude is less pronounced than those at low temperatures. Higher-resolution experiments could overcome current limitations and provide a more quantitative and robust understanding of this subtle linewidth feature at elevated temperatures.

\subsubsection{Stability of Dirac magnons}
\label{sec:CrBr3_Dirac_points}
\begin{figure}[tbh]
    \centering
    \includegraphics[width=0.50\textwidth]{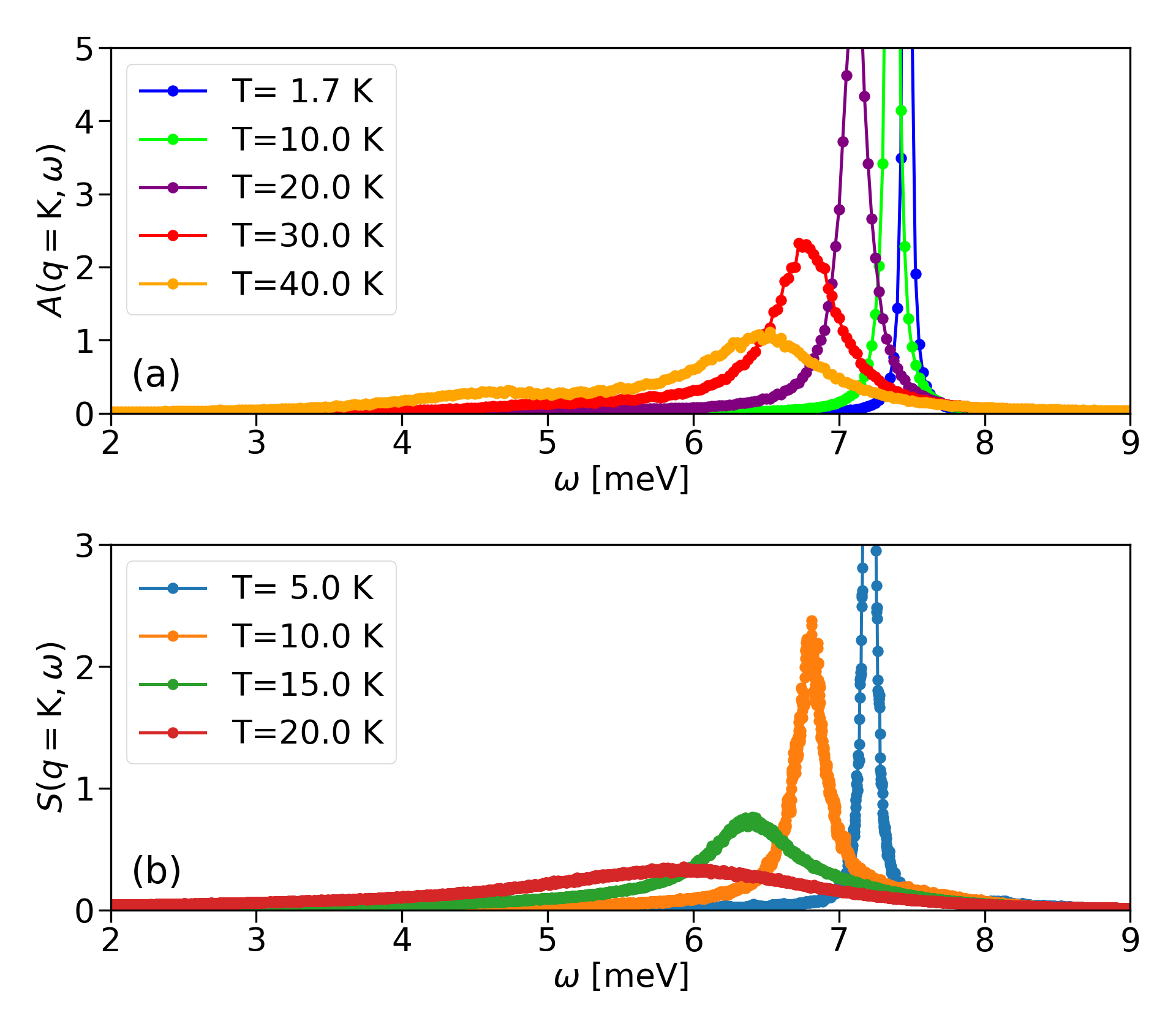}
    \caption{Linecuts of (a) the single-magnon spectral function $A(\mathbf{q}=\textrm{K},\omega)$ obtained with the off-shell T-matrix resummation at the K point of CrBr$_3$ at selected temperatures and (b) the dynamical spin-structure factor $S(\mathbf{q}=\textrm{K},\omega)$ obtained with the classical spin dynamics simulations.}
    \label{Fig10}
\end{figure}
We also investigate the damped Dirac magnons at the K point of CrBr$_3$ at finite temperatures. Figure~\ref{Fig10}(a) shows the temperature dependence of the single-magnon spectral function obtained with the \textit{off-shell} T-matrix resummation at the K point. The Dirac magnon, which exhibits a sharp single-peak structure at the lowest temperatures, acquires thermal-fluctuation-induced damping as the temperature increases, resulting in a broadening of the spectral shape. 
Notably, no peak splitting is observed at any temperature, indicating that the Dirac point remains intact under considerable thermal fluctuations. 
As is more clealy seen in Fig.~\ref{Fig08}, both magnon energy shifts and lifetimes of modes 1 and 2 are perfectly identical at K point at 1.7~K, 20~K, and 30~K, indicating Dirac point stays intact at all temperatures. Moreover, spectra shown in Fig.~\ref{Fig10}(b) obtained from the classical spin dynamics, whose theoretical and technical details are provided in Sec.~\ref{sec:CrBr3_classical_spin_dynamics}, also support the intactness of the Dirac magnon at elevated temperatures.
This is to be expected since the magnon-magnon interactions do not break a symmetry of the harmonic theory. This is in stark difference to the chiral honeycomb ferromagnets and antiferromagnets studied in Refs.~\cite{mook2021, Gohlke2023PRL}. At higher temperatures, such as 30~K and 40~K, a shoulder-like feature emerges as an incoherent background in the low-frequency region, suggesting nonperturbative corrections to the lineshape.

Before proceeding further, we briefly comment on Ref.~\cite{Banerjee2025}, which investigates the finite-temperature renormalization of magnon bands in CrBr$_3$ within a Hartree + full-sunset approximation. As can be inferred from Figs.~7(a) and 7(b) of Ref.~\cite{Banerjee2025}, the Dirac-point degeneracy appears to be lifted at finite temperatures. However, in the spin Hamiltonian used to describe CrBr$_3$ in Ref.~\cite{Banerjee2025} there is no mechanism that breaks the spin-space symmetry protecting the Dirac point, even at finite temperature. Consequently, on general symmetry grounds the Dirac-point degeneracy is expected to remain intact throughout the entire temperature range, independent of the specific level of approximation. This expectation is confirmed by our analysis, see Fig.~\ref{Fig08}. From this perspective, the finite-temperature lifting of the Dirac degeneracy reported in Ref.~\cite{Banerjee2025} should be interpreted with caution. By contrast, our Hartree + full-sunset calculations preserve all relevant symmetries, thereby providing a controlled second-order description of CrBr$_3$.

\subsubsection{Classical spin dynamics}
\label{sec:CrBr3_classical_spin_dynamics}
\begin{figure*}[htbp]
  \centering
  \includegraphics[width=0.95\textwidth]{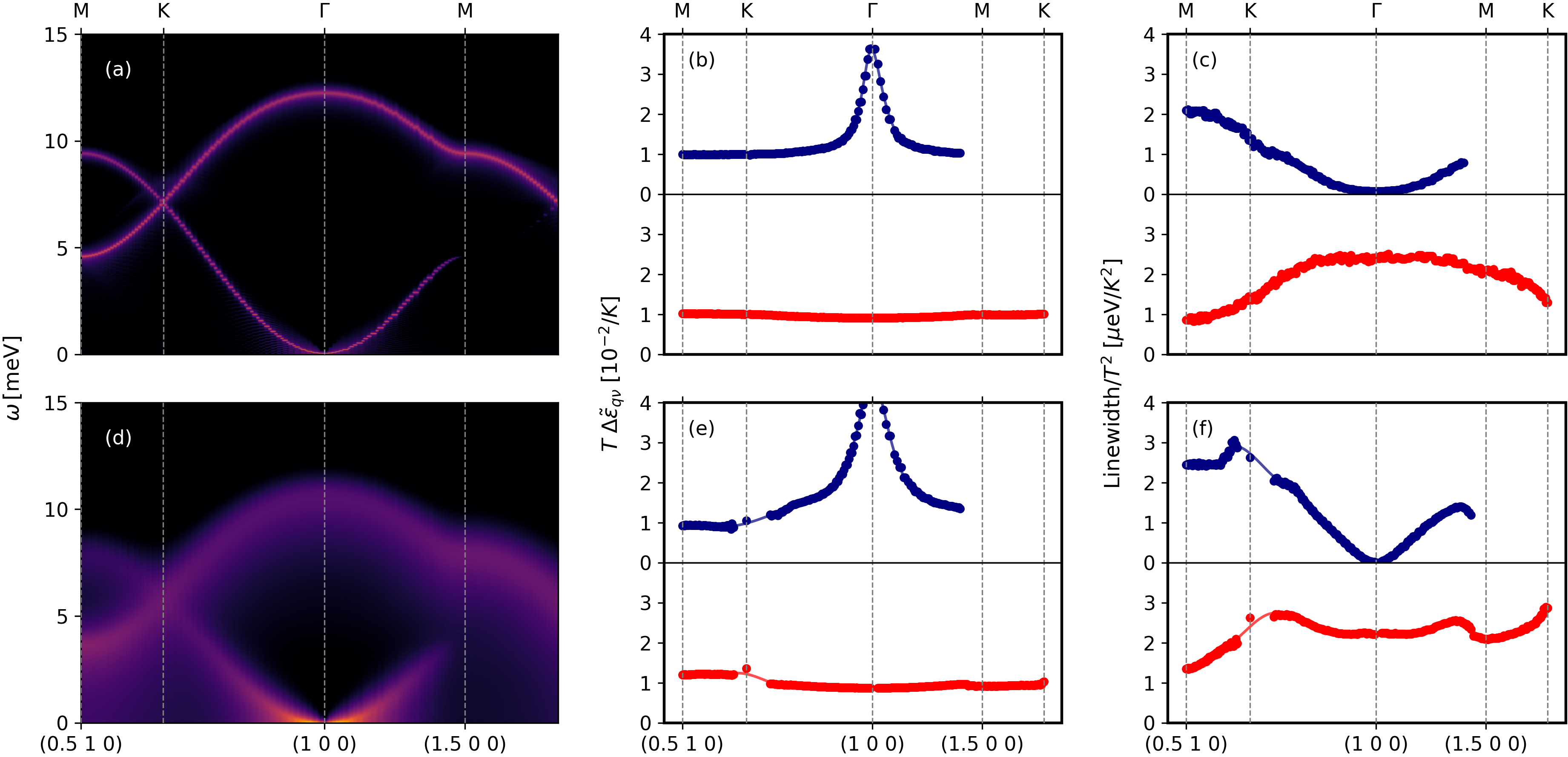}
  \caption{
  (a) Dynamical structure factor with a logarithmic intensity scale, (b) relative magnon energy shift, and (c) linewidth in CrBr$_3$ at 5~K obtained with classical spin-dynamics simulations. 
  (d--f) Those at 20~K. 
  For (b),~(c),~(e),~and~(f), red (blue) lines with markers in lower (upper) panels show the data of mode 1 (2), corresponding to those colors in Fig.~\ref{Fig02}(d).
  As seen in (b) and (d), magnon energy shift is shown in the units of $T\Delta\tilde{\epsilon}_{\mathbf{q}\nu}$. 
  Since $\Delta\tilde{\epsilon}_{\mathbf{q}\nu}$ given by Eq.~(\ref{eq:rescaled_energy_shift}) includes $T^2$-scaling, the presented unit $T\Delta\tilde{\epsilon}_{\mathbf{q}\nu}$ scales as $T$-linear.
  }
  \label{Fig11}
\end{figure*}

Classical \textit{atomistic} spin dynamics provides a complementary approach to evaluate the finite-temperature magnon spectrum. The interacting spin-wave framework developed in the previous section is based on the Holstein-Primakoff expansion of the spin Hamiltonian and incorporates magnon-magnon interactions through the Hartree and full sunset self-energies together with a non-perturbative T-matrix resummation. In contrast, spin-dynamics simulations treat the dynamics fully non-perturbatively by integrating the stochastic Landau-Lifshitz-Gilbert (LLG) equation, thereby including interaction effects to all orders through the real-time evolution of the spin configurations. For the large-spin $(S=3/2)$ systems considered here, this classical description is particularly well justified and provides an efficient way of accessing dynamical spectral functions. The resulting dynamical structure factor exhibits temperature-dependent renormalization and damping across the Brillouin zone, providing an independent benchmark for the predictions of the interacting spin-wave analysis.

The LLG equation~\cite{1353448} is given by
\begin{equation}
    \frac{d\boldsymbol{S}_i}{dt}=-\frac{\gamma}{1+\alpha^2_\textrm{G}} [\boldsymbol{S}_i \times \boldsymbol{B}_i^{\mathrm{eff}} + \alpha_\textrm{G} \boldsymbol{S}_i \times (\boldsymbol{S}_i \times \boldsymbol{B}_i^{\mathrm{eff}})],
    \label{eq:stochasticLLG}
\end{equation}
where $\gamma$ is the gyromagnetic ratio, $\alpha_\textrm{G}$ is the Gilbert damping, $\boldsymbol{S}_i$ is a unit-length spin vector and $\boldsymbol{B}_i^{\mathrm{eff}}$ is the effective magnetic field. We employed Langevin dynamics to incorporate the effect of thermal fluctuations. Accordingly, the effective field \cite{Evans2014} acting on each spin reads as
\begin{equation}
    \boldsymbol{B}_i^{\mathrm{eff}}=-\frac{1}{\mu_s} \frac{\partial \mathcal{H}}{\partial \boldsymbol{S}_i} + \boldsymbol{\Gamma}(t) \sqrt{\frac{2 \alpha_\textrm{G} k_{\mathrm{B}} T}{\gamma \mu_s \Delta t}}.
\end{equation}
The first term is derived from the Hamiltonian and accounts for the effect of interactions and external magnetic fields, with $\mu_s$ denoting the local spin moment. The second term represents thermal fluctuations via  a three-dimensional Gaussian white noise vector $\boldsymbol{\Gamma}(t)$, scaled such that the fluctuation-dissipation theorem is satisfied. Here, $k_{\mathrm{B}}$ is the Boltzmann constant, $T$ is the temperature, and $\Delta t$ is the time step of the simulation. For running the numerical calculations, we utilized the \textsc{Vampire} spin dynamics package \cite{Evans2014}.
The inclusion of the Gaussian noise field leads to classical thermal ensembles consistent with Boltzmann statistics. As a result, it tends to overestimate thermal fluctuations and artificially enhance the spectral weight of low-energy modes, a consequence of classical equipartition. To access spectral properties, we compute the dynamical spin structure factor (DSSF), which is defined as the space-time Fourier transform of spin-spin correlations and is directly observable in inelastic neutron scattering experiments,
\begin{equation}
    \mathcal{S}^{\mu\nu}(\mathbf{q},\omega) =
    \frac{1}{2\pi N} \sum_{i,j} \int_{-\infty}^{\infty} dt \,
    e^{\mathrm{i} \left[\omega t - \mathbf{q}\cdot (\mathbf{R}_i-\mathbf{R}_j)\right]}
    \langle s_i^\mu(t) s_j^\nu(0) \rangle .
\end{equation}
In order to partially correct the unphysical behavior at low energies, a classical-to-quantum rescaling \cite{PhysRevB.106.174428, PhysRevX.12.021015} is introduced,
\begin{equation}
    \mathcal{S}^{\text{quantum}}(\mathbf{q}, \omega) = 
    \frac{\beta \omega}{1 - e^{-\beta \omega}} 
    \mathcal{S}^{\text{classical}}(\mathbf{q}, \omega).
\end{equation}
This approach restores consistency with the quantum fluctuation-dissipation relation by suppressing the low-frequency spectral weight at low temperatures.  This correction modifies only the spectral distribution, not thermodynamic quantities such as the critical temperature, which remain driven by the classical dynamics.
\par Figure~\ref{Fig11} shows calculated DSSF as well as the rescaled magnon energy shift and linewidth at 5K and 20K, obtained from classical spin dynamics simulations, where the energy shifts and linewidths are extracted by fitting the DSSF spectra with a skewed Lorentzian,
\begin{equation}
\mathcal{S}^{\mathrm{fit}}(\mathbf{q},\omega) = A_\textrm{f} \frac{\left[1+\alpha(\omega-\omega_0)\right]^2}{(\omega-\omega_0)^2+\Gamma^2},
\label{eq:spindynamics_fitting}
\end{equation}
with fit parameters amplitude $A_\textrm{f}$, center frequency of the magnon peak $\omega_0$, linewidth $\Gamma$, and skewness $\alpha$.
At low temperatures the magnon peaks are well described by symmetric Lorentzians, whereas at elevated temperatures thermal fluctuations lead to asymmetric broadening, which we capture by employing a skewed Lorentzian line shape.
As a result of classical white noise, the temperature dependence of the data does not follow the $T^2$ behavior discussed in Refs.~\cite{Pershoguba2018,Nikitin2022}. In particular, as a consequence of classical equipartition, energy shifts in Figs.~\ref{Fig11}(b) and~\ref{Fig11}(e) exhibit linear-in-$T$ behavior [See vertical axes], rather than the expected $T^2$ scaling.
Partially missing data points close to the K point and along an M--K path in Figs.~\ref{Fig11}(b),~\ref{Fig11}(c),~\ref{Fig11}(e), and~\ref{Fig11}(f) are associated with failures of the fitting given by Eq.~(\ref{eq:spindynamics_fitting}) within a reasonable accuracy, specifically $R^2\geq0.99$. These failures occur either because the spectral weight vanishes along certain momentum paths, as seen in Figs.~\ref{Fig11}(a) and~\ref{Fig11}(d), or because, upon approaching the Dirac point, the peak positions of modes~1 and~2 become increasingly close. At higher temperatures, the substantial noise of both peaks further complicates the fitting of the spectrum as a sum of two skewed Lorentzians.

Setting aside the temperature dependence, several key features observed experimentally are well reproduced by the classical spin dynamics.
Notably, the energy shift of mode 1 in Figs.~\ref{Fig11}(b) and~\ref{Fig11}(e) shows minimal momentum dependence along the K--$\Gamma$--M--K path [(2/3 2/3 0)--(1 0 0)--(1.5 0 0)--(4/3 1/3 0)], consistent with both experimental results and interacting spin-wave theory.
The enhancement of the rescaled energy shift of the lower mode (mode 2 along the presented path) toward the $\Gamma$ point is also accurately reproduced.
Moreover, for those momentum paths and modes where statistically reliable data are obtained, the linewidth presented in Figs.~\ref{Fig11}(c) and~\ref{Fig11}(f) remains smooth across all momenta, with no spike-like anomalies, again in good agreement with both experiment and spin-wave theory.
Importantly, no linewidth broadening associated with single magnon-bound magnon interactions, as predicted by the T-matrix resummation approach, is observed in the classical dynamics. This absence underscores the quantum nature of the bound state contributions.
\subsection{Gapped Dirac magnons in CrI$_3$}
\label{sec:CrI3}
\begin{figure*}[tbh]
    \centering
    \includegraphics[width=\textwidth]{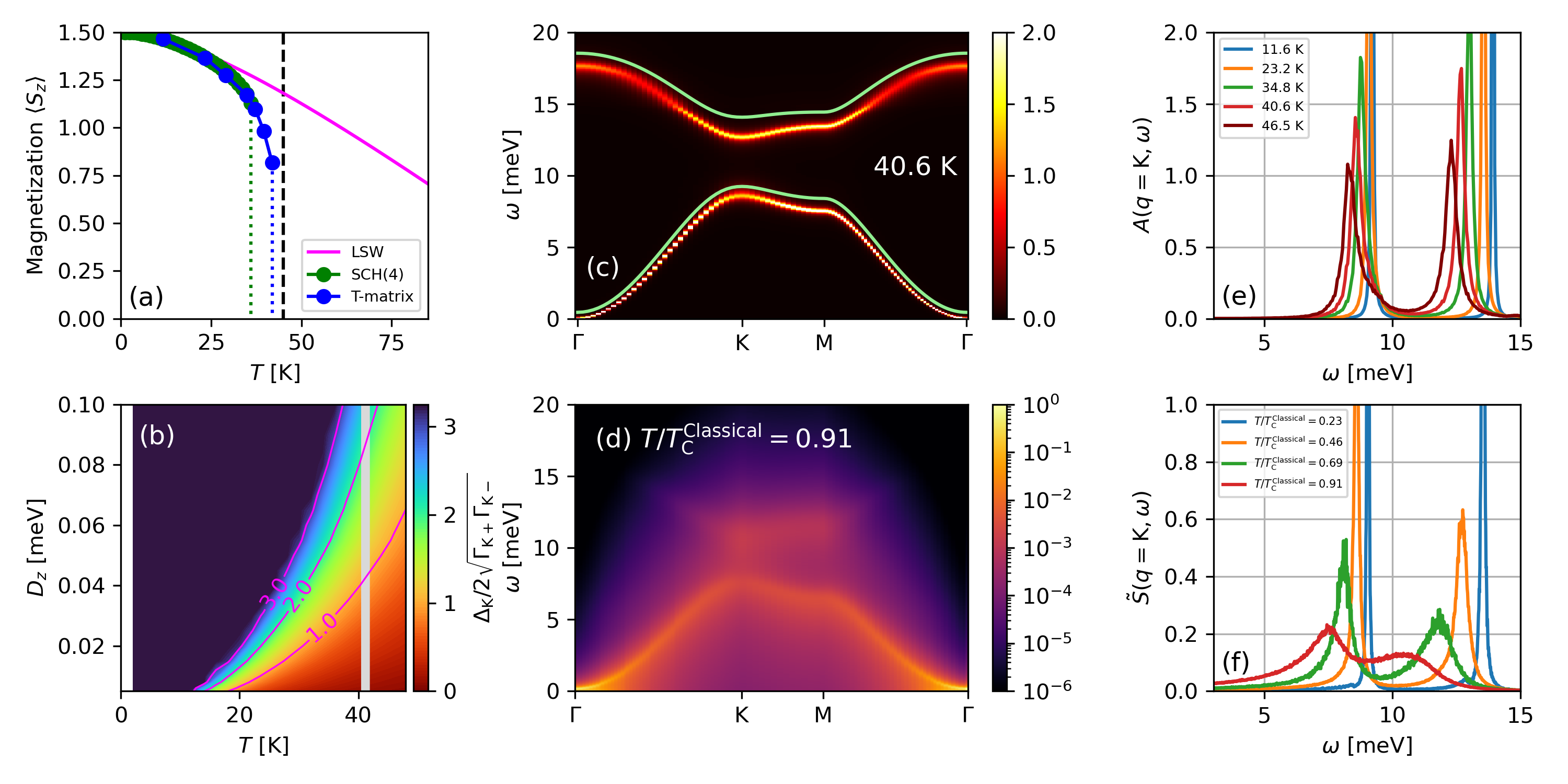}
    \caption{
    (a) Temperature-dependence of the magnetization $\ev{S_z}$ of CrI${}_3$ obtained with several different level approximations. Dashed vertical line shows the experimentally observed monolayer $T_\textrm{C}\approx$~45 K. Rightmost markers for each of different approximation schemes indicate maximum temperatures where the positive value of magnetization is obtained, and the actual predicted $T_\textrm{C}$s where magnetization vanishes are slightly above them. 
    (b) A ratio between the topological gap at K point and the magnon damping $\Delta_\textrm{K}/2\sqrt{\Gamma_{\textrm{K}+}\Gamma_{\textrm{K}-}}$ for CrI$_3$ obtained by the on-shell T-matrix resummation. Among several model parameters for CrI$_3$, only $D_z$ is artificially varied. White vertical line denotes the calculated $T_\textrm{C}^\text{T-matrix}$, which is almost independent of $D_z$. 
    (c) Single magnon spectrum $A(\mathbf{q},\omega)$ of CrI$_3$ at 40.6~K obtained with the off-shell resummation. Green lines show the LSW bands. 
    (d) The rescaled dynamical spin-structure factor $\tilde{S}(\mathbf{q},\omega)=S(\mathbf{q},\omega)/W(\mathbf{q})$ at $T/T_\textrm{C}^\text{Classical}=0.91$ obtained with classical spin-dynamics simulations. 
    (e) Linecuts of the single magnon spectrum at K point $A(\mathbf{q}=\textrm{K},\omega)$ obtained with the off-shell T-matrix resummation at varying temperatures.
    (f) Linecuts of the rescaled dynamical spin-structure factor $\tilde{S}(\mathbf{q}=\textrm{K},\omega)=S(\mathbf{q}=\textrm{K},\omega)/W(\mathbf{q}=\textrm{K})$ obtained with classical spin-dynamics simulations. 
    }
    \label{Fig12}
\end{figure*}
Here, using the resummation, we revisit the previously predicted thermal topological transition in CrI$_3$~\cite{Lu2021,CommentLu2021,ReplyLu2021}, associated with thermal fluctuation-induced topological gap closing and reopening at the K point.
\subsubsection{Recap of previous works}
\label{sec:CrI3_recap}
Ref.~\cite{Lu2021} theoretically studied magnon spectra in CrI$_3$ at finite temperatures, using the Hartree approximation (without self-consistency). The authors found an inversion of Chern numbers $\pm1$ between the lower and upper bands triggered by temperature-driven gap closing and reopening at K point. A successive comment~\cite{CommentLu2021} pointed out that the gap closing occurs around $T_\textrm{C}$, which is a limit where the assumptions of the Hartree approximation are questionable. However, within a self-consistent Hartree approximation incorporating not only four-magnon terms $\mathcal{H}^{(4)}$ but also a part of the six-magnon terms $\mathcal{H}^{(6)}$, the gap closing and reopening occurs  slightly below $T_\textrm{C}$ \cite{ReplyLu2021}.
This indicates that, as temperature increases, the gap closing and reopening—and the associated exotic phenomena such as the topological transition and a potential sign reversal of the thermal Hall conductivity—can occur just below $T_\textrm{C}$, before the system enters a paramagnetic phase.
\par Altogether, these studies demonstrate that significantly different predictions can arise depending on the approximation scheme employed. As a result, it has remained unclear whether such topological transitions are physically robust phenomena or artifacts introduced by the approximations inherent to interacting spin-wave theory. A definitive consensus has yet to be established.
\subsubsection{Predictions based on the resummation: Absence of the gap closing and reopening}
\label{sec:CrI3_absence_of_gap_closing}
Since the predicted topological transition is considered to be achieved at very high temperature just at $T_\textrm{C}$ or slightly below it, before evaluating the presence or absence of the gap closing and reopening, it is important to examine whether any particular approximation scheme predicts $T_\textrm{C}$.

To obtain the data presented in Figs.~\ref{Fig12}(a) and~\ref{Fig12}(b), we need to evaluate $\ev{S^z_\mathbf{r}}=S-\ev{\hat{a}^\dagger_{\mathbf{r}}\hat{a}_\mathbf{r}}$ numerically at varying temperatures. Within the linear spin-wave theory, the ensemble of magnon number is given by
\begin{equation}
    \ev{\hat{a}^\dagger_{\mathbf{r}=\mathbf{r}_\alpha}\hat{a}_{\mathbf{r}=\mathbf{r}_\alpha}} = \frac{1}{N_\textrm{muc}}\sum_\mathbf{k}\sum_\nu \left| \left[\mathcal{U}_{\mathbf{k}}\right]_{\alpha\nu} \right|^2 n^{(0)}_{\mathbf{k}\nu}.
    \label{eq:magnon_number_ensemble}
\end{equation}
In the self-consistent Hartree approximation [recall Eq.~\eqref{eq:selfconHartree}], the unitary matrix $\mathcal{U}_{\mathbf{k}}$ and the Bose factor $n^{(0)}_{\mathbf{k}\nu}$ in Eq.~(\ref{eq:magnon_number_ensemble}) are replaced by self-consistent solutions $\tilde{\mathcal{U}}_{\mathbf{k}}$ and $n^{(\textrm{c})}(\tilde{\omega}_{\mathbf{k}\nu})$, respectively. In the \textit{off-shell} approximations, the ensemble of magnon number is given by
\begin{equation}
    \ev{\hat{a}^\dagger_{\mathbf{r}=\mathbf{r}_\alpha}\hat{a}_{\mathbf{r}=\mathbf{r}_\alpha}}
    = \frac{1}{N_\textrm{muc}} \sum_\mathbf{k} \int n^{(\textrm{c})}(\omega)
    \left[\mathcal{U}_\mathbf{k}\mathcal{A}(\mathbf{k},\omega)\mathcal{U}^\dagger_\mathbf{k}\right]_{\alpha\alpha} d\omega.
\end{equation}
Note that $\ev{S^z} = \ev{S^z_{\mathbf{r}=\mathbf{r}_\mathrm{A}}} = \ev{S^z_{\mathbf{r}=\mathbf{r}_\mathrm{B}}}$ is satisfied in any cases.
It is also worth noting that, since we numerically found that the $T_\textrm{C}$ defined as the temperature at which $\ev{S^z}=0$ is satisfied is almost equivalent to that at which the onsite anisotropy-induced spin-gap vanishes within the on-shell approximation, namely, $\omega_{\mathbf{k}=\Gamma-}+\Sigma'_{\mathbf{k}=\Gamma--}(\omega_{\mathbf{k}=\Gamma-})=0$, we use the latter definition of $T_\textrm{C}$ only in Fig.~\ref{Fig12}(b) just to reduce computational cost.

Figure~\ref{Fig12}(a)  
presents $M$--$T$ curves obtained with different approximation schemes: linear spin-wave theory, self-consistent Hartree approximation with up to four-magnon terms (SCH(4)), and the \textit{off-shell} T-matrix resummation. Among these three approximation schemes, the resummation predicts $T_\textrm{C}$ most accurately, indicating that it remains highly reliable even near $T_\textrm{C}$.

\par We then reexamine the possibility of the temperature-driven gap closing and reopening and associated topological transitions using the resummation technique. Figure~\ref{Fig12}(c) shows the calculated single-magnon spectrum of CrI$_3$ at 40.6~K, which is slightly below the calculated $T_\textrm{C}^\textrm{T-matrix}\approx$ 41.8~K. 
While the gap size at the K point is slightly reduced from that at 0~K, it is still as large as 4~meV. 
In fact, we examined the gap size slightly above the calculated $T_\textrm{C}^\textrm{T-matrix}$ where the lower acoustic magnon sank down into the negative frequency region, and found the gap remains not closed, as seen in Fig.~\ref{Fig12}(e). 
This indicates the robustness of the topological gap up to $T_\textrm{C}$, and the topological transition predicted in Refs.~\cite{Lu2021,ReplyLu2021} could not be reproduced within the T-matrix resummation.

Robustness of the topological gap at the K point even in the vicinity of $T_\textrm{C}$ is corroborated by classical spin dynamics simulations. Figure~\ref{Fig12}(d) shows the rescaled dynamical spin-structure factor $\tilde{S}(\mathbf{q},\omega)=S(\mathbf{q},\omega)/W(\mathbf{q})$ at $T/T^\textrm{Classical}_\textrm{C}=0.91$, obtained by employing the stochastic Landau-Lifshitz-Gilbert equation.
Note that $W(\mathbf{q})=\int S(\mathbf{q},\omega)d\omega$, and $T^\textrm{Classical}_\textrm{C}\approx35$ K is also evaluated by the same methods.
While the spectral weight from the lower branch appears to be overestimated due to classical Boltzmann statistics, which tends to exaggerate thermal fluctuations, we can see that the topological gap at the K point is still present, as more clearly seen in the corresponding linecut presented in Fig.~\ref{Fig12}(f).

In summary, the T-matrix resummation and classical spin dynamics simulations suggest that the topological gap-closing behavior reported in Refs.~\cite{Lu2021,ReplyLu2021}, obtained within a self-consistent Hartree framework, arises from extending the Hartree approximation beyond the regime in which it can be expected to provide a reliable description of the system’s dynamics. While self-consistent Hartree approaches are well established to yield reasonable estimates for static quantities—such as Curie temperatures in ferromagnets—their application to dynamical properties, including spectral functions, generally calls for additional care.

\subsubsection{Observability of the topological gap}
\label{sec:CrI3_observability_of_gap}
So far, our T-matrix resummation scheme has shown that the topological gap in CrI$_3$ remains robust even in the high-temperature regime near $T_\textrm{C}$. While the robustness of the gap implies the absence of a topological transition involving the inversion of magnon Chern numbers, it does not necessarily guarantee the observability of the gapped Dirac magnons at finite temperatures, since the gap can be significantly smeared out by magnon lifetime effects~\cite{Habel2024}. To address this issue, we quantitatively analyze how magnon lifetime affects the visibility of gapped Dirac magnons near $T_\textrm{C}$.
\par Figure~\ref{Fig12}(b) shows a ratio between the gap size at the K point, $\Delta_\textrm{K}$, and a geometric mean of the lower- and upper-magnon branch linewidths $\Gamma_{\textrm{K}-}=-\textrm{Im}\Sigma_{\mathbf{k}=\textrm{K}--}^{(\textrm{T})}$ and $\Gamma_{\textrm{K}+}=-\textrm{Im}\Sigma_{\mathbf{k}=\textrm{K}++}^{(\textrm{T})}$ as a function of temperature $T$ and DMI $D_z$. The reference value for this ratio is set to $\Delta_\textrm{K}/2\sqrt{\Gamma_{\textrm{K}+}\Gamma_{\textrm{K}-}} = 1$, which corresponds to the threshold at which the dip between the two peaks in the spectral function becomes distinguishable. If the ratio exceeds 1, the gap is considered larger than the magnon linewidth, allowing the upper and lower branches to be observed as distinct features. Conversely, if the ratio falls below 1, the gap is effectively buried within the linewidth, making it impossible to resolve the two branches separately. In practice, considering that the resummation scheme may underestimate the linewidth by roughly a factor of two compared to experiment near $T_\mathrm{C}$, as seen in the case of CrBr$_3$ shown in Fig.~\ref{Fig07}(f), we adopt a more conservative criterion of $\Delta_\textrm{K}/2\sqrt{\Gamma_{\textrm{K}+}\Gamma_{\textrm{K}-}} = 2$.
As shown in Fig.~\ref{Fig12}(b), when $D_z=0.09$~meV, the ratio $\Delta_\textrm{K}/2\sqrt{\Gamma_{\textrm{K}+}\Gamma_{\textrm{K}-}}$ reaches approximately 2 near $T_\textrm{C}$. This means that the topological gap is observable in the vicinity of $T_\textrm{C}$ if $D_z$ exceeds the threshold $D_z^\star\approx0.09$~meV. In comparison, the actual value of $D_z$ in CrI$_3$ is 0.31~meV, and at 46~K---slightly above $T_\textrm{C}$---the ratio exceeds 3.5, while it is outside of the plotted region in Fig.~\ref{Fig12}(b). This suggests that the topological gap in CrI$_3$ remains clearly observable even in the vicinity of $T_\textrm{C}$. More generally, this result indicates that in honeycomb-lattice ferromagnets with a DMI to exchange ratio $D_z/J_1$ larger than $5\,\%$, the topological gap should be visible up to the ordering temperature if $J_2$ and further-neighbor exchange interactions are subleading, which is the case for CrSiTe$_3$ and CrGeTe$_3$.
\section{Discussion}
\label{sec:discussion}
\begin{figure}[tbh]
    \centering
    \includegraphics[scale=1.0]{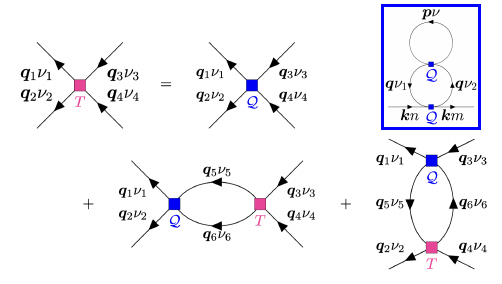}
    \caption{Diagrammatic representation of the two-channel parquet-type equation, which is not soluble in general. Inset shows the frequency-independent snowman diagram of order $1/S^2$.}
    \label{Fig13}
\end{figure}
\textit{Corrections in the long-wavelength limit.}-----
In the two-dimensional spin systems, the finite-temperature and isotropic limit lies in the regime where magnetic order is prohibited by the Mermin-Wagner theorem, making it a challenging task to accurately predict long-wavelength magnon energy shifts in such conditions.
We here discuss the breakdown of the interacting spin-wave perturbation theory in the long-wavelength and isotropic (Heisenberg) limits $(|\mathbf{k}|\rightarrow0, \ A\rightarrow0)$. The predictions of the thermal fluctuation-induced magnon energy shift in CrBr$_3$, shown in Fig.~\ref{Fig07}, exhibit a serious issue only in the long-wavelength limit: they overestimate the shift at high temperatures. In fact, both the Hartree+full sunset and the resummation underestimates $T_\textrm{C}$ compared to the experimentally observed value $T_\textrm{C}\approx32$~K, triggered by an overestimation of the energy shift of the lower-branch magnon in the long-wavelength limit.
\par A reason of this overestimation caused by the resummation is the short-wavelength limit employed when deriving the T-matrix self-energy given in Eq.~(\ref{eq:SE_T-matrix_SWL}). Unfortunately, since the real part of the T-matrix self-energy includes a principal value integral whose magnitude is hard to be evaluated both analytically and numerically, it is difficult to avoid this problem within the current BSE framework. It is also important to note that, contrary to the CrBr$_3$ case, in case of CrI$_3$ with a sizable monoaxial anisotropy of 0.22~meV, which is roughly eight times larger than that of CrBr$_3$ (0.028~meV), the resummation predicts $T_\textrm{C}$ quite accurately, indicating the resummation is free from such a breakdown except for the isotropic limit where the lower-branch magnon is nearly gapless.

The overestimation of energy shift caused by the Hartree+full sunset
is presumably because it lacks some of the frequency-independent real self-energy contributions of order $1/S^2$, for instance, those from the snowman diagram presented in the inset of Fig.~\ref{Fig13} or from $\mathcal{H}^{(6)}$. Incorporation of the snowman diagram into the interacting spin-wave theory is expected to relieve this problem. Moreover, this snowman diagram can be systematically incorporated into the BSE framework by extending it to the two-channel parquet-type equation presented in Fig.~\ref{Fig13}, which should be solved self-consistently. This may mitigate the overestimated energy shift of the long-wavelength acoustic mode in the original resummation scheme based on the BSE even without avoiding the aforementioned short-wavelength limit treatment given in Eq.~(\ref{eq:SE_T-matrix_SWL}). Such preciser analyses in the long-wavelength limit are left for future study.

\begin{figure}[tbh]
    \centering
    \includegraphics[scale=1.0]{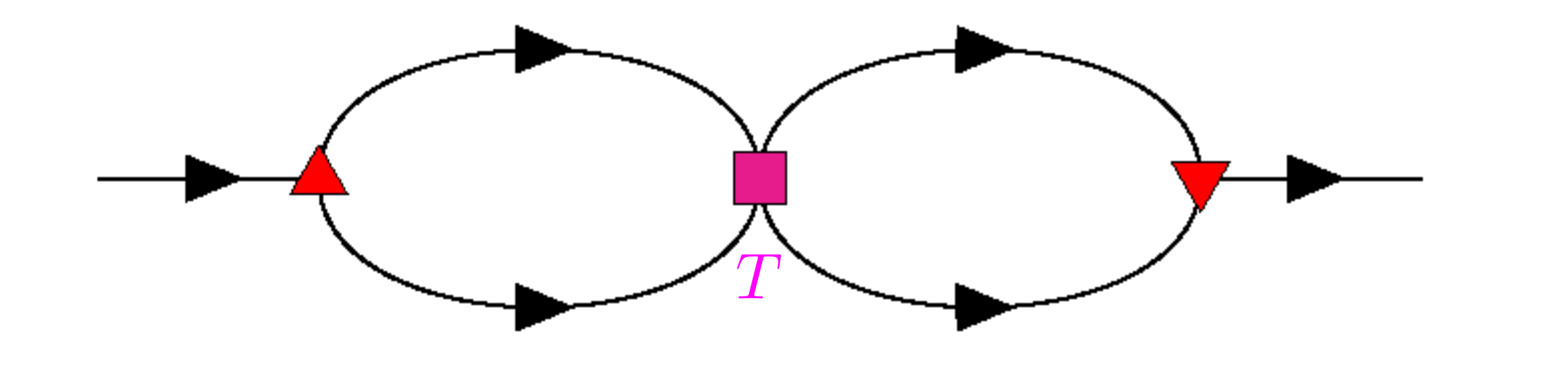}
    \caption{Diagrammatic representation of the \textit{quantum} magnon-bimagnon hybridization, which takes place in magnon-number-nonconserving magnets. The central four-magnon vertex denotes the T-matrix. The red triangular three-magnon vertices do not conserve magnon numbers.}
    \label{Fig14}
\end{figure}
\textit{Potential application of the T-matrix resummation for quantum hybridization between magnons and bimagnons.}----- 
Beyond the scope of this work, which focuses on thermal hybridization between single-particle and two-particle sectors in the presence of particle-number conservation, we point out that particle-number-nonconserving scattering processes containing the T-matrix potentially capture \textit{quantum} hybridization between them even at absolute zero.
For instance, a scattering process presented in Fig.~\ref{Fig14} describes \textit{quantum} magnon-bimagnon hybridization, where the central T-matrix encodes information of bimagnons, i.e., two-magnon bound states. It is important to mention that the left and right three-magnon vertices do not conserve magnon numbers, and thus directly correspond to quantum nature of the system considered. Self-energy of this diagrammatic process could provide an effective single-particle description of \textit{spin-multipolar topology}~\cite{Mook2023boundmagnon} associated with magnon-bimagnon hybridization.

\par \textit{Other material candidate.}----- In this study, we focused on Cr-based van der Waals honeycomb ferromagnets, which are well established to host spin-3/2 moments. However, to more clearly observe magnon excitation features such as linewidth broadening arising from thermally activated magnon-magnon interactions revealed in this work, materials with smaller spin magnitudes—more sensitive to quantum and thermal many-body effects—would be more suitable targets. Recently, several DFT studies~\cite{JZGeng2020,XRZou2025} have reported the emergence of out-of-plane ferromagnetism in monolayer TiBr$_3$, although its bulk counterpart tends to form a dimerized structure~\cite{Gapontsev2021}. In particular, Ref.~\cite{XRZou2025} derived a spin model to which the interacting spin-wave theory developed in this work can be directly applied. Studying this material thus constitutes a natural and promising extension of the present work.
\section{Summary}
\label{sec:summary}
We have developed a theoretical framework to describe the temperature-dependent behavior of magnons in honeycomb ferromagnets, focusing on van der Waals materials such as CrBr$_3$ and CrI$_3$. Moving beyond conventional truncated interacting spin-wave theory, we introduced a resummed perturbative approach that incorporates magnon-magnon interactions through a T-matrix formalism based on the Bethe-Salpeter equation. This enables us to capture the effects of two-magnon bound states—features largely overlooked in earlier studies but crucial for understanding magnon damping and spectral evolution, even at low temperatures.

Benchmarking against inelastic neutron scattering data for CrBr$_3$, our theory semiquantitatively reproduces the temperature-induced energy shifts and linewidth broadening across the Brillouin zone. It reveals clear signatures of many-body interactions, including distinct spectral fingerprints of bound-state scattering that elude lower-level approximations. Contrary to prior claims, our analysis of gapped magnon systems such as CrI$_3$ shows that the topological gap remains robust up to the magnetic transition temperature, with no evidence supporting a thermally driven topological transition. Additionally, we propose a practical criterion for assessing the observability of topological features in spectroscopic experiments.

Importantly, our resummation approach highlights the limitations of standard $1/S$ expansions, particularly near critical temperatures or in regimes where quantum effects persist despite large spin values. The results indicate that even systems with $S = 3/2$ can exhibit nontrivial quantum many-body phenomena when interaction channels are treated with care. These findings not only clarify how thermal fluctuations shape magnon spectra but also lay the groundwork for designing future spectroscopic experiments and magnonic devices that harness topological spin excitations in low-dimensional materials.

\section{Acknowledgment}

R.~E. and A.~M. are grateful to Stanislav E. Nikitin for allowing them to reproduce the data of inelastic neutron scattering experiment presented in Ref.~\cite{Nikitin2022}, and for valuable discussions with his experimental points of view. 
R.~E. also thanks Rhea Hoyer for discussions about construction methods of two-particle bases.

This work was supported by 
the Deutsche Forschungsgemeinschaft (DFG, German Research Foundation) - Project No.~504261060 (Emmy Noether Programme), 
JST~(Japan Science and Technology Agency) CREST (Grant No.~JPMJCR20T1), 
JSPS~(Japan Society for the Promotion of Science) KAKENHI (Grants No.~24H02231 and No.~25H00611), 
and Waseda University Grant for Special Research Projects (2025C-133). 
R.~E. was supported by Grant-in-Aid for JSPS Fellows (Grant No.~23KJ2047), JSPS Overseas Challenge Program for Young Researchers FY2024, and JSPS Overseas Fellowship. 
J.~K. acknowledges support from the Deutsche Forschungsgemeinschaft (DFG, German Research Foundation) under Germany’s Excellence Strategy- EXC-2111-390814868 and DFG Grants No. KN1254/1-2, KN1254/2-1 and TRR 360 - 492547816, as well as the Munich Quantum Valley, which is supported by the Bavarian state government with funds from the Hightech Agenda Bayern Plus. J.K. further acknowledges support from the Imperial-TUM flagship partnership.
A part of the numerical simulations was carried out at the Supercomputer Center, Institute for Solid State Physics, University of Tokyo.

Codes and data are available from the authors upon reasonable request.

\appendix
\onecolumngrid

\section{Spin-boson transformation}
\label{Appendix:SpinBosontransform}
As introduced in the main text, a standard form of the Holstein-Primakoff transformation is given by
\begin{equation}
\left\{
\begin{aligned}
    \hat{\mathsf{S}}^+_\mathbf{r} &= \frac{\hat{\mathsf{S}}^1_\mathbf{r}+i\hat{\mathsf{S}}^2_\mathbf{r}}{\sqrt{2}} = \sqrt{S-\frac{\hat{a}^\dagger_\mathbf{r}\hat{a}_\mathbf{r}}{2}}\hat{a}_\mathbf{r} = \sqrt{S}\left( \hat{a}_\mathbf{r} - \frac{1}{4S}\hat{a}^\dagger_\mathbf{r}\hat{a}_\mathbf{r}\hat{a}_\mathbf{r} + \cdots \right) \\
    \hat{\mathsf{S}}^-_\mathbf{r} &= \frac{\hat{\mathsf{S}}^1_\mathbf{r}-i\hat{\mathsf{S}}^2_\mathbf{r}}{\sqrt{2}} = \hat{a}^\dagger_\mathbf{r}\sqrt{S-\frac{\hat{a}^\dagger_\mathbf{r}\hat{a}_\mathbf{r}}{2}} = \sqrt{S}\left( \hat{a}^\dagger_\mathbf{r} - \frac{1}{4S}\hat{a}^\dagger_\mathbf{r}\hat{a}^\dagger_\mathbf{r}\hat{a}_\mathbf{r} + \cdots \right) \\
    \hat{\mathsf{S}}^0_\mathbf{r} &= S-\hat{a}^\dagger_\mathbf{r}\hat{a}_\mathbf{r}
\end{aligned}
\right. ,
\end{equation}
where $\hat{a}_\mathbf{r}$ $\left( \hat{a}^\dagger_\mathbf{r}\right)$ denotes the bosonic creation (annihilation) operators on the atomic site $\mathbf{r}$.
Using the Holstein-Primakoff transformation, bosonization of the spin-$S$ operator $\mathbf{S}_\mathbf{r}$ is performed as
\begin{align}
\begin{aligned}
    \mathbf{S}_\mathbf{r} &\rightarrow \hat{\mathsf{S}}^1_\mathbf{r}\mathbf{e}^1_\mathbf{r} + \hat{\mathsf{S}}^2_\mathbf{r}\mathbf{e}^2_\mathbf{r} + \hat{\mathsf{S}}^0_\mathbf{r}\mathbf{e}^0_\mathbf{r} \\
    &= \frac{\hat{\mathsf{S}}^+_\mathbf{r}+\hat{\mathsf{S}}^-_\mathbf{r}}{\sqrt{2}}\frac{\mathbf{e}^+_\mathbf{r}+\mathbf{e}^-_\mathbf{r}}{\sqrt{2}} + \frac{\hat{\mathsf{S}}^+_\mathbf{r}-\hat{\mathsf{S}}^-_\mathbf{r}}{\sqrt{2}i}\frac{\mathbf{e}^+_\mathbf{r}-\mathbf{e}^-_\mathbf{r}}{\sqrt{2}i} + \hat{\mathsf{S}}^0_\mathbf{r}\mathbf{e}^0_\mathbf{r} \\
    &= \hat{\mathsf{S}}^+_\mathbf{r}\mathbf{e}^-_\mathbf{r} + \hat{\mathsf{S}}^-_\mathbf{r}\mathbf{e}^+_\mathbf{r} + \hat{\mathsf{S}}^0_\mathbf{r}\mathbf{e}^0_\mathbf{r} \\
    &= \mathbf{e}^-_\mathbf{r} \sqrt{S} \left( \hat{a}_\mathbf{r}          
     - \frac{1}{4S} \hat{a}^\dagger_\mathbf{r}\hat{a}_\mathbf{r}\hat{a}_\mathbf{r}
     - \frac{1}{32S^2} \hat{a}^\dagger_\mathbf{r}\hat{a}_\mathbf{r}\hat{a}_\mathbf{r}
     - \frac{1}{32S^2} \hat{a}^\dagger_\mathbf{r}\hat{a}^\dagger_\mathbf{r}\hat{a}_\mathbf{r}\hat{a}_\mathbf{r}\hat{a}_\mathbf{r}
     + \cdots \right) \\
    &\quad + \mathbf{e}^+_\mathbf{r} \sqrt{S} \left( \hat{a}^\dagger_\mathbf{r}
     - \frac{1}{4S} \hat{a}^\dagger_\mathbf{r}\hat{a}^\dagger_\mathbf{r}\hat{a}_\mathbf{r}
     - \frac{1}{32S^2} \hat{a}^\dagger_\mathbf{r}\hat{a}^\dagger_\mathbf{r}\hat{a}_\mathbf{r}
     - \frac{1}{32S^2} \hat{a}^\dagger_\mathbf{r}\hat{a}^\dagger_\mathbf{r}\hat{a}^\dagger_\mathbf{r}\hat{a}_\mathbf{r}\hat{a}_\mathbf{r}
     + \cdots \right) \\
    &\quad + \mathbf{e}^0_\mathbf{r} \left( S-\hat{a}^\dagger_\mathbf{r}\hat{a}_\mathbf{r} \right).
\end{aligned}
\end{align}
Note that, for the out-of-plane ferromagnetic ground states, $\mathbf{e}^1_\mathbf{r}=(1,0,0)$, $\mathbf{e}^2_\mathbf{r}=(0,1,0)$, $\mathbf{e}^0_\mathbf{r}=(0,0,1)$, and $\mathbf{e}^\pm_\mathbf{r}=\left(\mathbf{e}^1_\mathbf{r}\pm i\mathbf{e}^2_\mathbf{r}\right)/\sqrt{2}$, independent of $\mathbf{r}$.
A boson reprentation of general two-body spin-spin interaction is given by
\begin{align}
\begin{aligned}
    &\quad \mathbf{S}_\mathbf{r}^\top \mathbb{J}_{\mathbf{r},\mathbf{r}'} \mathbf{S}_{\mathbf{r}'} \\
    &= \mathcal{J}^{+-}_{\mathbf{r},\mathbf{r}'} \left( S\hat{a}^\dagger_{\mathbf{r}}\hat{a}_{\mathbf{r}'}
     - \frac{1}{4} \left(1+\frac{1}{8S}\right) \hat{a}^\dagger_{\mathbf{r}}\hat{a}^\dagger_{\mathbf{r}}\hat{a}_{\mathbf{r}}\hat{a}_{\mathbf{r}'}
     - \frac{1}{4} \left(1+\frac{1}{8S}\right) \hat{a}^\dagger_{\mathbf{r}}\hat{a}^\dagger_{\mathbf{r}'}\hat{a}_{\mathbf{r}'}\hat{a}_{\mathbf{r}'}
    \right. \\ &\quad\quad\quad\quad\quad\quad \left.
     - \frac{1}{32S} \hat{a}^\dagger_{\mathbf{r} }\hat{a}^\dagger_{\mathbf{r}'}\hat{a}^\dagger_{\mathbf{r}'}\hat{a}        _{\mathbf{r}'}\hat{a}        _{\mathbf{r}'}\hat{a}        _{\mathbf{r}'}
     - \frac{1}{32S} \hat{a}^\dagger_{\mathbf{r} }\hat{a}^\dagger_{\mathbf{r} }\hat{a}^\dagger_{\mathbf{r} }\hat{a}        _{\mathbf{r} }\hat{a}        _{\mathbf{r} }\hat{a}        _{\mathbf{r}'}
     + \frac{1}{16S} \hat{a}^\dagger_{\mathbf{r} }\hat{a}^\dagger_{\mathbf{r} }\hat{a}^\dagger_{\mathbf{r}'}\hat{a}        _{\mathbf{r} }\hat{a}        _{\mathbf{r}'}\hat{a}        _{\mathbf{r}'}
     + \cdots \right) \\
    &+ \mathcal{J}^{-+}_{\mathbf{r},\mathbf{r}'} \left( S\hat{a}_{\mathbf{r}}\hat{a}^\dagger_{\mathbf{r}'}
     - \frac{1}{4} \left(1+\frac{1}{8S}\right) \hat{a}^\dagger_{\mathbf{r}'}\hat{a}^\dagger_{\mathbf{r}'}\hat{a}_{\mathbf{r}}\hat{a}_{\mathbf{r}'}
     - \frac{1}{4} \left(1+\frac{1}{8S}\right) \hat{a}^\dagger_{\mathbf{r} }\hat{a}^\dagger_{\mathbf{r}'}\hat{a}_{\mathbf{r}}\hat{a}_{\mathbf{r} }
    \right. \\ &\quad\quad\quad\quad\quad\quad \left.
     - \frac{1}{32S} \hat{a}^\dagger_{\mathbf{r}'}\hat{a}^\dagger_{\mathbf{r}'}\hat{a}^\dagger_{\mathbf{r}'}\hat{a}        _{\mathbf{r} }\hat{a}        _{\mathbf{r}'}\hat{a}        _{\mathbf{r}'}
     - \frac{1}{32S} \hat{a}^\dagger_{\mathbf{r} }\hat{a}^\dagger_{\mathbf{r} }\hat{a}^\dagger_{\mathbf{r}'}\hat{a}        _{\mathbf{r} }\hat{a}        _{\mathbf{r} }\hat{a}        _{\mathbf{r} }
     + \frac{1}{16S} \hat{a}^\dagger_{\mathbf{r} }\hat{a}^\dagger_{\mathbf{r}'}\hat{a}^\dagger_{\mathbf{r}'}\hat{a}        _{\mathbf{r} }\hat{a}        _{\mathbf{r} }\hat{a}        _{\mathbf{r}'}
     + \cdots \right) \\
    &+\mathcal{J}^{00}_{\mathbf{r},\mathbf{r}'} \left( S^2 - S\hat{a}^\dagger_{\mathbf{r}}\hat{a}_{\mathbf{r}} - S\hat{a}^\dagger_{\mathbf{r}'}\hat{a}_{\mathbf{r}'}
     + \hat{a}^\dagger_{\mathbf{r}}\hat{a}^\dagger_{\mathbf{r}'}\hat{a}_{\mathbf{r}}\hat{a}_{\mathbf{r}'} \right) \\
    &+ \mathcal{J}^{+0}_{\mathbf{r},\mathbf{r}'} \left( \cdots \right) 
     + \mathcal{J}^{0+}_{\mathbf{r},\mathbf{r}'} \left( \cdots \right)
     + \mathcal{J}^{-0}_{\mathbf{r},\mathbf{r}'} \left( \cdots \right)
     + \mathcal{J}^{0-}_{\mathbf{r},\mathbf{r}'} \left( \cdots \right)
     + \mathcal{J}^{++}_{\mathbf{r},\mathbf{r}'} \left( \cdots \right)
     + \mathcal{J}^{--}_{\mathbf{r},\mathbf{r}'} \left( \cdots \right),
\end{aligned}
\end{align}
where $\mathbb{J}_{\mathbf{r},\mathbf{r}'}$ is a 3$\times$3 matrix. Note that $\mathcal{J}^{\chi\chi'}_{\mathbf{r},\mathbf{r}'}=\left[\mathbf{e}^\chi_\mathbf{r}\right]^\top\mathbb{J}_{\mathbf{r},\mathbf{r}'}\left[\mathbf{e}^{\chi'}_{\mathbf{r}'}\right]$ $(\chi,\chi'=\pm,0)$. Here, we do not show magnon-number nonconserving terms explicitly, since these contributions are canceled due to U(1) symmetry of the Hamiltonian given by Eq.~(\ref{eq:SpinHm}) with the out-of-plane ferromagnetic order.
\section{Derivation of the T-matrix self-energy}
\label{Appendix:T-matrix}
\subsection{Solution of the Bethe-Salpeter equation}
In the following, we denote the decomposition of the four-point (two-in-two-out) vertex as
\begin{equation}
    \mathcal{Q}_{\mathbf{q}_1,\mathbf{q}_2\leftrightarrow\mathbf{q}_3,\mathbf{q}_4}^{\nu_1,\nu_2\leftrightarrow\nu_3,\nu_4} = 
    \mathbf{v}^\mathrm{L}_{\nu_1,\nu_2}\left( \mathbf{q}_1,\mathbf{q}_2 \right) 
    \frac{\Gamma}{4N_\textrm{muc}} 
    \mathbf{v}^\mathrm{R}_{\nu_3,\nu_4}\left( \mathbf{q}_3,\mathbf{q}_4 \right).
\end{equation}
where $\Gamma$ is a $5N_\mathrm{sub}^2N_\mathrm{bond}\times5N_\mathrm{sub}^2N_\mathrm{bond}$-dimensional matrix, and $\mathbf{v}^\mathrm{L}_{\nu_1,\nu_2}\left( \mathbf{q}_1,\mathbf{q}_2 \right)$ $\left(\mathbf{v}^\mathrm{R}_{\nu_3,\nu_4}\left( \mathbf{q}_3,\mathbf{q}_4 \right)\right)$ is a $5N_\mathrm{sub}^2N_\mathrm{bond}$-dimensional row (column) vector. The factor 5 comes from the number of elements in a vector $\mathfrak{Y}(\mathbf{q}_a,\mathbf{q}_b;\bm{\delta})$ given by Eq.~(\ref{eq:vector_Y}).
$N_\textrm{sub}$ and $N_\textrm{bond}$ denote the number of sublattices in a magnetic unit cell and that of bonds connecting different sites $\{\mathbf{r},\mathbf{r}'\}$ with nonzero spin-spin interaction $\mathbb{J}_{{\mathbf{r},\mathbf{r}'}}$, respectively.
\par A solution of the BSE given by Eq.~(\ref{eq:BSE}) has the form~\cite{Silberglitt1967}:
\begin{equation}
    T_{\mathbf{q}_1,\mathbf{q}_2\leftrightarrow\mathbf{q}_3,\mathbf{q}_4}^{\nu_1,\nu_2\leftrightarrow\nu_3,\nu_4}(z) = \mathbf{v}^\mathrm{L}_{\nu_1,\nu_2}\left( \mathbf{q}_1,\mathbf{q}_2 \right) \left[ \mathsf{1}-A_{\mathbf{q}_1+\mathbf{q}_2}(z) \right]^{-1} \frac{\Gamma}{4N_\textrm{muc}} \mathbf{v}^\mathrm{R}_{\nu_3,\nu_4}\left( \mathbf{q}_3,\mathbf{q}_4 \right).
\end{equation}
Using the relation $\left( \mathsf{1}-A \right)^{-1} = \mathsf{1}+A+A^2+\cdots$, the $(n+1)$-th order term of the left side with respect to $A$ corresponds to the $n$-th order term of the right side. For $n=0$, one gets the following relation:
\begin{align}
\begin{aligned}
    &\quad \ \mathbf{v}^\mathrm{L}_{\nu_1,\nu_2}\left( \mathbf{q}_1,\mathbf{q}_2 \right) A_\mathbf{K}(z) \frac{\Gamma}{4N_\textrm{muc}} \mathbf{v}^\mathrm{R}_{\nu_3,\nu_4}\left( \mathbf{q}_3,\mathbf{q}_4\right) \\
    &= \frac{2}{\hbar} \sum_{\mathbf{q}_5,\mathbf{q}_6} \sum_{\nu_5,\nu_6}
    \left[ \mathbf{v}^\mathrm{L}_{\nu_1,\nu_2}\left( \mathbf{q}_1,\mathbf{q}_2 \right) \frac{\Gamma}{4N_\textrm{muc}} \mathbf{v}^\mathrm{R}_{\nu_5,\nu_6}\left( \mathbf{q}_5,\mathbf{q}_6\right) \right]
    \frac{1+n^{(0)}_{\mathbf{q}_5\nu_5}+n^{(0)}_{\mathbf{q}_6\nu_6}}{z-\omega_{\mathbf{q}_5\nu_5}-\omega_{\mathbf{q}_6\nu_6}}
    \left[ \mathbf{v}^\mathrm{L}_{\nu_5,\nu_6}\left( \mathbf{q}_5,\mathbf{q}_6 \right) \frac{\Gamma}{4N_\textrm{muc}} \mathbf{v}^\mathrm{R}_{\nu_3,\nu_4}\left( \mathbf{q}_3,\mathbf{q}_4\right) \right].
    \label{eq:BSE_n=0}
\end{aligned}
\end{align}
Eq.~(\ref{eq:BSE_n=0}) immediately gives the following solution:
\begin{align}
\begin{aligned}
    A_\mathbf{K}(z) &= \frac{2}{\hbar} \sum_{\mathbf{q}_5,\mathbf{q}_6} \sum_{\nu_5,\nu_6}
    \left[ \frac{\Gamma}{4N_\textrm{muc}} \mathbf{v}^\mathrm{R}_{\nu_5,\nu_6}\left( \mathbf{q}_5,\mathbf{q}_6\right) \otimes \mathbf{v}^\mathrm{L}_{\nu_5,\nu_6}\left( \mathbf{q}_5,\mathbf{q}_6 \right) \right]
    \frac{1+n^{(0)}_{\mathbf{q}_5\nu_5}+n^{(0)}_{\mathbf{q}_6\nu_6}}{z-\omega_{\mathbf{q}_5\nu_5}-\omega_{\mathbf{q}_6\nu_6}} \\
    &= \frac{2}{\hbar} \sum_{\mathbf{q}_5} \sum_{\nu_5,\nu_6}
    \left[ \frac{\Gamma}{4N_\textrm{muc}} \mathbf{v}^\mathrm{R}_{\nu_5,\nu_6}\left( \mathbf{q}_5,\mathbf{K}-\mathbf{q}_5\right) \otimes \mathbf{v}^\mathrm{L}_{\nu_5,\nu_6}\left( \mathbf{q}_5,\mathbf{K}-\mathbf{q}_5 \right) \right]
    \frac{1+n^{(0)}_{\mathbf{q}_5\nu_5}+n^{(0)}_{\mathbf{K}-\mathbf{q}_5\nu_6}}{z-\omega_{\mathbf{q}_5\nu_5}-\omega_{\mathbf{K}-\mathbf{q}_5\nu_6}}.
\end{aligned}
\end{align}
\subsection{Decomposition of the vertex}
The $5N_\mathrm{sub}^2N_\mathrm{bond}\times5N_\mathrm{sub}^2N_\mathrm{bond}$-dimensional matrix $\Gamma$ can be block-diagonalizable into $N_\mathrm{bond}$ irreducible $5N_\mathrm{sub}^2\times5N_\mathrm{sub}^2$-dimensional matrices $\Gamma_{\{\mathbf{r},\mathbf{r}'\}}$. This block-diagonalization is explicitly given in the form
\begin{align}
\begin{aligned}
    \mathcal{Q}_{{\mathbf{q}_1,\mathbf{q}_2\leftrightarrow\mathbf{q}_3,\mathbf{q}_4}}^{\nu_1,\nu_2\leftrightarrow\nu_3,\nu_4} 
    &= \mathbf{v}^\mathrm{L}_{\nu_1,\nu_2}(\mathbf{q}_1,\mathbf{q}_2) \frac{\Gamma}{4N_\textrm{muc}} \mathbf{v}^\mathrm{R}_{\nu_3,\nu_4}(\mathbf{q}_3,\mathbf{q}_4) \\
    &= \frac{1}{4N_\textrm{muc}} \mathbf{v}^\mathrm{L}_{\nu_1,\nu_2}(\mathbf{q}_1,\mathbf{q}_2)
    \left[ \bigoplus_{\{\mathbf{r},\mathbf{r}'\}_\textrm{pair}} \Gamma_{\{\mathbf{r},\mathbf{r}'\}} \right] \mathbf{v}^\mathrm{R}_{\nu_3,\nu_4}(\mathbf{q}_3,\mathbf{q}_4) \\
    &= \frac{1}{4N_\textrm{muc}} \sum_{\{\mathbf{r},\mathbf{r}'\}_\textrm{pair}} \mathbf{v}^\mathrm{L}_{\nu_1,\nu_2}(\mathbf{q}_1,\mathbf{q}_2;{\mathbf{r}'-\mathbf{r}})
    \Gamma_{\{\mathbf{r},\mathbf{r}'\}} \mathbf{v}^\mathrm{R}_{\nu_3,\nu_4}(\mathbf{q}_3,\mathbf{q}_4;{\mathbf{r}'-\mathbf{r}}) \\
    &= \frac{1}{4N_\textrm{muc}} \sum_{\{\mathbf{r},\mathbf{r}'\}_\textrm{pair}} \sum_{\alpha\beta\gamma\delta}
    \left[ \mathbf{v}^\mathrm{L}_{\nu_1,\nu_2}(\mathbf{q}_1,\mathbf{q}_2;{\mathbf{r}'-\mathbf{r}}) \right]_{(\alpha\beta)}
    \left[ \Gamma_{\{\mathbf{r},\mathbf{r}'\}} \right]_{(\alpha\beta)(\gamma\delta)} 
    \left[ \mathbf{v}^\mathrm{R}_{\nu_3,\nu_4}(\mathbf{q}_3,\mathbf{q}_4;{\mathbf{r}'-\mathbf{r}}) \right]_{(\gamma\delta)},
\end{aligned}
\end{align}
where $\alpha$, $\beta$, $\gamma$, and $\delta$ $(=1,2,\cdots,N_\textrm{sub})$ are sublattice indices. 
$\oplus_{\{\mathbf{r},\mathbf{r}'\}_\textrm{pair}}$ and $\sum_{\{\mathbf{r},\mathbf{r}'\}_\textrm{pair}}$ sum over all pairs $\{\mathbf{r},\mathbf{r}'\}$ with nonzero $\mathbb{J}_{\mathbf{r},\mathbf{r}'}$ in a magnetic unit cell.
Here we choose the basis of the four-point vertex decomposition with the following functional form:
\begin{align}
\begin{aligned}
    \left[ \mathbf{v}^\mathrm{L}_{\nu_1,\nu_2}(\mathbf{q}_1,\mathbf{q}_2;\bm{\delta}) \right]_{\left(\alpha\beta\right)}^\top
    &= \left[\mathcal{U}_{\mathbf{q}_1}\right]_{\alpha\nu_1}^*
       \left[\mathcal{U}_{\mathbf{q}_2}\right]_{\beta \nu_2}^*
       \mathfrak{Y}(\mathbf{q}_1,\mathbf{q}_2;\bm{\delta}), \\
    \left[ \mathbf{v}^\mathrm{R}_{\nu_3,\nu_4}(\mathbf{q}_3,\mathbf{q}_4;\bm{\delta}) \right]_{\left(\gamma\delta\right)}
    &= \left[\mathcal{U}_{\mathbf{q}_3}\right]_{\gamma\nu_3}
       \left[\mathcal{U}_{\mathbf{q}_4}\right]_{\delta\nu_4}
       \mathfrak{Y}(\mathbf{q}_3,\mathbf{q}_4;\bm{\delta}),
\end{aligned}
\end{align}
where $\mathfrak{Y}(\mathbf{q}_a,\mathbf{q}_b;\bm{\delta})$ is a five-component vector given by
\begin{equation}
    \mathfrak{Y}(\mathbf{q}_a,\mathbf{q}_b;\bm{\delta}) = \left( 
    1,
    e^{ i\mathbf{q}_a\cdot\bm{\delta}},
    e^{-i\mathbf{q}_a\cdot\bm{\delta}},
    e^{ i\mathbf{q}_b\cdot\bm{\delta}},
    e^{-i\mathbf{q}_b\cdot\bm{\delta}} \right)^\top.
    \label{eq:vector_Y}
\end{equation}
A symmetrized $5\times5$ block element of the $5N_\mathrm{sub}^2\times5N_\mathrm{sub}^2$ matrix $\Gamma_{\ev{\mathbf{r},\mathbf{r}'}}$ is given by
\begin{equation}
    \left[\Gamma_{\{\mathbf{r}_i,\mathbf{r}_j\}}\right]_{\left(\alpha\beta\right)\left(\gamma\delta\right)} = \frac{1}{2} \left( \begin{array}{ccccc}
          0 
        &-\mathcal{J}^{-+}_{\gamma\delta} \delta_{\alpha i}\delta_{\beta i}\delta_{\gamma j}\delta_{\delta i}
        &-\mathcal{J}^{-+}_{\gamma\delta} \delta_{\alpha j}\delta_{\beta j}\delta_{\gamma i}\delta_{\delta j}
        &-\mathcal{J}^{+-}_{\gamma\delta} \delta_{\alpha i}\delta_{\beta i}\delta_{\gamma i}\delta_{\delta j} 
        &-\mathcal{J}^{+-}_{\gamma\delta} \delta_{\alpha j}\delta_{\beta j}\delta_{\gamma j}\delta_{\delta i}
        \\
         -\mathcal{J}^{+-}_{\alpha\beta } \delta_{\alpha i}\delta_{\beta j}\delta_{\gamma j}\delta_{\delta j} 
        & 0 
        &+\mathcal{J}^{00}_{\alpha\beta } \delta_{\alpha i}\delta_{\beta j}\delta_{\gamma i}\delta_{\delta j} 
        & 0 
        &+\mathcal{J}^{00}_{\alpha\beta } \delta_{\alpha i}\delta_{\beta j}\delta_{\gamma j}\delta_{\delta i} 
        \\
         -\mathcal{J}^{+-}_{\alpha\beta } \delta_{\alpha j}\delta_{\beta i}\delta_{\gamma i}\delta_{\delta i}
        &+\mathcal{J}^{00}_{\alpha\beta } \delta_{\alpha j}\delta_{\beta i}\delta_{\gamma j}\delta_{\delta i}
        & 0 
        &+\mathcal{J}^{00}_{\alpha\beta } \delta_{\alpha j}\delta_{\beta i}\delta_{\gamma i}\delta_{\delta j}
        & 0 
        \\
         -\mathcal{J}^{-+}_{\alpha\beta } \delta_{\alpha j}\delta_{\beta i}\delta_{\gamma j}\delta_{\delta j}
        & 0 
        &+\mathcal{J}^{00}_{\beta\alpha } \delta_{\alpha j}\delta_{\beta i}\delta_{\gamma i}\delta_{\delta j} 
        & 0 
        &+\mathcal{J}^{00}_{\beta\alpha } \delta_{\alpha j}\delta_{\beta i}\delta_{\gamma j}\delta_{\delta i} 
        \\
         -\mathcal{J}^{-+}_{\alpha\beta } \delta_{\alpha i}\delta_{\beta j}\delta_{\gamma i}\delta_{\delta i}
        &+\mathcal{J}^{00}_{\beta\alpha } \delta_{\alpha i}\delta_{\beta j}\delta_{\gamma j}\delta_{\delta i}
        & 0 
        &+\mathcal{J}^{00}_{\beta\alpha } \delta_{\alpha i}\delta_{\beta j}\delta_{\gamma i}\delta_{\delta j} 
        & 0
    \end{array}\right).
\end{equation}
Note that, just for concreteness, additional sublattice indices $i,j=1,2,\cdots,N_\textrm{sub}$ are introduced.
\subsection{Short-wavelength limit}
\label{Appendix:shortwavelengthlimit}
The T-matrix self-energy~\cite{Silberglitt1967} is given by
\begin{equation}
    \Sigma^{\mathrm{(T)}}_{\mathbf{k},mn}(i\omega_n) = -\frac{4}{\hbar}\sum_\mathbf{p}\sum_\nu \frac{1}{\beta\hbar} \sum_{i\omega_s} \frac{1}{i\omega_s-\omega_{\mathbf{p}\nu}} T_{{\mathbf{k},\mathbf{p}\leftrightarrow\mathbf{k},\mathbf{p}}}^{m,\nu\leftrightarrow n,\nu}(i\omega_n+i\omega_s).
\end{equation}
Using the Kramers-Kronig relation (dispersion relation)~\cite{Rastelli2011}:
\begin{align}
\begin{split}
    &\quad \left\{
    \begin{aligned}
    \mathrm{Re} \left[ T_{{\mathbf{k},\mathbf{p}\leftrightarrow\mathbf{k},\mathbf{p}}}^{m,\nu\leftrightarrow m,\nu}(\omega) \right] &= \mathcal{Q}_{{\mathbf{k},\mathbf{p}\leftrightarrow\mathbf{k},\mathbf{p}}}^{m,\nu\leftrightarrow m,\nu} - \frac{\mathrm{P}}{\pi} \int_{-\infty}^\infty d\omega' \frac{\mathrm{Im} \left[ T_{{\mathbf{k},\mathbf{p}\leftrightarrow\mathbf{k},\mathbf{p}}}^{m,\nu\leftrightarrow m,\nu}(\omega') \right]}{\omega-\omega'} \\
    T_{{\mathbf{k},\mathbf{p}\leftrightarrow\mathbf{k},\mathbf{p}}}^{m,\nu\leftrightarrow m,\nu}(\omega+i\epsilon) &= \mathrm{Re}\left[ T_{{\mathbf{k},\mathbf{p}\leftrightarrow\mathbf{k},\mathbf{p}}}^{m,\nu\leftrightarrow m,\nu}(\omega) \right] + i\mathrm{Im}\left[ T_{{\mathbf{k},\mathbf{p}\leftrightarrow\mathbf{k},\mathbf{p}}}^{m,\nu\leftrightarrow m,\nu}(\omega) \right]
    \end{aligned} \right. \\
    &\Longleftrightarrow T_{{\mathbf{k},\mathbf{p}\leftrightarrow\mathbf{k},\mathbf{p}}}^{m,\nu\leftrightarrow m,\nu}(z) = \mathcal{Q}_{{\mathbf{k},\mathbf{p}\leftrightarrow\mathbf{k},\mathbf{p}}}^{m,\nu\leftrightarrow m,\nu} - \frac{1}{\pi} \int_{-\infty}^\infty d\omega' \frac{\mathrm{Im} \left[ T_{{\mathbf{k},\mathbf{p}\leftrightarrow\mathbf{k},\mathbf{p}}}^{m,\nu\leftrightarrow m,\nu}(\omega') \right]}{z-\omega'}
    \label{eq:KramersKronig_diag}
\end{split}
\end{align}
where $z=\omega+i0^+$, the T-matrix is rewritten as follows:
\begin{align}
\begin{aligned}
    \Sigma^{\mathrm{(T)}}_{\mathbf{k},mm}(i\omega_n) &= -\frac{4}{\hbar}\sum_\mathbf{p}\sum_\nu \frac{1}{\beta\hbar} \sum_s \frac{1}{i\omega_s-\omega_{\mathbf{p}\nu}} \left\{ \mathcal{Q}_{{\mathbf{k},\mathbf{p}\leftrightarrow\mathbf{k},\mathbf{p}}}^{m,\nu\leftrightarrow m,\nu} - \frac{1}{\pi} \int_{-\infty}^\infty d\omega' \frac{\mathrm{Im} \left[ T_{{\mathbf{k},\mathbf{p}\leftrightarrow\mathbf{k},\mathbf{p}}}^{m,\nu\leftrightarrow m,\nu}(\omega') \right]}{i\omega_s+i\omega_n-\omega'} \right\} \\
    &= \frac{4}{\hbar}\sum_\mathbf{p}\sum_\nu \mathcal{Q}_{{\mathbf{k},\mathbf{p}\leftrightarrow\mathbf{k},\mathbf{p}}}^{m,\nu\leftrightarrow m,\nu} n^{(0)}_{\mathbf{p}\nu} \\
    &\quad\quad + \frac{4}{\hbar}\sum_\mathbf{p}\sum_\nu \frac{1}{\pi} \int_{-\infty}^\infty d\omega' \frac{\mathrm{Im} \left[ T_{{\mathbf{k},\mathbf{p}\leftrightarrow\mathbf{k},\mathbf{p}}}^{m,\nu\leftrightarrow m,\nu}(\omega') \right]}{i\omega_n-\omega'+\omega_{\mathbf{p}\nu}} \frac{1}{\beta\hbar} \sum_s \left( \frac{1}{i\omega_s-\omega_{\mathbf{p}\nu}} - \frac{1}{i\omega_s+i\omega_n-\omega'} \right) \\
    &= \frac{4}{\hbar}\sum_\mathbf{p}\sum_\nu n^{(0)}_{\mathbf{p}\nu} \left\{ \mathcal{Q}_{{\mathbf{k},\mathbf{p}\leftrightarrow\mathbf{k},\mathbf{p}}}^{m,\nu\leftrightarrow m,\nu} - \frac{1}{\pi} \int_{-\infty}^\infty d\omega' \frac{\mathrm{Im} \left[ T_{{\mathbf{k},\mathbf{p}\leftrightarrow\mathbf{k},\mathbf{p}}}^{m,\nu\leftrightarrow m,\nu}(\omega') \right]}{i\omega_n-\omega'+\omega_{\mathbf{p}\nu}} \right\} \\
    &\quad\quad + \frac{4}{\hbar}\sum_\mathbf{p}\sum_\nu \frac{1}{\pi} \int_{-\infty}^\infty d\omega' \frac{\mathrm{Im} \left[ T_{{\mathbf{k},\mathbf{p}\leftrightarrow\mathbf{k},\mathbf{p}}}^{m,\nu\leftrightarrow m,\nu}(\omega') \right]}{i\omega_n-\omega'+\omega_{\mathbf{p}\nu}} n^{(\textrm{c})}(\omega') \\
    &= \frac{4}{\hbar}\sum_\mathbf{p}\sum_\nu T_{{\mathbf{k},\mathbf{p}\leftrightarrow\mathbf{k},\mathbf{p}}}^{m,\nu\leftrightarrow m,\nu}(i\omega_n+\omega_{\mathbf{p}\nu}) n^{(0)}_{\mathbf{p}\nu}
     + \frac{4}{\hbar}\sum_\mathbf{p}\sum_\nu \frac{1}{\pi} \int_{-\infty}^\infty d\omega' \frac{\mathrm{Im} \left[ T_{{\mathbf{k},\mathbf{p}\leftrightarrow\mathbf{k},\mathbf{p}}}^{m,\nu\leftrightarrow m,\nu}(\omega') \right]}{i\omega_n-\omega'+\omega_{\mathbf{p}\nu}} n^{(\textrm{c})}(\omega').
    \label{eq:Tmat_diag_separation}
\end{aligned}
\end{align}
$\textrm{P}$ denotes the principal value. In the short-wavelength limit, one neglects the second term because it is proportional to the ``non-elementary" Bose factor $n^{(\textrm{c})}(\omega)$, which is negligible compared to the ``elementary" Bose factor $n^{(0)}_{\mathbf{p}\nu}$. As a result, the diagonal part of the approximated T-matrix self-energy is given by
\begin{equation}
    \Sigma^{\mathrm{(T)}}_{\mathbf{k},mm}(i\omega_n) \approx \frac{4}{\hbar}\sum_\mathbf{p}\sum_\nu T_{{\mathbf{k},\mathbf{p}\leftrightarrow\mathbf{k},\mathbf{p}}}^{m,\nu\leftrightarrow m,\nu}(i\omega_n+\omega_{\mathbf{p}\nu}) n^{(0)}_{\mathbf{p}\nu}. \label{eq:approxTmat_diag}
\end{equation}
\begin{figure}[tbh]
    \centering
    \includegraphics[scale=1.0]{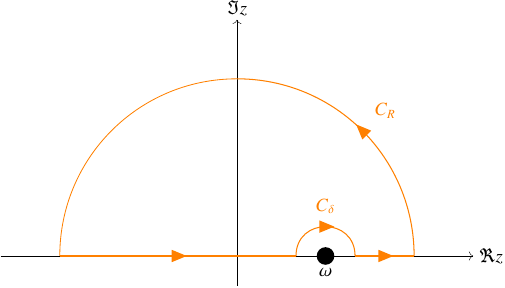}
    \caption{Closed path of integration $C$. As a consequence of the causality, the numerator of the integrand $T_{{\mathbf{k},\mathbf{p}\leftrightarrow\mathbf{k},\mathbf{p}}}^{m,\nu\leftrightarrow n,\nu}(z)-\mathcal{Q}_{{\mathbf{k},\mathbf{p}\leftrightarrow\mathbf{k},\mathbf{p}}}^{m,\nu\leftrightarrow n,\nu}$ is analytic in the upper half plane.}
    \label{fig:path}
\end{figure}
Next we consider the generalization of the above discussion to the off-diagonal components. One considers the following complex integral along the closed path $C$ presented in Fig.~\ref{fig:path}:
\begin{align}
\begin{aligned}
    0 &= \oint dz \frac{T_{{\mathbf{k},\mathbf{p}\leftrightarrow\mathbf{k},\mathbf{p}}}^{m,\nu\leftrightarrow n,\nu}(z)-\mathcal{Q}_{{\mathbf{k},\mathbf{p}\leftrightarrow\mathbf{k},\mathbf{p}}}^{m,\nu\leftrightarrow n,\nu}}{z-\omega} \\
    &= \mathrm{P} \int_{-R}^R d\omega' \frac{T_{{\mathbf{k},\mathbf{p}\leftrightarrow\mathbf{k},\mathbf{p}}}^{m,\nu\leftrightarrow n,\nu}(\omega')-\mathcal{Q}_{{\mathbf{k},\mathbf{p}\leftrightarrow\mathbf{k},\mathbf{p}}}^{m,\nu\leftrightarrow n,\nu}}{\omega'-\omega} 
    + \int_{C_\delta} dz \frac{T_{{\mathbf{k},\mathbf{p}\leftrightarrow\mathbf{k},\mathbf{p}}}^{m,\nu\leftrightarrow n,\nu}(z)-\mathcal{Q}_{{\mathbf{k},\mathbf{p}\leftrightarrow\mathbf{k},\mathbf{p}}}^{m,\nu\leftrightarrow n,\nu}}{z-\omega} 
    + \int_{C_R}      dz \frac{T_{{\mathbf{k},\mathbf{p}\leftrightarrow\mathbf{k},\mathbf{p}}}^{m,\nu\leftrightarrow n,\nu}(z)-\mathcal{Q}_{{\mathbf{k},\mathbf{p}\leftrightarrow\mathbf{k},\mathbf{p}}}^{m,\nu\leftrightarrow n,\nu}}{z-\omega} \\
    &\rightarrow -\mathrm{P} \int_{-\infty}^\infty d\omega' \frac{T_{{\mathbf{k},\mathbf{p}\leftrightarrow\mathbf{k},\mathbf{p}}}^{m,\nu\leftrightarrow n,\nu}(\omega')-\mathcal{Q}_{{\mathbf{k},\mathbf{p}\leftrightarrow\mathbf{k},\mathbf{p}}}^{m,\nu\leftrightarrow n,\nu}}{\omega-\omega'} - i\pi \left[ T_{{\mathbf{k},\mathbf{p}\leftrightarrow\mathbf{k},\mathbf{p}}}^{m,\nu\leftrightarrow n,\nu}(\omega)-\mathcal{Q}_{{\mathbf{k},\mathbf{p}\leftrightarrow\mathbf{k},\mathbf{p}}}^{m,\nu\leftrightarrow n,\nu} \right] \ \ (\delta\rightarrow0,R\rightarrow\infty)
\end{aligned}
\end{align}
A natural extension of Eq.~(\ref{eq:KramersKronig_diag}) to the full matrix elements, not only the diagonal components but also the off-diagonal ones, is given by
\begin{align}
\begin{aligned}
    T_{{\mathbf{k},\mathbf{p}\leftrightarrow\mathbf{k},\mathbf{p}}}^{m,\nu\leftrightarrow n,\nu}(\omega) &= \mathcal{Q}_{{\mathbf{k},\mathbf{p}\leftrightarrow\mathbf{k},\mathbf{p}}}^{m,\nu\leftrightarrow n,\nu} - \frac{\mathrm{P}}{i\pi} \int_{-\infty}^\infty d\omega' \frac{T_{{\mathbf{k},\mathbf{p}\leftrightarrow\mathbf{k},\mathbf{p}}}^{m,\nu\leftrightarrow n,\nu}(\omega') - \mathcal{Q}_{{\mathbf{k},\mathbf{p}\leftrightarrow\mathbf{k},\mathbf{p}}}^{m,\nu\leftrightarrow n,\nu} }{\omega-\omega'} \\
    \Longleftrightarrow T_{{\mathbf{k},\mathbf{p}\leftrightarrow\mathbf{k},\mathbf{p}}}^{m,\nu\leftrightarrow n,\nu}(z) &= \mathcal{Q}_{{\mathbf{k},\mathbf{p}\leftrightarrow\mathbf{k},\mathbf{p}}}^{m,\nu\leftrightarrow n,\nu} - \frac{1}{i\pi} \int_{-\infty}^\infty d\omega' \frac{T_{{\mathbf{k},\mathbf{p}\leftrightarrow\mathbf{k},\mathbf{p}}}^{m,\nu\leftrightarrow n,\nu}(\omega') - \mathcal{Q}_{{\mathbf{k},\mathbf{p}\leftrightarrow\mathbf{k},\mathbf{p}}}^{m,\nu\leftrightarrow n,\nu} }{z-\omega'}.
    \label{eq:KramersKronig_offdiag}
\end{aligned}
\end{align}
Indeed, we can easily check that the diagonal part of Eq.~(\ref{eq:KramersKronig_offdiag}) reduces to Eq.~(\ref{eq:KramersKronig_diag}) by the following calculations:
\begin{align}
\begin{aligned}
    T_{{\mathbf{k},\mathbf{p}\leftrightarrow\mathbf{k},\mathbf{p}}}^{m,\nu\leftrightarrow m,\nu}(\omega) 
    &= \mathcal{Q}_{{\mathbf{k},\mathbf{p}\leftrightarrow\mathbf{k},\mathbf{p}}}^{m,\nu\leftrightarrow m,\nu} - \frac{\mathrm{P}}{i\pi} \int_{-\infty}^\infty d\omega' \frac{ \mathrm{Re}\left[T_{{\mathbf{k},\mathbf{p}\leftrightarrow\mathbf{k},\mathbf{p}}}^{m,\nu\leftrightarrow m,\nu}(\omega')\right] + i\mathrm{Im}\left[T_{{\mathbf{k},\mathbf{p}\leftrightarrow\mathbf{k},\mathbf{p}}}^{m,\nu\leftrightarrow m,\nu}(\omega')\right] - \mathcal{Q}_{{\mathbf{k},\mathbf{p}\leftrightarrow\mathbf{k},\mathbf{p}}}^{m,\nu\leftrightarrow m,\nu} }{\omega-\omega'} \\
    &= \mathcal{Q}_{{\mathbf{k},\mathbf{p}\leftrightarrow\mathbf{k},\mathbf{p}}}^{m,\nu\leftrightarrow m,\nu} - \frac{\mathrm{P}}{\pi} \int_{-\infty}^\infty d\omega' \frac{ \mathrm{Im}\left[T_{{\mathbf{k},\mathbf{p}\leftrightarrow\mathbf{k},\mathbf{p}}}^{m,\nu\leftrightarrow m,\nu}(\omega')\right]}{\omega-\omega'}
     - \frac{\mathrm{P}}{i\pi} \int_{-\infty}^\infty d\omega' \frac{ \mathrm{Re}\left[T_{{\mathbf{k},\mathbf{p}\leftrightarrow\mathbf{k},\mathbf{p}}}^{m,\nu\leftrightarrow m,\nu}(\omega')\right] - \mathcal{Q}_{{\mathbf{k},\mathbf{p}\leftrightarrow\mathbf{k},\mathbf{p}}}^{m,\nu\leftrightarrow m,\nu} }{\omega-\omega'} \\
    &= \mathcal{Q}_{{\mathbf{k},\mathbf{p}\leftrightarrow\mathbf{k},\mathbf{p}}}^{m,\nu\leftrightarrow m,\nu} - \frac{1}{\pi} \left\{ \mathrm{P} \int_{-\infty}^\infty d\omega' \frac{ \mathrm{Im}\left[T_{{\mathbf{k},\mathbf{p}\leftrightarrow\mathbf{k},\mathbf{p}}}^{m,\nu\leftrightarrow m,\nu}(\omega')\right]}{\omega-\omega'} - i\pi\mathrm{Im}\left[T_{{\mathbf{k},\mathbf{p}\leftrightarrow\mathbf{k},\mathbf{p}}}^{m,\nu\leftrightarrow m,\nu}(\omega)\right] \right\} \\
    &= \mathcal{Q}_{{\mathbf{k},\mathbf{p}\leftrightarrow\mathbf{k},\mathbf{p}}}^{m,\nu\leftrightarrow m,\nu} - \frac{1}{\pi} \int_{-\infty}^\infty d\omega' \frac{ \mathrm{Im}\left[T_{{\mathbf{k},\mathbf{p}\leftrightarrow\mathbf{k},\mathbf{p}}}^{m,\nu\leftrightarrow m,\nu}(\omega')\right]}{\omega-\omega'}.
\end{aligned}
\end{align}
Using steps very similar to Eq.~(\ref{eq:Tmat_diag_separation}), one derives
\begin{equation}
    \Sigma^{\mathrm{(T)}}_{\mathbf{k},mn}(i\omega_n)
    = \frac{4}{\hbar}\sum_\mathbf{p}\sum_\nu T_{{\mathbf{k},\mathbf{p}\leftrightarrow\mathbf{k},\mathbf{p}}}^{m,\nu\leftrightarrow n,\nu}(i\omega_n+\omega_{\mathbf{p}\nu}) n^{(0)}_{\mathbf{p}\nu}
    + \frac{4}{\hbar}\sum_\mathbf{p}\sum_\nu \frac{1}{i\pi} \int_{-\infty}^\infty d\omega' \frac{T_{{\mathbf{k},\mathbf{p}\leftrightarrow\mathbf{k},\mathbf{p}}}^{m,\nu\leftrightarrow n,\nu}(\omega') - \mathcal{Q}_{{\mathbf{k},\mathbf{p}\leftrightarrow\mathbf{k},\mathbf{p}}}^{m,\nu\leftrightarrow n,\nu}}{i\omega_n-\omega'+\omega_{\mathbf{p}\nu}} n^{(\textrm{c})}(\omega').
    \label{eq:TmatSE_general}
\end{equation}
For diagonal components, Eq.~(\ref{eq:TmatSE_general}) reduces to Eq.~(\ref{eq:Tmat_diag_separation}) by the following procedure:
\begin{equation}
\begin{aligned}
\Sigma^{\mathrm{(T)}}_{\mathbf{k},mm}(i\omega_n)
    &= \frac{4}{\hbar}\sum_\mathbf{p}\sum_\nu T_{{\mathbf{k},\mathbf{p}\leftrightarrow\mathbf{k},\mathbf{p}}}^{m,\nu\leftrightarrow m,\nu}(i\omega_n+\omega_{\mathbf{p}\nu}) n^{(0)}_{\mathbf{p}\nu} \\
    &\quad\quad + \frac{4}{\hbar}\sum_\mathbf{p}\sum_\nu \frac{1}{\pi} \int_{-\infty}^\infty d\omega' \left\{ -\frac{1}{\pi} \int_{-\infty}^\infty d\omega'' \frac{\textrm{Im}\left[ T_{{\mathbf{k},\mathbf{p}\leftrightarrow\mathbf{k},\mathbf{p}}}^{m,\nu\leftrightarrow m,\nu}(\omega'') \right]}{\omega'-\omega''} \right\} \frac{1}{i\omega_n-\omega'+\omega_{\mathbf{p}\nu}} n^{(\textrm{c})}(\omega') \\
    &\approx \frac{4}{\hbar}\sum_\mathbf{p}\sum_\nu T_{{\mathbf{k},\mathbf{p}\leftrightarrow\mathbf{k},\mathbf{p}}}^{m,\nu\leftrightarrow m,\nu}(i\omega_n+\omega_{\mathbf{p}\nu}) n^{(0)}_{\mathbf{p}\nu}
    + \frac{4}{\hbar}\sum_\mathbf{p}\sum_\nu \frac{1}{\pi} \int_{-\infty}^\infty d\omega' \frac{\textrm{Im}\left[ T_{{\mathbf{k},\mathbf{p}\leftrightarrow\mathbf{k},\mathbf{p}}}^{m,\nu\leftrightarrow m,\nu}(\omega') \right]}{i\omega_n-\omega'+\omega_{\mathbf{p}\nu}} n^{(\textrm{c})}(\omega').
\end{aligned}
\end{equation}
Within the short-wavelength limit, the approximated T-matrix self-energy is given by
\begin{equation}
    \Sigma^{\mathrm{(T)}}_{\mathbf{k},mn}(i\omega_n) \approx \frac{4}{\hbar}\sum_\mathbf{p}\sum_\nu T_{{\mathbf{k},\mathbf{p}\leftrightarrow\mathbf{k},\mathbf{p}}}^{m,\nu\leftrightarrow n,\nu}(i\omega_n+\omega_{\mathbf{p}\nu}) n^{(0)}_{\mathbf{p}\nu}. \label{eq:approxTmat_full}
\end{equation}

\begin{figure*}[tbh]
    \centering
    \includegraphics[width=\textwidth]{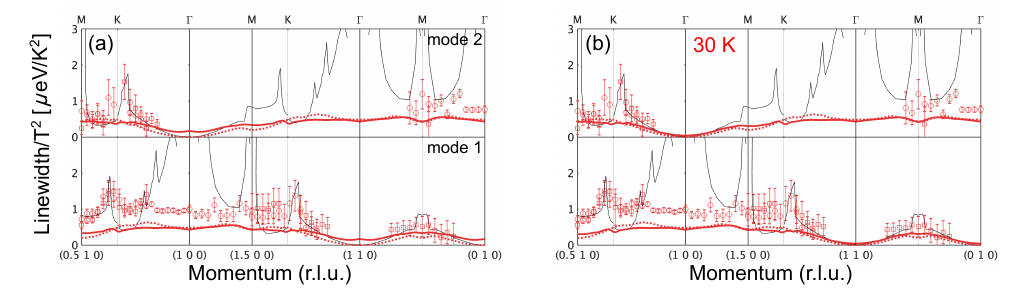}
    \caption{
    Magnon band linewidth of CrBr${}_3$ at 30~K. Red solid lines in (b) are obtained using the full T-matrix self-energy without the short-wavelength approximation, while those in (a) are evaluated by the 
    T-matrix self-energy within the approximation. Note that (b) is the same as Fig.~\ref{Fig07}(f).
    }
    \label{FigS02}
\end{figure*}

Taking the magnon band linewidth in CrBr${}_3$ as an example, we examine the quantitative accuracy of the short-wavelength approximation. Figure~\ref{FigS02} compares the results obtained with and without this approximation. Except for the long-wavelength lower-branch magnons, the linewidth calculated from the T-matrix self-energy within the short-wavelength approximation accurately reproduces the one obtained from the full T-matrix self-energy, underscoring the quantitative validity of the approximation.

\section{Hartree approximation}
\subsection{Expectation values}
The unitary transformation from the atomic bases to the normal bases, which is originally introduced in the main text, is explicitly given by
\begin{equation}
    \left\{
    \begin{array}{c}
        \hat{a}_{ \mathbf{k}\alpha} = \sum_{\nu=1}^{N_\mathrm{sub}} 
        \left[ \mathcal{U}_\mathbf{k} \right]_{\alpha\nu} \hat{b}_{ \mathbf{k}\nu} \\
        \hat{a}^\dagger_{\mathbf{k}\alpha} = \sum_{\nu=1}^{N_\mathrm{sub}} 
        \left[ \mathcal{U}_\mathbf{k} \right]^*_{\alpha\nu}
        \hat{b}^\dagger_{\mathbf{k}\nu}
    \end{array}
    \right. .
\end{equation}
Expectation values in terms of the normal bases are given by
\begin{equation}
    \left\{
    \begin{array}{c}
        \ev{\hat{b}^\dagger_{\mathbf{k}\nu}\hat{b}_{\mathbf{k}'\nu'}} = \delta_{\mathbf{k},\mathbf{k}'}\delta_{\nu\nu'} n^{(0)}_{\mathbf{k}\nu} \\
        \ev{\hat{b}_{\mathbf{k}\nu}\hat{b}^\dagger_{\mathbf{k}'\nu'}} = \delta_{\mathbf{k},\mathbf{k}'}\delta_{\nu\nu'} \left(1+n^{(0)}_{\mathbf{k}\nu}\right) \\
        \ev{\hat{b}^\dagger_{\mathbf{k}\nu}\hat{b}^\dagger_{\mathbf{k}'\nu'}} = \ev{\hat{b}_{\mathbf{k}\nu}\hat{b}_{\mathbf{k}'\nu'}} = 0
    \end{array}
    \right. .
\end{equation}
Using this, an arbitrary ensemble in terms of the atomic bases are given in the form
\begin{equation}
\begin{aligned}
    \ev{\hat{a}^\dagger_{\mathbf{k}\alpha}\hat{a}_{\mathbf{k}'\beta}} = 
    \delta_{\mathbf{k},\mathbf{k}'} \sum_\nu 
    \left[ \mathcal{U}_\mathbf{k} \right]^*_{\alpha\nu}
    \left[ \mathcal{U}_\mathbf{k} \right]  _{\beta \nu}
    \ev{\hat{b}^\dagger_{\mathbf{k}\nu}\hat{b}_{\mathbf{k}\nu}}
    = \delta_{\mathbf{k},\mathbf{k}'} \sum_\nu 
    \left[ \mathcal{U}_\mathbf{k} \right]^*_{\alpha\nu}
    \left[ \mathcal{U}_\mathbf{k} \right]  _{\beta \nu}
    n^{(0)}_{\mathbf{k}\nu}.
\end{aligned}
\end{equation}
\subsection{Four-magnon mean-fields}
\label{Appendix:H4MF}
The four-magnon Hamiltonian in terms of the atomic bases are given by
\begin{align}
\begin{aligned}
    \mathcal{H}^{(4)} &= \sum_{\alpha,\beta,\gamma,\delta} \sum_{\mathbf{q}_1,\mathbf{q}_2,\mathbf{q}_3,\mathbf{q}_4} \mathfrak{Q}_{{\bm q_1},\mathbf{q}_2\leftrightarrow\mathbf{q}_3,\mathbf{q}_4}^{\alpha,\beta\leftrightarrow\gamma,\delta} 
    \hat{a}^\dagger_{\mathbf{q}_1,\alpha}\hat{a}^\dagger_{\mathbf{q}_2,\beta}\hat{a}_{\mathbf{q}_3,\gamma}\hat{a}_{\mathbf{q}_4,\delta}, \\
    \mathfrak{Q}_{{\bm q_1},\mathbf{q}_2\leftrightarrow\mathbf{q}_3,\mathbf{q}_4}^{\alpha,\beta,\leftrightarrow\gamma,\delta} 
    &= \frac{1}{2} \frac{1}{4N} \delta({\bm q_1}+\mathbf{q}_2-\mathbf{q}_3-\mathbf{q}_4) \\
    &\quad\quad \times \left[ \left\{ 
        \delta_{\alpha\gamma}\delta_{\beta\delta} \mathcal{J}^{00}_{\mathbf{q}_1-\mathbf{q}_3,\alpha\beta} + 
        \delta_{\alpha\delta}\delta_{\beta\gamma} \mathcal{J}^{00}_{\mathbf{q}_1-\mathbf{q}_4,\alpha\beta} + 
        \delta_{\alpha\delta}\delta_{\beta\gamma} \mathcal{J}^{00}_{\mathbf{q}_2-\mathbf{q}_3,\beta\alpha} + 
        \delta_{\alpha\gamma}\delta_{\beta\delta} \mathcal{J}^{00}_{\mathbf{q}_2-\mathbf{q}_4,\beta\alpha} 
    \right\} \right. \\ 
    &\quad\quad\quad\quad 
    \left. - \left\{ 
        \delta_{\alpha\beta} \delta_{\alpha\gamma} \mathcal{J}^{+-}_{ \mathbf{q}_4,\gamma\delta} + 
        \delta_{\beta\delta} \delta_{\gamma\delta} \mathcal{J}^{+-}_{ \mathbf{q}_1,\alpha\beta}  + 
        \delta_{\alpha\gamma}\delta_{\alpha\delta} \mathcal{J}^{-+}_{-\mathbf{q}_2,\alpha\beta}  + 
        \delta_{\alpha\beta} \delta_{\alpha\delta} \mathcal{J}^{-+}_{-\mathbf{q}_3,\gamma\delta} 
    \right\} \right],
\end{aligned}
\label{eq:FourmagnonHm}
\end{align}
where $\mathcal{J}^{\chi\chi'}_{\mathbf{q},\alpha\beta} = \mathcal{J}^{\chi\chi'}_{\alpha\beta} e^{i\mathbf{q}\cdot\left( \mathbf{r}_\beta-\mathbf{r}_\alpha \right)}$ $(\chi,\chi'=\pm,0)$.
We consider the mean-field decoupling of general four-magnon terms given in the form
\begin{align}
\begin{aligned}
    \hat{a}^\dagger_{\mathbf{q}_1,\alpha}\hat{a}^\dagger_{\mathbf{q}_2,\beta}\hat{a}_{\mathbf{q}_3,\gamma}\hat{a}_{\mathbf{q}_4,\delta} &\approx 
    \ev{ \hat{a}^\dagger_{\mathbf{q}_1,\alpha} \hat{a}_{\mathbf{q}_3,\gamma} } \hat{a}^\dagger_{\mathbf{q}_2,\beta } \hat{a}_{\mathbf{q}_4,\delta} +
    \ev{ \hat{a}^\dagger_{\mathbf{q}_2,\beta } \hat{a}_{\mathbf{q}_4,\delta} } \hat{a}^\dagger_{\mathbf{q}_1,\alpha} \hat{a}_{\mathbf{q}_3,\gamma} \\
    &+
    \ev{ \hat{a}^\dagger_{\mathbf{q}_1,\alpha} \hat{a}_{\mathbf{q}_4,\delta} } \hat{a}^\dagger_{\mathbf{q}_2,\beta } \hat{a}_{\mathbf{q}_3,\gamma} +
    \ev{ \hat{a}^\dagger_{\mathbf{q}_2,\beta } \hat{a}_{\mathbf{q}_3,\gamma} } \hat{a}^\dagger_{\mathbf{q}_1,\alpha} \hat{a}_{\mathbf{q}_4,\delta} \\
    &+
    \ev{ \hat{a}^\dagger_{\mathbf{q}_1,\alpha}\hat{a}^\dagger_{\mathbf{q}_2,\beta} } \hat{a}_{\mathbf{q}_3,\gamma}\hat{a}_{\mathbf{q}_4,\delta} +
    \ev{ \hat{a}_{\mathbf{q}_3,\gamma}\hat{a}_{\mathbf{q}_4,\delta} } \hat{a}^\dagger_{\mathbf{q}_1,\alpha}\hat{a}^\dagger_{\mathbf{q}_2,\beta}.
\end{aligned}
\label{eq:FourmagnonMF_Hartree}
\end{align}
Applying Eq.~(\ref{eq:FourmagnonMF_Hartree}) to Eq.~(\ref{eq:FourmagnonHm}), one obtains the decoupled four-magnon Hamiltonian given by
\begin{align}
\begin{aligned}
    \mathcal{H}^{(4)}_\textrm{MF} \left( \left\{ \ev{\hat{a}^\dagger_{\mathbf{p},\alpha} \hat{a}_{\mathbf{p},\beta}} \right\} \right) =
    \sum_\mathbf{k} \mathcal{H}^{(4)}_{\mathbf{k},\textrm{MF}} \left( \left\{ \ev{\hat{a}^\dagger_{\mathbf{p},\alpha} \hat{a}_{\mathbf{p},\beta}} \right\} \right) &=
    \sum_{\mathbf{k}}\sum_{\mathbf{p}} \sum_{\alpha\beta\gamma\delta} \left[
        \mathfrak{Q}_{\mathbf{p},\mathbf{k}\leftrightarrow\mathbf{p},\mathbf{k}}^{\alpha,\beta\leftrightarrow\gamma,\delta} 
        \ev{\hat{a}^\dagger_{\mathbf{p},\alpha}\hat{a}_{\mathbf{p},\gamma}} \hat{a}^\dagger_{\mathbf{k},\beta }\hat{a}_{\mathbf{k},\delta} +
        \mathfrak{Q}_{\mathbf{k},\mathbf{p}\leftrightarrow\mathbf{k},\mathbf{p}}^{\alpha,\beta\leftrightarrow\gamma,\delta}
        \ev{\hat{a}^\dagger_{\mathbf{p},\beta }\hat{a}_{\mathbf{p},\delta}} \hat{a}^\dagger_{\mathbf{k},\alpha}\hat{a}_{\mathbf{k},\gamma} \right. \\
    &\quad\quad\quad\quad\quad\quad \left. +
        \mathfrak{Q}_{\mathbf{p},\mathbf{k}\leftrightarrow\mathbf{k},\mathbf{p}}^{\alpha,\beta\leftrightarrow\gamma,\delta} 
        \ev{\hat{a}^\dagger_{\mathbf{p},\alpha}\hat{a}_{\mathbf{p},\delta}} \hat{a}^\dagger_{\mathbf{k},\beta }\hat{a}_{\mathbf{k},\gamma} + 
        \mathfrak{Q}_{\mathbf{k},\mathbf{p}\leftrightarrow\mathbf{p},\mathbf{k}}^{\alpha,\beta\leftrightarrow\gamma,\delta} 
        \ev{\hat{a}^\dagger_{\mathbf{p},\beta }\hat{a}_{\mathbf{p},\gamma}} \hat{a}^\dagger_{\mathbf{k},\alpha}\hat{a}_{\mathbf{k},\delta} \right]
\end{aligned}
\end{align}
Note that, the pairing expectation values $\ev{\hat{a}^\dagger_{\mathbf{p},\alpha} \hat{a}^\dagger_{\mathbf{p},\beta}}$ and $\ev{\hat{a}_{\mathbf{p},\alpha} \hat{a}_{\mathbf{p},\beta}}$ vanish exactly for U(1)-symmetric systems with ferromagnetic ground states, as the honeycomb ferromagnet discussed in the main text.
\subsection{Self-consistent Hartree approximation}
To obtain self-consistent solutions $\ev{\hat{a}^\dagger_{\mathbf{p},\alpha} \hat{a}_{\mathbf{p},\beta}}$, we iteratively solve the eigenvalue problem given by
\begin{equation}
    \left( S\mathcal{H}^{(2)}_{\mathbf{k}} + \mathcal{H}^{(4)}_{\mathbf{k},\textrm{MF}} \left( \left\{ \ev{\hat{a}^\dagger_{\mathbf{p},\alpha} \hat{a}_{\mathbf{p},\beta}} \right\} \right) \right) \tilde{\mathcal{U}}_{\mathbf{k}}
    = \tilde{\mathcal{U}}_{\mathbf{k}}
    \tilde{\mathcal{W}}_\mathbf{k},
\end{equation}
where $\tilde{\mathcal{W}}_\mathbf{k}=\textrm{diag}\left(\tilde{\omega}_{\mathbf{k}-},\tilde{\omega}_{\mathbf{k}+}\right)$.
Eigenvalues $\left\{ \tilde{\omega}_{\mathbf{k}\pm} \right\}$, eigenvectors $\left\{ \tilde{\mathcal{U}}_\mathbf{k} \right\}$, and expectation values $\left\{ \ev{\hat{a}^\dagger_{\mathbf{p},\alpha} \hat{a}_{\mathbf{p},\beta}} \right\}$ are determined self-consistently. The self-consistent loop is iterated until the error in the expectation value $\ev{\hat{a}^\dagger_{\mathbf{p},\alpha} \hat{a}_{\mathbf{p},\beta}}$ falls below $10^{-6}$ for all crystal momenta $\mathbf{p}$ and sublattice indices $\alpha$ and $\beta$.
\section{Construction of two-magnon Hamiltonian on translational invariant non-Bravais lattices}
\label{Appendix:2magnonHilbertspace}
Let us first consider a spin Hamiltonian $\mathcal{H}$ on a Bravais lattice with collinear ferromagnetic order along the $z$-direction. This Hamiltonian holds U(1) symmetry around $z$-axis, ensuring the following commutation relation
\begin{equation}
    \left[ \mathcal{H},S^z \right] = 0, \ \ S^z=\sum_\mathbf{r}S_\mathbf{r}^z,
\end{equation}
and thus, the total $z$ spin is a good quantum number. Eigenstates can be labeled by their $z$ spin $\Delta S=S-m=0,1,2,3,\cdots$ relative to the fully-polarized ground state,
\begin{equation}
    \ket{0} = \bigotimes_\mathbf{r} \ket{S_\mathbf{r}^z=S,m_\mathbf{r}=S},
\end{equation}
i.e., $\Delta S=0$. It is a tensor product of local $\ket{S_\mathbf{r}^z=S,m_\mathbf{r}=S}$ states, which are eigenstates of the $z$-spin operator: 
\begin{equation}
    S_\mathbf{r}^z\ket{S_\mathbf{r}^z=S,m_\mathbf{r}}=m_\mathbf{r}\ket{S_\mathbf{r}^z=S,m_\mathbf{r}},
\end{equation}
with spin quantum number $S=\frac{1}{2},1,\frac{3}{2},2,\cdots$. $m_\mathbf{r}$ and $m=\sum_\mathbf{r}m_\mathbf{r}$ are the local and total magnetic quantum numbers, respectively. Here we consider $\Delta S=2$ sector, whose basis can be written as
\begin{equation}
    \ket{\mathbf{k},\mathbf{r}} = 
    \frac{1}{\sqrt{N_\textrm{muc}}}
    \sum_{\mathbf{r}'} e^{i\mathbf{k}\cdot\mathbf{r}'} e^{i\mathbf{k}\cdot\mathbf{r}/2}\ket{\mathbf{r}',\mathbf{r}+\mathbf{r}'},
    \label{eq:2magnonbasis_Bravais}
\end{equation}
where states with two spin flips $\ket{\mathbf{r}',\mathbf{r}+\mathbf{r}'}=\mathcal{N}_\mathbf{r}^{-1}S_{\mathbf{r}'}^-S_{\mathbf{r}+\mathbf{r}'}^-\ket{0}$ are superimposed. Under the translational invariance acting on each site, the double spin-flip state is labeled by its center-of-mass crystal momentum $\mathbf{k}$ and relative distance vector connecting the two spin-flipped sites $\mathbf{r}$. $S_\mathbf{r}^-=S_\mathbf{r}^x-iS_\mathbf{r}^y$ denotes the ladder operator satisfying the following relation:
\begin{equation}
    S^\pm_\mathbf{r} \ket{S,m_\mathbf{r}} = \sqrt{S(S+1)-m(m\pm1)} \ket{S,m_\mathbf{r}\pm1}.
\end{equation}
The normalization factor originating from the Clebsch-Gordan coefficients reads $\mathcal{N}_\mathbf{r}=2S$ for $\mathbf{r}\neq\mathbf{0}$, and $\mathcal{N}_\mathbf{O}=\sqrt{2S}\sqrt{4S-2}$ for $\mathbf{r}=\mathbf{O}=\mathbf{0}$.
\begin{figure*}[tbh]
    \centering
    \includegraphics[width=\textwidth]{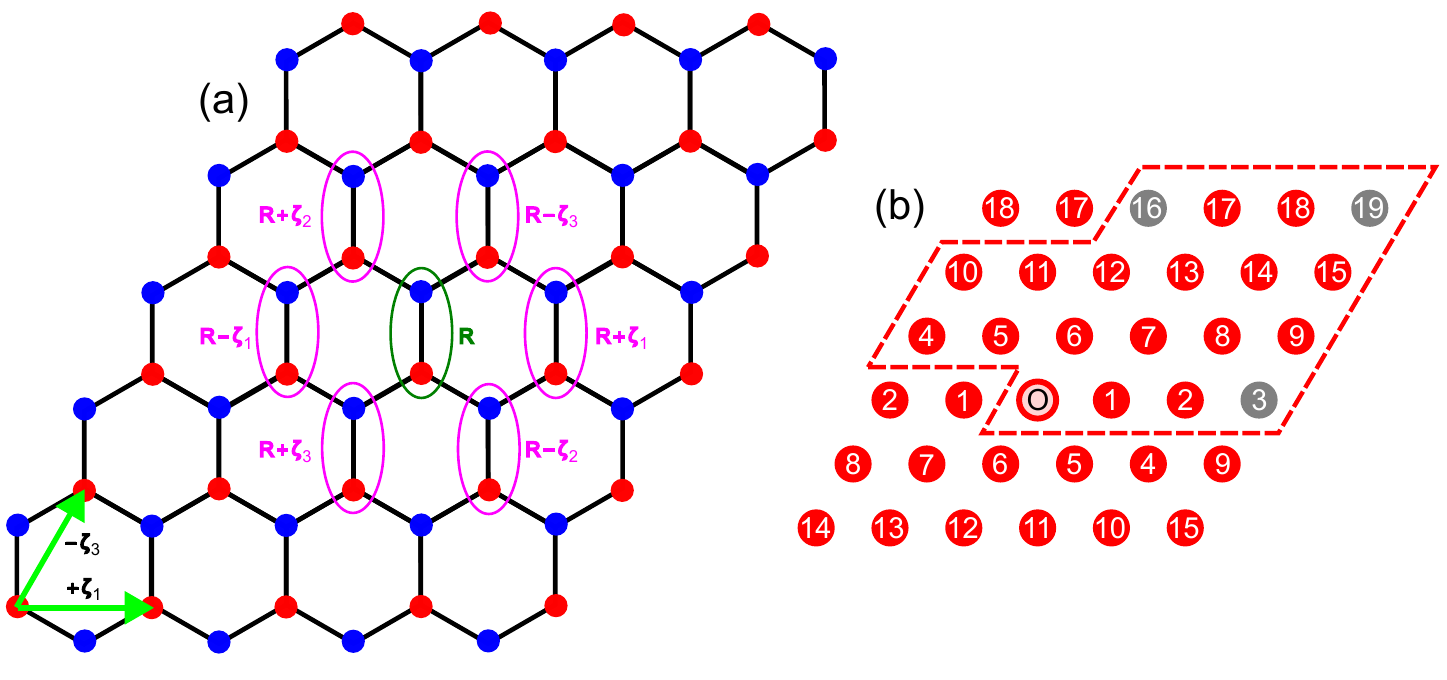}
    \caption{
    (a) Schematic of the honeycomb lattice. $\bm{\zeta}_i$ $(i=1,2,3)$ denotes primitive lattice vectors. Red (blue) circles indicate the sublattice A (B).
    (b) Schematic of the first redundancy given by Eq.~(\ref{eq:redundancy1}) under inversion symmetry and periodic boundary condition. Numbers show indices of the unit cells of the honeycomb lattice when $L=6$. O denotes an unit cell at the origin. A pair of unit cells indicated by red circles with the same numbers, which are connected by spatial inversion with each other, is regarded to be identical in the presence of translational symmetry. Unit cells indicated by gray circles do not have their counterparts under inversion symmetry.
    } 
    \label{FigS03}
\end{figure*}
\par The methodology to construct the two-magnon Hilbert space and the spin-$S$ Heisenberg (+anisotropy) Hamiltonian matrix acting on it under the translational invariance for Bravais-lattice systems is discussed in Refs.~\cite{Reklis1974,Kecke2007,Mook2023boundmagnon}. However, they do not contain sublattice degrees of freedom in non-Bravais lattice systems and are not straightforwardly applicable to them. We here extend this to non-Bravais lattices containing multiple inequivalent sublattices with $N=N_\mathrm{sub}L^2$ sites under periodic boundary conditions, where $N_\mathrm{sub}$ is the number of sublattices in a magnetic unit cell and $N_\mathrm{muc}=L^2$ is the number of magnetic unit cells. We assume a restriction that $L$ is an even positive integer. Since the two-spin-flip basis for Bravais lattices given by Eq.~(\ref{eq:2magnonbasis_Bravais}) cannot span an entire two-spin-flip space for non-Bravais lattices, we instead introduce the following basis set
\begin{equation}
    \ket{\mathbf{k},\mathbf{R}}_{\gamma\gamma'} = \frac{1}{\sqrt{N_\mathrm{muc}}} \sum_{\mathbf{R}'} 
    e^{i\mathbf{k}\cdot\mathbf{R}'} 
    e^{i\mathbf{k}\cdot(\mathbf{R}+\mathbf{r}_{\gamma'}-\mathbf{r}_\gamma)/2}
    \ket{\mathbf{R}',\mathbf{R}+\mathbf{R}'}_{\gamma\gamma'},
    \label{eq:2magnonbasis_nonBravais}
\end{equation}
where an arbitrary lattice vector $\mathbf{R}$ ($\mathbf{R}'$) indicating a specific magnetic unit cell is given by linear combinations of the primitive lattice vectors $\bm{\zeta}_1$ and $-\bm{\zeta}_3$ [see Fig.~\ref{FigS03}(a)], namely
\begin{gather}
\begin{aligned}
    \mathbf{R} = m\bm{\zeta}_1 - n\bm{\zeta}_3.
\end{aligned}
\end{gather}
$m$ and $n$ are integers satisfying $m,n\in[-L/2+1,L/2]$. $\mathbf{r}_\gamma$ $(\mathbf{r}_{\gamma'})$ denotes an internal coordinate of the $\gamma$- ($\gamma'$-)th sublattice in magnetic unit cells. The state $\ket{\mathbf{R}',\mathbf{R}+\mathbf{R}'}_{\gamma\gamma'}=\mathcal{N}_{\mathbf{R}+\mathbf{r}_{\gamma'}-\mathbf{r}_\gamma}^{-1}S^-_{\mathbf{R}'+\mathbf{r}_\gamma}S^-_{\mathbf{R}+\mathbf{R}'+\mathbf{r}_{\gamma'}}\ket{0}$ is that with two-spin flips on the $\gamma$-th sublattice in a magnetic unit cell $\mathbf{R}'$ and the $\gamma'$-th sublattice in a magnetic unit cell $\mathbf{R}+\mathbf{R}'$. Since the two-spin-flip basis set given by Eq.~(\ref{eq:2magnonbasis_nonBravais}) holds unphysical redundancies given in the from
\begin{align}
    \ket{\mathbf{k},\mathbf{R}}_{\gamma\gamma } &= \ket{\mathbf{k},-\mathbf{R}}_{\gamma \gamma}, \label{eq:redundancy1} \\
    \ket{\mathbf{k},\mathbf{R}}_{\gamma\gamma'} &= \ket{\mathbf{k},-\mathbf{R}}_{\gamma'\gamma} \ \ (\mathrm{for} \ \gamma\neq\gamma'), \label{eq:redundancy2}
\end{align}
we need to take them into account when constructing the two-spin-flip Hilbert space under translational invariance. Note that these redundancies do not apply to the following four specific $\mathbf{R}$s with $(m,n)=(0,0)$, $(L/2,0)$, $(0,L/2)$, and $(L/2,L/2)$ since they are invariant under $\mathbf{R}\leftrightarrow-\mathbf{R}$ transformation as indicated in Fig.~\ref{FigS03}(b). The first redundancy given by Eq.~(\ref{eq:redundancy1}) is the same one as that in Bravais lattice systems. This reduces the space dimension of the $\gamma\gamma$-sector from $|\mathbf{R}|=L^2$ to $|\tilde{\mathbf{R}}|=L^2/2+2$ (from $|\mathbf{R}|=L^2-1$ to $|\tilde{\mathbf{R}}|=L^2/2+1$) when $S\geq1$ ($S=1/2$), as roughly indicated in Fig.~\ref{FigS03}(b). We practically introduce a \textit{reduced relative distance lattice vector} $\tilde{\mathbf{R}}\in\{\mathbf{R},-\mathbf{R}\}$, where the two bases connected by the equality in Eq.~(\ref{eq:redundancy1}) is identified as
\begin{equation}
    \ket{\mathbf{k},\tilde{\mathbf{R}}}_{\gamma\gamma}^\mathsf{r} \coloneqq \left\{
    \begin{array}{ll}
    \ket{\mathbf{k}, \mathbf{R}}_{\gamma\gamma} \equiv \
    \ket{\mathbf{k},-\mathbf{R}}_{\gamma\gamma} &
    \left(\mathbf{R}\neq\mathbf{O}, \ (L/2)\bm{\zeta}_1, \ -(L/2)\bm{\zeta}_3, \ (L/2)\bm{\zeta}_1-(L/2)\bm{\zeta}_3\right), \\
    \ket{\mathbf{k}, \mathbf{R}}_{\gamma\gamma} & 
    \textrm{otherwise.}
    \end{array} \right.
\end{equation}
The second redundancy given by Eq.~(\ref{eq:redundancy2}) is unique in non-Bravais lattices. It indicates that, in the presence of translational symmetry, only the $\gamma\gamma'$-sector $(\gamma\neq\gamma')$ should be considered, and its counterpart, the $\gamma'\gamma$-sector, must be projected onto the $\gamma\gamma'$-sector by basis-dependent appropriate spatial translation. We introduce the projected basis given by
\begin{align}
    \ket{\mathbf{k},\mathbf{R}}_{\gamma\nu\gamma'}^\mathsf{p} &\coloneqq \left\{
    \begin{array}{ll}
    \ket{\mathbf{k},\mathbf{R}}_{\gamma\gamma'} \equiv \ \ket{\mathbf{k},-\mathbf{R}}_{\gamma'\gamma}
    & \left(\mathbf{R}\neq(L/2)\bm{\zeta}_1, \ -(L/2)\bm{\zeta}_3, \ (L/2)\bm{\zeta}_1-(L/2)\bm{\zeta}_3, \ \textrm{regardless of} \ \nu=\leftarrow,\rightarrow \right) \\
    \ket{\mathbf{k},\mathbf{R}}_{\gamma\gamma'} & \left(\mathbf{R}=(L/2)\bm{\zeta}_1, \ -(L/2)\bm{\zeta}_3, \ (L/2)\bm{\zeta}_1-(L/2)\bm{\zeta}_3 \ \textrm{and} \ \nu=\rightarrow\right) \\
    \ket{\mathbf{k},\mathbf{R}}_{\gamma'\gamma} & \left(\mathbf{R}=(L/2)\bm{\zeta}_1, \ -(L/2)\bm{\zeta}_3, \ (L/2)\bm{\zeta}_1-(L/2)\bm{\zeta}_3 \ \textrm{and} \ \nu=\leftarrow\right)
    \end{array} \right.
\end{align}
Note that, when $\mathbf{R}\neq(L/2)\bm{\zeta}_1$, $-(L/2)\bm{\zeta}_3$, nor $(L/2)\bm{\zeta}_1-(L/2)\bm{\zeta}_3$, $\nu$ is fixed to $\rightarrow$.
This projection reduces the corresponding space dimension from $2|\mathbf{R}|$ $=$$2L^2$ to $|\mathbf{R}|+3=L^2+3$ for an arbitrary spin length. Thus, the set of bases $\left\{\ket{\mathbf{k},\mathbf{R}}^\mathsf{p}_{\gamma\leftrightarrow\gamma'}\right\}=\left\{\ket{\mathbf{k},\mathbf{R}}^\mathsf{p}_{\gamma\rightarrow\gamma'},\ket{\mathbf{k},\mathbf{R}}^\mathsf{p}_{\gamma\leftarrow\gamma'}\right\}$ spans $(L^2+3)$-dimensional subspace.
\par Finally, we consider a matrix form of the two-spin-flip Hamiltonian on the honeycomb lattice system with two inequivalent sublattices A and B. The Hamiltonian has the following $3\times3$ block form:
\begin{equation}
   \hat{H}_{\bm{\mathsf{2}}}(\mathbf{k}) = 2S\left(
   \begin{array}{ccc}
       \hat{h}^{\mathrm{AA},\mathrm{AA}}(\mathbf{k}) & \hat{\mathsf{0}} & \hat{h}^{\mathrm{AA},\mathrm{AB}}(\mathbf{k}) \\
       \hat{\mathsf{0}} & \hat{h}^{\mathrm{BB},\mathrm{BB}}(\mathbf{k}) & \hat{h}^{\mathrm{BB},\mathrm{AB}}(\mathbf{k}) \\
       \left[\hat{h}^{\mathrm{AA},\mathrm{AB}}(\mathbf{k})\right]^\dagger & \left[\hat{h}^{\mathrm{BB},\mathrm{AB}}(\mathbf{k})\right]^\dagger & \hat{h}^{\mathrm{AB},\mathrm{AB}}(\mathbf{k})
   \end{array} \right).
\end{equation}
$\textrm{AA}$, $\textrm{BB}$, $\textrm{AB}$ sectors are $(L^2/2+2)$-, $(L^2/2+2)$-, and $(L^2+3)$-dimensional subspaces, respectively. Matrix elements belonging to each block are respectively given by
\begin{equation}
\begin{aligned}
    \hat{h}^{\mathrm{AA},\mathrm{AA}}_{\tilde{\mathbf{R}}\tilde{\mathbf{R}}'}
    &= {}_{\mathrm{AA}\vphantom{\mathrm{AA}}}^{\ \ \ \mathsf{r}} \langle \mathbf{k},\tilde{\mathbf{R}}|\mathcal{H}|\mathbf{k},\tilde{\mathbf{R}}' \rangle_{\mathrm{AA}}^\mathsf{r}, \\
    \hat{h}^{\mathrm{BB},\mathrm{BB}}_{\tilde{\mathbf{R}}\tilde{\mathbf{R}}'}
    &= {}_{\mathrm{BB}\vphantom{\mathrm{BB}}}^{\ \ \ \mathsf{r}} \langle \mathbf{k},\tilde{\mathbf{R}}|\mathcal{H}|\mathbf{k},\tilde{\mathbf{R}}' \rangle_{\mathrm{BB}}^\mathsf{r}, \\
    \hat{h}^{\mathrm{AB},\mathrm{AB}}_{(\mathbf{R}\nu')(\mathbf{R}'\nu')}
    &= {}_{\mathrm{A}\nu\mathrm{B}}\vphantom{\mathrm{A}\nu\mathrm{B}}^{\mathsf{p}} \langle \mathbf{k},\mathbf{R}|\mathcal{H}|\mathbf{k},\mathbf{R}' \rangle_{\mathrm{A}\nu'\mathrm{B}}^\mathsf{p}, \\
    \hat{h}^{\mathrm{AA},\mathrm{AB}}_{\tilde{\mathbf{R}}(\mathbf{R}'\nu)}
    &= {}_{\mathrm{AA}\vphantom{\mathrm{AA}}}^{\ \ \ \mathsf{r}} \langle \mathbf{k},\tilde{\mathbf{R}}|\mathcal{H}|\mathbf{k},\mathbf{R}' \rangle_{\mathrm{A}\nu\mathrm{B}}^\mathsf{p}, \\
    \hat{h}^{\mathrm{BB},\mathrm{AB}}_{\tilde{\mathbf{R}}(\mathbf{R}'\nu)}
    &= {}_{\mathrm{BB}\vphantom{\mathrm{BB}}}^{\ \ \ \mathsf{r}} \langle \mathbf{k},\tilde{\mathbf{R}}|\mathcal{H}|\mathbf{k},\mathbf{R}' \rangle_{\mathrm{A}\nu\mathrm{B}}^\mathsf{p},
\end{aligned}
\end{equation}
where $\nu,\nu'=\rightarrow,\leftarrow$. To be concrete, we present matrix elements for the spin-$S$ $J_1$-$J_2$-$J_3$-$A$ Heisenberg+onsite anisotropy model for CrBr$_3$, whose Hamiltonian is given by
\begin{equation}
\begin{aligned}
    \mathcal{H} &= 
      J_1\sum_{\ev{\mathbf{r},\mathbf{r}'}\in\mathrm{1NN}} \mathbf{S}_\mathbf{r}\cdot\mathbf{S}_{\mathbf{r}'}
    + J_2\sum_{\ev{\mathbf{r},\mathbf{r}'}\in\mathrm{2NN}} \mathbf{S}_\mathbf{r}\cdot\mathbf{S}_{\mathbf{r}'}
    + J_3\sum_{\ev{\mathbf{r},\mathbf{r}'}\in\mathrm{3NN}} \mathbf{S}_\mathbf{r}\cdot\mathbf{S}_{\mathbf{r}'}
    - A\sum_\mathbf{r} \left(S^z_\mathbf{r}\right)^2 \\
    &= \underbrace{
    \left\{
      \frac{J_1}{2}\sum_{\ev{\mathbf{r},\mathbf{r}'}\in\mathrm{1NN}}
    + \frac{J_2}{2}\sum_{\ev{\mathbf{r},\mathbf{r}'}\in\mathrm{2NN}}
    + \frac{J_3}{2}\sum_{\ev{\mathbf{r},\mathbf{r}'}\in\mathrm{3NN}}
    \right\}
    \left( S^+_\mathbf{r}S^-_{\mathbf{r}'} + S^-_\mathbf{r}S^+_{\mathbf{r}'} \right)}_{\equiv\mathcal{H}_\pm}
    + \underbrace{
    \left\{
      J_1\sum_{\ev{\mathbf{r},\mathbf{r}'}\in\mathrm{1NN}}
    + J_2\sum_{\ev{\mathbf{r},\mathbf{r}'}\in\mathrm{2NN}}
    + J_3\sum_{\ev{\mathbf{r},\mathbf{r}'}\in\mathrm{3NN}}
    \right\} S^z_\mathbf{r}S^z_{\mathbf{r}'}
    - A\sum_\mathbf{r} \left(S^z_\mathbf{r}\right)^2}_{\equiv\mathcal{H}_z}.
\end{aligned}
\end{equation}
Note that $S^+_\mathbf{r}=\left(S^-_\mathbf{r}\right)^\dagger$. $\mathcal{H}_\pm$ ($\mathcal{H}_z$) denotes the hopping (diagonal) term. The explicit form of the matrix elements for this $J_1$-$J_2$-$J_3$-$A$ model is given by
\begin{align}
    &\hat{h}^{\mathrm{AA},\mathrm{AA}}_{\tilde{\mathbf{R}}\tilde{\mathbf{R}}'}(\mathbf{k}) = 
    \delta_{\tilde{\mathbf{R}},\tilde{\mathbf{R}}'} \left[ -3J_{1z}-6J_{2z}-3J_{3z}+\frac{A(2S-1)}{S} + \frac{J_{2z}}{2S}\sum_{n=1,2,3}\delta_{\tilde{\mathbf{R}},\bm{\zeta}_n}-\frac{A}{S}\delta_{\tilde{\mathbf{R}},\mathbf{O}} \right]
    + \sum_{\mathbf{R}\in\{\tilde{\mathbf{R}}\}} \sum_{\mathbf{R}'\in\{\tilde{\mathbf{R}}'\}}
    {}_{\mathrm{AA}\vphantom{\mathrm{AA}}}\langle \mathbf{k},\mathbf{R}|\mathcal{H}_\pm|\mathbf{k},\mathbf{R}' \rangle_{\mathrm{AA}}, \label{eq:2magnonHm_matrixelement_No1}\\
    &\hat{h}^{\mathrm{BB},\mathrm{BB}}_{\tilde{\mathbf{R}}\tilde{\mathbf{R}}'}(\mathbf{k})
    = \hat{h}^{\mathrm{AA},\mathrm{AA}}_{\tilde{\mathbf{R}}\tilde{\mathbf{R}}'}(\mathbf{k}), \\
    &\begin{aligned}
    \hat{h}^{\mathrm{AB},\mathrm{AB}}_{(\mathbf{R}\nu)(\mathbf{R}'\nu')}(\mathbf{k})
    &= \delta_{(\mathbf{R}\nu),(\mathbf{R}'\nu')} \left[ -3J_{1z}-6J_{2z}-3J_{3z}+\frac{A(2S-1)}{S}
    + \frac{J_{1z}}{2S} \sum_{n=1,2,3} \left( \delta_{\mathbf{R},\mathbf{O}} + \delta_{\mathbf{R},\bm{\zeta}_3} + \delta_{\mathbf{R},-\bm{\zeta}_2} \right)
    + \frac{J_{3z}}{2S} \sum_{n=1,2,3} \left( \delta_{\mathbf{R},\bm{\zeta}_1} + \delta_{\mathbf{R},-\bm{\zeta}_1} + \delta_{\mathbf{R},\bm{\zeta}_3-\bm{\zeta}_2} \right)
    \right] \\
    &\quad\quad + \delta_{\nu,\rightarrow} \delta_{\nu',\rightarrow}
    {}_{\mathrm{A}\mathrm{B}\vphantom{\mathrm{A}\mathrm{B}}}\langle \mathbf{k},\mathbf{R}|\mathcal{H}_\pm|\mathbf{k},\mathbf{R}' \rangle_{\mathrm{A}\mathrm{B}}
    + \delta_{\nu,\rightarrow} \delta_{\nu',\leftarrow}
    {}_{\mathrm{A}\mathrm{B}\vphantom{\mathrm{A}\mathrm{B}}}\langle \mathbf{k},\mathbf{R}|\mathcal{H}_\pm|\mathbf{k},\mathbf{R}' \rangle_{\mathrm{B}\mathrm{A}} \\
    &\quad\quad + \delta_{\nu,\leftarrow} \delta_{\nu',\rightarrow}
    {}_{\mathrm{B}\mathrm{A}\vphantom{\mathrm{B}\mathrm{A}}}\langle \mathbf{k},\mathbf{R}|\mathcal{H}_\pm|\mathbf{k},\mathbf{R}' \rangle_{\mathrm{A}\mathrm{B}}
    + \delta_{\nu,\leftarrow} \delta_{\nu',\leftarrow}
    {}_{\mathrm{B}\mathrm{A}\vphantom{\mathrm{B}\mathrm{A}}}\langle \mathbf{k},\mathbf{R}|\mathcal{H}_\pm|\mathbf{k},\mathbf{R}' \rangle_{\mathrm{B}\mathrm{A}}
    \end{aligned} \\
    &\hat{h}^{\mathrm{AA},\mathrm{AB}}_{\tilde{\mathbf{R}}(\mathbf{R}'\nu)}(\mathbf{k})
    = \sum_{\mathbf{R}\in\{\tilde{\mathbf{R}}\}} \left[
    \delta_{\nu,\rightarrow} {}_{\mathrm{AA}\vphantom{\mathrm{AA}}}\langle \mathbf{k},\mathbf{R}|\mathcal{H}_\pm|\mathbf{k},\mathbf{R}' \rangle_{\mathrm{AB}} +
    \delta_{\nu,\leftarrow} {}_{\mathrm{AA}\vphantom{\mathrm{AA}}}\langle \mathbf{k},\mathbf{R}|\mathcal{H}_\pm|\mathbf{k},\mathbf{R}' \rangle_{\mathrm{BA}} 
    \right], \\
    &\hat{h}^{\mathrm{BB},\mathrm{AB}}_{\tilde{\mathbf{R}}(\mathbf{R}'\nu)}(\mathbf{k})
    = \sum_{\mathbf{R}\in\{\tilde{\mathbf{R}}\}} \left[
    \delta_{\nu,\rightarrow} {}_{\mathrm{BB}\vphantom{\mathrm{BB}}}\langle \mathbf{k},\mathbf{R}|\mathcal{H}_\pm|\mathbf{k},\mathbf{R}' \rangle_{\mathrm{BA}} +
    \delta_{\nu,\leftarrow} {}_{\mathrm{BB}\vphantom{\mathrm{BB}}}\langle \mathbf{k},\mathbf{R}|\mathcal{H}_\pm|\mathbf{k},\mathbf{R}' \rangle_{\mathrm{AB}}
    \right].
    \label{eq:2magnonHm_matrixelement_No5}
\end{align}
Matrix elements in Eqs.~(\ref{eq:2magnonHm_matrixelement_No1})--(\ref{eq:2magnonHm_matrixelement_No5}) are evaluated using the following algorithms \ref{alg:1},~\ref{alg:2},~\ref{alg:3}, and~\ref{alg:4}:
\begin{figure}[H]
\begin{algorithm}[H]
\caption{Calculate ${}_{\mathrm{AA}\vphantom{\mathrm{AA}}}\langle \mathbf{k},\mathbf{R}|\mathcal{H}_\pm|\mathbf{k},\mathbf{R}' \rangle_{\mathrm{AB}/\mathrm{BA}}$}
\label{alg:1}
\begin{algorithmic}[1]
    \FOR{all $\mathbf{R}$ and $\mathbf{R}'$}
    \STATE ${}_{\mathrm{AA}\vphantom{\mathrm{AA}}}\langle \mathbf{k},\mathbf{R}|\mathcal{H}_\pm|\mathbf{k},\mathbf{R}' \rangle_{\mathrm{AB}/\mathrm{BA}}$ $\leftarrow$ 0
    \ENDFOR
    \FOR{$\mathbf{R}$ with $(m,n)=[-L/2+1,L/2]^2$}
    \STATE ${}_{\mathrm{AA}\vphantom{\mathrm{AA}}}\langle \mathbf{k},\mathbf{R}|\mathcal{H}_\pm|\mathbf{k},\mathbf{R} \rangle_{\mathrm{AB}/\mathrm{BA}}$ $\pluseq$ 
    $(J_1/2)e^{i\mathbf{k}\cdot\bm{\delta}_1/2}$
    $\left\{1+\delta_{\mathbf{R},\mathbf{O}}\left( \sqrt{2-\frac{1}{S}}-1\right)\right\}$
    \STATE ${}_{\mathrm{AA}\vphantom{\mathrm{AA}}}\langle \mathbf{k},\mathbf{R}|\mathcal{H}_\pm|\mathbf{k},\mathbf{R}+\bm{\zeta}_3 \rangle_{\mathrm{AB}/\mathrm{BA}}$ $\pluseq$ 
    $(J_1/2)e^{i\mathbf{k}\cdot\bm{\delta}_2/2}$
    $\left\{1+\delta_{\mathbf{R},\mathbf{O}}\left( \sqrt{2-\frac{1}{S}}-1\right)\right\}$
    \STATE ${}_{\mathrm{AA}\vphantom{\mathrm{AA}}}\langle \mathbf{k},\mathbf{R}|\mathcal{H}_\pm|\mathbf{k},\mathbf{R}-\bm{\zeta}_2 \rangle_{\mathrm{AB}/\mathrm{BA}}$ $\pluseq$ 
    $(J_1/2)e^{i\mathbf{k}\cdot\bm{\delta}_3/2}$
    $\left\{1+\delta_{\mathbf{R},\mathbf{O}}\left( \sqrt{2-\frac{1}{S}}-1\right)\right\}$
    \STATE ${}_{\mathrm{AA}\vphantom{\mathrm{AA}}}\langle \mathbf{k},\mathbf{R}|\mathcal{H}_\pm|\mathbf{k},\mathbf{R}+\bm{\zeta}_1 \rangle_{\mathrm{AB}/\mathrm{BA}}$ $\pluseq$ 
    $(J_3/2)e^{i\mathbf{k}\cdot\bm{\xi}_1/2}$
    $\left\{1+\delta_{\mathbf{R},\mathbf{O}}\left( \sqrt{2-\frac{1}{S}}-1\right)\right\}$
    \STATE ${}_{\mathrm{AA}\vphantom{\mathrm{AA}}}\langle \mathbf{k},\mathbf{R}|\mathcal{H}_\pm|\mathbf{k},\mathbf{R}-\bm{\zeta}_1 \rangle_{\mathrm{AB}/\mathrm{BA}}$ $\pluseq$ 
    $(J_3/2)e^{i\mathbf{k}\cdot\bm{\xi}_2/2}$
    $\left\{1+\delta_{\mathbf{R},\mathbf{O}}\left( \sqrt{2-\frac{1}{S}}-1\right)\right\}$
    \STATE ${}_{\mathrm{AA}\vphantom{\mathrm{AA}}}\langle \mathbf{k},\mathbf{R}|\mathcal{H}_\pm|\mathbf{k},\mathbf{R}+\bm{\zeta}_3-\bm{\zeta}_2 \rangle_{\mathrm{AB}/\mathrm{BA}}$ $\pluseq$ 
    $(J_3/2)e^{i\mathbf{k}\cdot\bm{\xi}_3/2}$
    $\left\{1+\delta_{\mathbf{R},\mathbf{O}}\left( \sqrt{2-\frac{1}{S}}-1\right)\right\}$
    \ENDFOR
\end{algorithmic}
\end{algorithm}
\end{figure}
\begin{figure}[H]
\begin{algorithm}[H]
\caption{Calculate ${}_{\mathrm{BB}\vphantom{\mathrm{BB}}}\langle \mathbf{k},\mathbf{R}|\mathcal{H}_\pm|\mathbf{k},\mathbf{R}' \rangle_{\mathrm{BA}/\mathrm{AB}}$}
\label{alg:2}
\begin{algorithmic}[1]
    \FOR{all $\mathbf{R}$ and $\mathbf{R}'$}
    \STATE ${}_{\mathrm{BB}\vphantom{\mathrm{BB}}}\langle \mathbf{k},\mathbf{R}|\mathcal{H}_\pm|\mathbf{k},\mathbf{R}' \rangle_{\mathrm{BA}/\mathrm{AB}}$ $\leftarrow$ 0
    \ENDFOR
    \FOR{$\mathbf{R}$ with $(m,n)=[-L/2+1,L/2]^2$}
    \STATE ${}_{\mathrm{BB}\vphantom{\mathrm{BB}}}\langle \mathbf{k},\mathbf{R}|\mathcal{H}_\pm|\mathbf{k},\mathbf{R} \rangle_{\mathrm{BA}/\mathrm{AB}}$ $\pluseq$ 
    $(J_1/2)e^{-i\mathbf{k}\cdot\bm{\delta}_1/2}$
    $\left\{1+\delta_{\mathbf{R},\mathbf{O}}\left( \sqrt{2-\frac{1}{S}}-1\right)\right\}$
    \STATE ${}_{\mathrm{BB}\vphantom{\mathrm{BB}}}\langle \mathbf{k},\mathbf{R}|\mathcal{H}_\pm|\mathbf{k},\mathbf{R}+\bm{\zeta}_3 \rangle_{\mathrm{BA}/\mathrm{AB}}$ $\pluseq$ 
    $(J_1/2)e^{-i\mathbf{k}\cdot\bm{\delta}_2/2}$
    $\left\{1+\delta_{\mathbf{R},\mathbf{O}}\left( \sqrt{2-\frac{1}{S}}-1\right)\right\}$
    \STATE ${}_{\mathrm{BB}\vphantom{\mathrm{BB}}}\langle \mathbf{k},\mathbf{R}|\mathcal{H}_\pm|\mathbf{k},\mathbf{R}-\bm{\zeta}_2 \rangle_{\mathrm{BA}/\mathrm{AB}}$ $\pluseq$ 
    $(J_1/2)e^{-i\mathbf{k}\cdot\bm{\delta}_3/2}$
    $\left\{1+\delta_{\mathbf{R},\mathbf{O}}\left( \sqrt{2-\frac{1}{S}}-1\right)\right\}$
    \STATE ${}_{\mathrm{BB}\vphantom{\mathrm{BB}}}\langle \mathbf{k},\mathbf{R}|\mathcal{H}_\pm|\mathbf{k},\mathbf{R}+\bm{\zeta}_1 \rangle_{\mathrm{BA}/\mathrm{AB}}$ $\pluseq$ 
    $(J_3/2)e^{-i\mathbf{k}\cdot\bm{\xi}_1/2}$
    $\left\{1+\delta_{\mathbf{R},\mathbf{O}}\left( \sqrt{2-\frac{1}{S}}-1\right)\right\}$
    \STATE ${}_{\mathrm{BB}\vphantom{\mathrm{BB}}}\langle \mathbf{k},\mathbf{R}|\mathcal{H}_\pm|\mathbf{k},\mathbf{R}-\bm{\zeta}_1 \rangle_{\mathrm{BA}/\mathrm{AB}}$ $\pluseq$ 
    $(J_3/2)e^{-i\mathbf{k}\cdot\bm{\xi}_2/2}$
    $\left\{1+\delta_{\mathbf{R},\mathbf{O}}\left( \sqrt{2-\frac{1}{S}}-1\right)\right\}$
    \STATE ${}_{\mathrm{BB}\vphantom{\mathrm{BB}}}\langle \mathbf{k},\mathbf{R}|\mathcal{H}_\pm|\mathbf{k},\mathbf{R}+\bm{\zeta}_3-\bm{\zeta}_2 \rangle_{\mathrm{BA}/\mathrm{AB}}$ $\pluseq$ 
    $(J_3/2)e^{-i\mathbf{k}\cdot\bm{\xi}_3/2}$
    $\left\{1+\delta_{\mathbf{R},\mathbf{O}}\left( \sqrt{2-\frac{1}{S}}-1\right)\right\}$
    \ENDFOR
\end{algorithmic}
\end{algorithm}
\end{figure}
\begin{figure}[H]
\begin{algorithm}[H]
\caption{Calculate ${}_{\mathrm{AA}\vphantom{\mathrm{AA}}}\langle \mathbf{k},\mathbf{R}|\mathcal{H}_\pm|\mathbf{k},\mathbf{R}' \rangle_{\mathrm{AA}}$}
\label{alg:3}
\begin{algorithmic}[1]
    \FOR{all $\mathbf{R}$ and $\mathbf{R}'$}
    \STATE ${}_{\mathrm{AA}\vphantom{\mathrm{AA}}}\langle \mathbf{k},\mathbf{R}|\mathcal{H}_\pm|\mathbf{k},\mathbf{R}' \rangle_{\mathrm{AA}}$ $\leftarrow$ 0
    \ENDFOR
    \FOR{$\mathbf{R}$ with $(m,n)=[-L/2+1,L/2]^2$}
    \STATE ${}_{\mathrm{AA}\vphantom{\mathrm{AA}}}\langle \mathbf{k},\mathbf{R}|\mathcal{H}_\pm|\mathbf{k},\mathbf{R}\pm\bm{\zeta}_1 \rangle_{\mathrm{AA}}$ $\pluseq$ 
    $(J_2/4)e^{\pm i\mathbf{k}\cdot\bm{\zeta}_1/2}$
    $\left\{1+ \left( \delta_{\mathbf{R},\mathbf{O}} + \delta_{\mathbf{R}',\mathbf{O}} - \delta_{\mathbf{R},\mathbf{O}}\delta_{\mathbf{R}',\mathbf{O}} \right)\left( \sqrt{2-\frac{1}{S}}-1\right)\right\}$
    \STATE ${}_{\mathrm{AA}\vphantom{\mathrm{AA}}}\langle \mathbf{k},\mathbf{R}|\mathcal{H}_\pm|\mathbf{k},\mathbf{R}\pm\bm{\zeta}_2 \rangle_{\mathrm{AA}}$ $\pluseq$ 
    $(J_2/4)e^{\pm i\mathbf{k}\cdot\bm{\zeta}_2/2}$
    $\left\{1+ \left( \delta_{\mathbf{R},\mathbf{O}} + \delta_{\mathbf{R}',\mathbf{O}} - \delta_{\mathbf{R},\mathbf{O}}\delta_{\mathbf{R}',\mathbf{O}} \right)\left( \sqrt{2-\frac{1}{S}}-1\right)\right\}$
    \STATE ${}_{\mathrm{AA}\vphantom{\mathrm{AA}}}\langle \mathbf{k},\mathbf{R}|\mathcal{H}_\pm|\mathbf{k},\mathbf{R}\pm\bm{\zeta}_3 \rangle_{\mathrm{AA}}$ $\pluseq$ 
    $(J_2/4)e^{\pm i\mathbf{k}\cdot\bm{\zeta}_3/2}$
    $\left\{1+ \left( \delta_{\mathbf{R},\mathbf{O}} + \delta_{\mathbf{R}',\mathbf{O}} - \delta_{\mathbf{R},\mathbf{O}}\delta_{\mathbf{R}',\mathbf{O}} \right)\left( \sqrt{2-\frac{1}{S}}-1\right)\right\}$
    \ENDFOR
\end{algorithmic}
\end{algorithm}
\end{figure}
\begin{figure}[H]
\begin{algorithm}[H]
\caption{Calculate ${}_{\mathrm{AB}\vphantom{\mathrm{AB}}}\langle \mathbf{k},\mathbf{R}|\mathcal{H}_\pm|\mathbf{k},\mathbf{R}' \rangle_{\mathrm{AB}}$}
\label{alg:4}
\begin{algorithmic}[1]
    \FOR{all $\mathbf{R}$ and $\mathbf{R}'$}
    \STATE ${}_{\mathrm{AB}\vphantom{\mathrm{AB}}}\langle \mathbf{k},\mathbf{R}|\mathcal{H}_\pm|\mathbf{k},\mathbf{R}' \rangle_{\mathrm{AB}}$ $\leftarrow$ 0
    \ENDFOR
    \FOR{$\mathbf{R}$ with $(m,n)=[-L/2+1,L/2]^2$}
    \STATE ${}_{\mathrm{AB}\vphantom{\mathrm{AB}}}\langle \mathbf{k},\mathbf{R}|\mathcal{H}_\pm|\mathbf{k},\mathbf{R}\pm\bm{\zeta}_1 \rangle_{\mathrm{AB}}$ $\pluseq$ 
    $(J_2/4)e^{\pm i\mathbf{k}\cdot\bm{\zeta}_1/2}$
    \STATE ${}_{\mathrm{AB}\vphantom{\mathrm{AB}}}\langle \mathbf{k},\mathbf{R}|\mathcal{H}_\pm|\mathbf{k},\mathbf{R}\pm\bm{\zeta}_2 \rangle_{\mathrm{AB}}$ $\pluseq$ 
    $(J_2/4)e^{\pm i\mathbf{k}\cdot\bm{\zeta}_2/2}$
    \STATE ${}_{\mathrm{AB}\vphantom{\mathrm{AB}}}\langle \mathbf{k},\mathbf{R}|\mathcal{H}_\pm|\mathbf{k},\mathbf{R}\pm\bm{\zeta}_3 \rangle_{\mathrm{AB}}$ $\pluseq$ 
    $(J_2/4)e^{\pm i\mathbf{k}\cdot\bm{\zeta}_3/2}$
    \ENDFOR
\end{algorithmic}
\end{algorithm}
\end{figure}
\noindent ${}_{\mathrm{AB}\vphantom{\mathrm{AB}}}\langle \mathbf{k},\mathbf{R}|\mathcal{H}_\pm|\mathbf{k},\mathbf{R}' \rangle_{\mathrm{BA}}$,
${}_{\mathrm{BA}\vphantom{\mathrm{BA}}}\langle \mathbf{k},\mathbf{R}|\mathcal{H}_\pm|\mathbf{k},\mathbf{R}' \rangle_{\mathrm{AB}}$, and
${}_{\mathrm{BA}\vphantom{\mathrm{BA}}}\langle \mathbf{k},\mathbf{R}|\mathcal{H}_\pm|\mathbf{k},\mathbf{R}' \rangle_{\mathrm{BA}}$ are also evaluated by the Algorithm 4 or its complex conjugate form.
\par It is important to note that, due to the finite-size effect, long-wavelength two-magnon spectrum has a tiny positive finite gap, which vanishes in the thermodynamic limit.
\section{Details of classical spin dynamics simulations}
We summarize the key parameters used in the spin dynamics simulations reported in the main text. All simulations were performed on two-dimensional honeycomb lattices with 80 $\times$ 80 unit cells with 2 atoms per unit cell and periodic boundary conditions in both directions. For CrBr$_3$, the time step size was set $\Delta t = 8.8768 \cdot 10^{-16} \;\mathrm{s}$, and samples were equilibrated with $n_{\mathrm{eq}} = 1-5 \cdot 10^7$ steps, depending on the temperature of the simulation. During the data collection phase, $n_{\mathrm{rec}} = 1 \cdot 10^7$ steps were simulated, with configurations recorded every 50 steps.
In the case of the CrI$_3$, $\Delta t = 3.279 \cdot 10^{-16} \;\mathrm{s}$, while the equilibration and recording steps were equal to those used for CrBr$_3$. For both cases, we averaged over 1000 independent stochastic trajectories to extract the dynamical structure factor.

The equilibration time steps were simulated with damping $\alpha_\textrm{G} = 0.01$ at finite temperature, but the recording phase was evolved microcanonically with $\alpha_\textrm{G} = 0$ and $T = 0$. To model higher-spin systems within the constraint of unit-length vectors, interactions were rescaled by a factor of $S^2$ and set $\mu_s = 2 \sqrt{S (S + 1)}$. The dynamical structure factor was evaluated directly in reciprocal space by Fourier transforming the spin configurations $s$, thus avoiding the explicit construction of real-space spin-spin correlators (see Supplemental Material of Ref.~\cite{PhysRevLett.122.167203}). Specifically,
\begin{equation}
    \tilde{s}^\mu(\mathbf{q}, \omega) = \frac{1}{\sqrt{N}} \sum_i \sum_t 
e^{-\mathrm{i} \mathbf{q} \cdot \mathbf{r}_i + \mathrm{i} \omega t} s^\mu(\mathbf{r}_i, t),
\end{equation}
where $\mu \in \{x, y, z\}$, $N$ is the total number of spins, and the sum runs over all lattice sites $i$ and discrete time steps $t$. The dynamical structure factor $\mathcal{S}^{\mu \nu}(\mathbf{q}, \omega)$ is the given by
\begin{equation}
    \mathcal{S}^{\mu \nu}(\mathbf{q}, \omega) \propto
\left\langle \tilde{s}^\mu(\mathbf{q}, \omega) \tilde{s}^\nu(-\mathbf{q}, -\omega) \right\rangle.
\end{equation}

\section{Temperature evolution of the Hartree+full sunset self-energy}
\label{Appendix:Hartree_fullSunset_tempev}

\begin{figure*}[tbh]
    \centering
    \includegraphics[width=\textwidth]{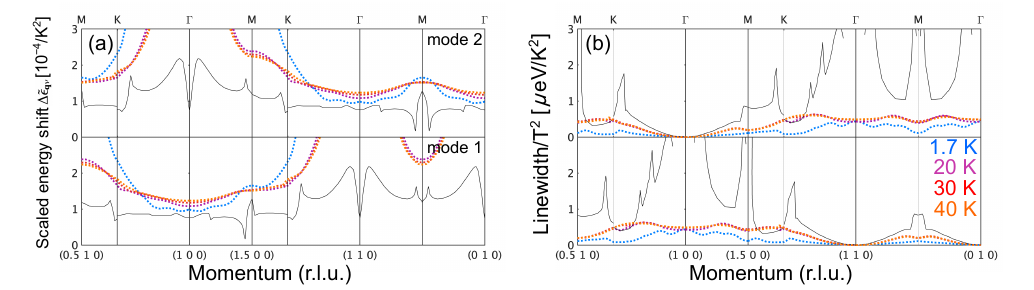}
    \caption{
    Calculated (a) magnon energy shift and (b) linewidth of CrBr${}_3$ obtained with the Hartree+full sunset approximation.
    Thin black lines show energy shift and linewidth obtained from the Hatree+reduced sunset self-energy, reproduced from Ref.~\cite{Nikitin2022}.
    } 
    \label{FigS04}
\end{figure*}

Magnon energy shift and linewidth of CrBr${}_3$ at various temperatures estimated using the Hartree+full sunset approximation are presented in Fig.~\ref{FigS04}. Data at 20~K, 30~K, and 40~K follow the $T^2$-behavior discussed in Ref.~\cite{Nikitin2022}, while that at 1.7~K deviates from it.

\twocolumngrid
\bibliography{bib}

\end{document}